\documentclass[]{aa}  

\usepackage{array, multirow, graphicx}
\usepackage{xcolor}
\usepackage{wasysym}[integrals]
\usepackage{siunitx}
\usepackage{booktabs, tabularx}

\usepackage{txfonts}
\usepackage[]{hyperref}
\hypersetup{colorlinks,citecolor=blue, linkcolor=blue, urlcolor=blue}

\newcounter{magicrownumbers}

\usepackage{cellspace, makecell} 

\begin{document} 
%%%%%%%%%%%%%%%%%%%%%% my newcommands
\newcommand{\bcdot}{\boldsymbol{\cdot}}
\newcommand{\w}[1]{\mathbf{#1}}
\newcommand{\bra}[1]{\left\langle #1 \middle | \right.}
\newcommand{\ket}[1]{\left.\middle | #1\right\rangle}
\newcommand{\braket}[2]{\left\langle #1  \middle | #2 \right\rangle}
\newcommand{\oplin}[2]{\left.\middle | #1 \right \rangle \left \langle #2 \middle | \right.}
\newcommand{\lf}{\left}
\newcommand{\rg}{\right}
\newcommand{\tonda}[1]{\!\left ( #1 \right )}
\newcommand{\quadra}[1]{\left [ #1 \right ]}
\newcommand{\graffa}[1]{\left \{ #1 \right \}}
\newcommand{\dirac}[1]{\updelta\!\left( #1 \right )}
\newcommand{\deltak}[1]{\updelta_{ #1 }}
\newcommand{\D}[2]{\frac{d #1}{d  #2}}
\newcommand{\DD}[2]{\frac{d^2 #1}{d #2^2}}
\newcommand{\pd}[2]{\frac{\partial #1}{\partial #2}}
\newcommand{\pdd}[2]{\frac{\partial^2 #1}{\partial #2^2}}
\newcommand{\medio}[1]{\left\langle #1 \right\rangle}
\newcommand{\abs}[1]{\left | #1 \right |}
\newcommand{\norma}[1]{\left \| #1 \right \|}
\newcommand{\sca}[2]{\left \langle #1 , #2 \right\rangle}
\newcommand{\ndiv}[1]{{\boldsymbol{\nabla}}\boldsymbol{\cdot}\w{#1}}
\newcommand{\ndivx}[1]{{\boldsymbol{\nabla}_{\w{x}}}\boldsymbol{\cdot}\w{#1}}
\newcommand{\ndivv}[1]{{\boldsymbol{\nabla}_{\w{v}}}\boldsymbol{\cdot}\w{#1}}
\newcommand{\grad}{{\boldsymbol{\nabla}}}
\newcommand{\gradx}{{\boldsymbol{\nabla}}_{\w{x}}}
\newcommand{\gradv}{{\boldsymbol{\nabla}}_{\w{v}}}
\newcommand{\lap}[1]{\boldsymbol{\nabla}^2 #1}
\newcommand{\rot}[1]{\boldsymbol{\nabla}\times\w{#1}}
\newcommand{\pois}[2]{\Bigl \{ #1 , #2 \Bigr\}}
\newcommand{\com}[2]{\left [ #1 , #2 \right ]}
\newcommand{\sistemai}{\lf\{\begin{aligned}}
\newcommand{\sistemaf}{\end{aligned}\rg.}
\newcommand{\rn}{\mathtt{I}}
\newcommand{\rnn}{\mathtt{I\!I}}
\newcommand{\rnnn}{\mathtt{I\!I\!I}}
\newcommand{\implica}{\Longrightarrow}
\newcommand{\sse}{\Longleftrightarrow}
\newcommand{\id}{\mathbb{I}}
\newcommand{\sopra}[1]{\overline{#1}}
\newcommand{\sotto}[1]{\underline{#1}}
\newcommand{\gv}[1]{\ensuremath{\mbox{\boldmath$ #1 $}}}
\newcommand{\calc}[3]{\lf.{#1}\rg |_{#2}^{#3}}
\newcommand{\cro}{\dagger}
\newcommand{\eq}[1]{\begin{equation} #1 \end{equation}}
\newcommand{\eqn}[1]{\begin{equation*} #1 \end{equation*}}
\newcommand{\virg}[1]{``#1''}
\newcommand{\dmat}[1]{\frac{\mathcal{D} #1}{\mathcal{D}t}}
\newcommand{\mc}[1]{\mathcal{#1}}
\newcommand{\bxi}{\boldsymbol{\xi}}
\newcommand{\vx}{\hat{\w{x}}}
\newcommand{\vy}{\hat{\w{y}}}
\newcommand{\vz}{\hat{\w{z}}}
\newcommand{\vn}{\hat{\w{n}}}
\newcommand{\disc}[1]{\biggl[ #1 \biggr]}
\newcommand{\td}{\tau_{\text{diff}}}
\newcommand{\tc}{\tau_{\text{conv}}}
\newcommand{\ft}{\tilde{f}_1}
\newcommand{\fth}{\widehat{f}_1}
\newcommand{\et}{\tilde{\w{E}}_1}
\newcommand{\eet}{\tilde{E}_1}
\newcommand{\eeth}{\widehat{E}_1}
\newcommand{\asun}{\astrosun}
\newcommand{\al}[1]{\begin{aligned} #1 \end{aligned}}
\newcommand{\ftonda}[2]{\!\left ( \frac{#1}{#2} \right )}
\newcommand{\mum}{\unit{\mu m}}
\newcommand{\hh}{H$_{2}$ }
\newcommand{\Mbh}{M_\text{BH}}
\newcommand{\bs}[1]{\boldsymbol{#1}}
\newcommand{\fit}{\emph{fit}\xspace}
\newcommand{\pixel}{\emph{pixel}\xspace}
%newcommand to write the line labels
\newcommand{\cii}{{\rm [CII]}}
\newcommand{\nii}{{\rm [NII]}}
\newcommand{\ci}{{\rm [CI]}}
\newcommand{\co}{{\rm CO}}
\newcommand{\hdo}{{\rm H$_2$O}}
\newcommand{\oh}{{\rm OH}}
\newcommand{\hdue}{{\rm H$_2$}}
\newcommand{\red}[1]{\textcolor{red}{#1}}

   \title{An ALMA multi-line survey of the ISM in two quasar host--companion galaxy pairs at $z>6$}

%   \subtitle{I. Overviewing the $\kappa$-mechanism}

   \author{A.~Pensabene
          \inst{\ref{difa},\ref{inaf-bo}}
%          \thanks{\email{antonio.pensabene2@unibo.it}}
          	\and
          R.~Decarli
          \inst{\ref{inaf-bo}}
          	\and
          E.~Ba{\~{n}}ados
          \inst{\ref{mpia}}
         	 \and
          B.~Venemans
          \inst{\ref{mpia}}
                 \and
	  F.~Walter
	  \inst{\ref{mpia}, \ref{nrao}}
	  	\and
          F.~Bertoldi
          \inst{\ref{uni-bonn}}
          	\and
          X.~Fan
          \inst{\ref{uni-az}}
        	  	\and
          E.~P.~Farina
          \inst{\ref{mpia-garching}}
          	\and
	  J.~Li
	  \inst{\ref{uni-china},\ref{kavli-china}}
	  	\and
          C.~Mazzucchelli
          \inst{\ref{eso-chile}}
                 \and
          M.~Novak
          \inst{\ref{mpia}}
          	\and
	 D.~Riechers
	 \inst{\ref{cornell}}
	 	\and
	 H.-W.~Rix
	 \inst{\ref{mpia}}
	 	\and
          M.~A.~Strauss
          \inst{\ref{princeton}}
          	\and        
          R.~Wang
          \inst{\ref{kavli-china}}
          	\and
	  A.~Wei{\ss}
	 \inst{\ref{mpi-bonn}}
	 	\and
          J.~Yang
          \inst{\ref{uni-az}}
          	\and
          Y.~Yang
          \inst{\ref{kassi}}
%          \fnmsep\thanks{Just to show the usage
%          of the elements in the author field}
          }

   \institute{Dipartimento di Fisica e Astronomia, Alma Mater Studiorum, Universit\`a di Bologna, Via Gobetti 93/2, I-40129 Bologna, Italy\\\email{antonio.pensabene2@unibo.it}\label{difa}
		\and
   	INAF-Osservatorio di Astrofisica e Scienza dello Spazio, Via Gobetti 93/3, I-40129 Bologna, Italy\label{inaf-bo}
		\and
	Max-Planck-Institut f\"{u}r Astronomie, K\"{o}nigstuhl 17, D-69117 Heidelberg, Germany\label{mpia}
		\and
	National Radio Astronomy Observatory, Pete V. Domenici Array Science Center, P.O. Box O, Socorro, NM 87801, USA\label{nrao}
		\and
	Argelander-Institut f\"ur Astronomie, University at Bonn, Auf dem H\"ugel 71, D-53121 Bonn, Germany\label{uni-bonn}
		\and
	Steward Observatory, University of Arizona, 933 North Cherry Avenue, Tucson, AZ 85721, USA\label{uni-az}
		\and
	Max Planck Institut f\"ur Astrophysik, Karl--Schwarzschild--Stra{\ss}e 1, D-85748, Garching bei M\"unchen, Germany\label{mpia-garching}
		\and
	Department of Astronomy, School of Physics, Peking University, 5 Yiheyuan Road, Haidian District, Beijing, 10087, China\label{uni-china}
		\and
	Kavli Institute for Astronomy and Astrophysics, Peking University, 5 Yiheyuan Road, Haidian District, Beijing, 100871, China\label{kavli-china}
		\and
	European Southern Observatory, Alonso de C\'ordova 3107, Casilla 19001, Vitacura, Santiago 19, Chile\label{eso-chile}
		\and
	Department of Astronomy, Cornell University, Space Sciences Building, Ithaca, NY 14853, USA\label{cornell}
		\and
	Department of Astrophysical Sciences, Princeton University, Princeton, New Jersey 08544, USA\label{princeton}
		\and
	Max-Planck-Institut f\"ur Radioastronomie, Auf dem H\"ugel 69, D-53121 Bonn, Germany\label{mpi-bonn}
		\and
	Korea Astronomy and Space Science Institute, 776 Daedeokdae-ro, Yuseong-gu, Daejeon 34055, Korea\label{kassi}
             }

   \date{Received XXX accepted YYY}

\abstract{%
We present a multi-line survey of the interstellar medium (ISM) in two $z>6$ quasar (QSO) host galaxies, PJ231-20 ($z=6.59$) and PJ308-21 ($z=6.23$), and their two companion galaxies. Observations were carried out using the Atacama Large (sub-)Millimeter Array (ALMA). We targeted eleven transitions including atomic fine structure lines (FSLs) and molecular lines: [NII]$_{\rm 205\mu m}$, [CI]$_{\rm 369\mu m}$, CO ($J_{\rm up} = 7, 10, 15, 16$), H$_2$O $3_{12}-2_{21}$, $3_{21}-3_{12}$, $3_{03}-2_{12}$, and the OH$_{\rm 163\mu m}$ doublet. The underlying far-infrared (FIR) continuum samples the Rayleigh-Jeans tail of the respective dust emission. By combining this information with our earlier ALMA [CII]$_{\rm 158\mu m}$ observations, we explore the effects of star formation and black hole feedback on the galaxies' ISM using the CLOUDY radiative transfer models. We estimate dust masses, spectral indexes, IR luminosities, and star-formation rates from the FIR continuum. The analysis of the FSLs indicates that the [CII]$_{\rm 158\mu m}$ and [CI]$_{\rm 369\mu m}$ emission arises predominantly from the neutral medium in photodissociation regions (PDRs). We find that line deficits are in agreement with those of local luminous infrared galaxies. The CO spectral line energy distributions (SLEDs), reveal significant high-$J$ CO excitation in both quasar hosts. Our CO SLED modeling of the quasar PJ231-20 shows that PDRs dominate the molecular mass and CO luminosities for $J_{\rm up}\le 7$, while the $J_{\rm up}\ge10$ CO emission is likely driven by X-ray dissociation regions produced by the active galactic nucleus (AGN) at the very center of the quasar host. The $J_{\rm up}>10$ lines are undetected in the other galaxies in our study. The H$_2$O $3_{21}-3_{12}$ line detection in the same quasar places this object on the $L_{\rm H_2O}-L_{\rm TIR}$ relation found for low-$z$ sources, thus suggesting that this water vapor transition is predominantly excited by IR pumping. Models of the H$_2$O SLED and of the H$_2$O/OH$_{\rm 163\mu m}$ ratio point to PDR contributions with high volume and column density ($n_{\rm H}\sim 0.8\times10^5\,{\rm cm^{-3}}$, $N_{\rm H}=10^{24}\,{\rm cm^{-2}}$), in an intense radiation field. The presented analysis suggests a less excited medium in the companion galaxies. However, the current data do not allow us to definitively rule out the presence of an AGN in these sources as suggested by previous studies of the same objects. This work demonstrates the power of multi-line studies of FIR diagnostics in order to dissect the physical conditions in the first massive galaxies emerging from cosmic dawn.}

               
   \titlerunning{An ALMA multi-line survey of the ISM in two $z>6$ QSO host--companion galaxy pairs}
   \authorrunning{A. Pensabene et al.}
   \maketitle
   
%-------------------------------------------------------------------

\section{Introduction}
\label{sect:introduction}
Quasars (or QSOs) beyond redshift $z\sim 6$ (when the age of the Universe was less than $1\,{\rm Gyr}$) are the most luminous non-transient sources in the Universe, and are believed to be the progenitors of present-day early-type massive galaxies. The very high luminosities ($L_{\rm bol} > 10^{13}L_{\astrosun}$) from their active nuclei (AGN) are powered by rapid accretion of matter ($> 10\,M_{\astrosun}\,{\rm yr^{-1}}$) onto central supermassive black holes (BHs, $M_{\rm BH}\apprge 10^{8}\,M_{\astrosun}$; \citealt{Jiang+2007, DeRosa+2011, Mazzucchelli+2017}) and often accompanied by rapid consumption of huge gas reservoirs through vigorous episodes of star formation (${\rm SFR} > 100\,M_{\astrosun}\,{\rm yr^{-1}}$; see e.g. \citealt{Bertoldi+2003b, Bertoldi+2003a, Walter+2003, Walter+2009nat, Venemans+2018}). Therefore, the host galaxies of $z>6$ quasars are ideal laboratories to characterize the physical properties of the interstellar medium (ISM) in such extreme conditions, and they provide crucial insights on the interplay between star formation and black hole accretion in massive galaxies emerging from cosmic dawn.

The redshift $z>6$ range is of particular interest as galaxies emerge from the last phase transition of the Universe, marked by a rapid shift in the ionization properties of the intergalactic medium \citep{Fan+2006, Becker+2015}.
Thanks to the efforts of various groups using wide field surveys (\mbox{see e.g. \citealt{Fan+2006b}}, \citealt{Banados+2016, Jiang+2016, Matsuoka+2018}, and references therein), $\sim 200$ quasars have been discovered at $z>6$, including the three most distant quasars known at $z>7.5$ \citep{Banados+2018, YangJ+2020, Wang+2021}. The very presence of such BHs at these early epochs requires rapid build-up of black holes and galaxy assembly, challenging our understanding of how BHs form in the first place \citep[e.g.][]{Volonteri2012, Sijacki+2015,Habouzit+2019,vanderVlugt+2019}.

Models and numerical simulations of massive black hole formation predicts that $z>6$ quasars reside the extreme peaks of the large-scale density structure \citep[e.g][]{Begelman+2006, Overzier+2009, Bonoli+2009, Bonoli+2014, Angulo+2012, Costa+2014}, and therefore, $z>6$ quasars can be used to identify the first galaxy overdensities -- however the question whether there are in fact overdensities of galaxies around $z\sim6$ quasars is contested -- \citep[e.g.][]{Overzier+2009, Morselli+2014, Balmaverde+2017, Farina+2017, Mazzucchelli+2017b, Decarli+2019b}. On smaller scale ($\sim 100 \,{\rm pkpc}$), the exceptional sensitivity and imaging power of the Atacama Large (sub-)Millimeter Array (ALMA) has allowed us to image the dust and cold gas reservoir of galaxies in the early Universe, leading to strong evidence of overdensities around $z>5-6$ quasars \citep[e.g.][]{Decarli+2017, Trakhtenbrot+2017, Willott+2017, Neeleman+2019, Venemans+2019}, supported by the intriguingly serendipitous discoveries of starburst and sub-millimeter galaxies (SMGs) in the close environment of the quasars, with high star formation rate (${\rm SFR} > 500-1000\,M_{\astrosun}\,{\rm yr^{-1}}$) and showing no evidence of BH accretion (\citealt{Decarli+2017, Mazzucchelli+2019}; but see also \citealt{Connor+2019, Connor+2020, Vito+2019} for tentative evidence of X-ray emission from these companions).

In this context, these quasar host--companion galaxy pairs represent unique testing ground to study star formation, AGN feedback and structure growth in galaxies at $z>6$. Multi-line surveys of the ISM in these systems provide key diagnostics of the physical properties of the ISM, including densities, and the source of the powering radiation field \citep[see][for a review]{CarilliWalter2013}. For high-redshift galaxies, many atomic and molecular far-infrared (FIR) emission lines fall in the (sub-)mm bands, thus making them accessible to sensitive ground-based facilities such as ALMA or the Northern Extended Millimeter Array (NOEMA). Indeed, previous studies detected galaxy dust continuum and gas emission lines even in highest redshift quasars. These studies have mainly targeted the brightest emitters such as the fine-structure line (FSL) of singly-ionized carbon [CII]$_{\rm 158\mu m}$, and the rotational transitions of carbon monoxide \citep[CO; see e.g.][]{Walter+2003,Walter+2004, Walter+2009nat, Maiolino+2005, Maiolino+2009, Maiolino+2015,Wang+2010,Wang+2013, Willott+2015mbh, Willott+2015, Gallerani+2014, Carniani+2017, Carniani+2019, Decarli+2017, Decarli+2018, Venemans+2017b, Venemans+2017c, Venemans+2017a, Venemans+2019, Feruglio+2018}, and have revealed the ubiquitous presence of massive cold gas ($M_{\rm gas}\apprge 10^{10}\,M_{\astrosun}$) and dust ($M_{\rm dust}\apprge 10^{9}\,M_{\astrosun}$) reservoirs in high-$z$ quasar host galaxies. Indeed, [CII]$_{\rm 158\mu m}$ and the low-/mid-$J$ CO transitions (rotational quantum number of the CO molecule upper levels $J_{\rm up}< 7-8$) are the major coolants of the atomic and molecular ISM, respectively. %
The C$^+$ ion has an ionization potential energy slightly lower than that of atomic hydrogen (H), thus it traces both the atomic neutral and ionized phase of the ISM, and its $158\,{\rm \mu m}$ transition arises predominantly from photodissociation regions (PDRs, see \citealt{Hollenbach+1999} for a review) at the interface between the atomic and molecular media in the outskirts of molecular clouds in galaxy star-forming regions \citep[see e.g.][]{Diaz-Santos+2017}. On the other hand, the CO is the most abundant molecule of the ISM after H$_2$, and requires little energy to be excited, thus its low- and mid-$J$ upper levels can be easily populated by collisions with H$_2$. It is thus representing the best observational tracer of cold molecular ISM embedded in clouds (with a hydrogen density of $n_{\rm H}\sim 10^{3}\,{\rm cm^{-3}}$, and a kinetic gas temperature of $T<100\,{\rm K}$).

Multiple atomic FSLs and molecular transitions arising from different ISM phases provide both direct and indirect information on the ionized phase in the proximity of young (O and B) stars (e.g. the doubly-ionized oxygen [OIII]$_{\rm 88\mu m}$ and the singly-ionized nitrogen [NII]$_{\rm 122\mu m,\,205\mu m}$) or the denser and warm phases buried in the cores of molecular clouds (such e.g. the high-$J$ CO, the water vapor H$_2$O, and the hydroxyl and its anion (hydroxide) OH, OH$^+$, rotational transitions). Recent studies have started to probe such diagnostics in few $z>6$ systems \citep[e.g.][]{Riechers+2009, Riechers+2013, Strandet+2017, Venemans+2017c, Venemans+2017a, Walter+2018, Hashimoto+2019a, Hashimoto+2019b, Novak+2019, WangF+2019, YangJ+2019,Li+2020b, Li+2020}.

This work is focused on the ISM characterization of two $z>6$ quasar host galaxies, \object{PJ231-20} at $z\sim6.59$, and \object{PJ308-21} at $z\sim6.24$, and the nearby companions which were serendipitously discovered with ALMA \citep{Decarli+2017, Neeleman+2019}. These companions are among the most star-forming galaxies known to date (with [CII]--based ${\rm SFR} \apprge 200-500\,M_{\astrosun}\,{\rm yr^{-1}}$) at $z>6$ that do not show evidence of AGN activity in the rest-frame optical/UV \citep{Decarli+2017, Mazzucchelli+2019} and in the X-rays \citep{Connor+2019, Connor+2020}. For these objects, the [CII]$_{\rm 158\mu m}$ line and dust continuum have been clearly detected with previous ALMA observations \citep{Decarli+2017}, while higher-resolution [CII]$_{\rm 158\mu m}$ follow-up observations have allowed study of their kinematics \citep[see][]{Decarli+2019, Neeleman+2019}. The companion galaxies in the two systems, have a projected separation of $\sim 8.5\,{\rm kpc}$ and $\sim 14\,{\rm kpc}$, with a velocity offset of $+137\,{\rm km\,s^{-1}}$ and $+591\,{\rm km\,s^{-1}}$ for PJ231-20 and PJ308-21, respectively. Remarkably, the [CII]$_{\rm 158\mu m}$ emission of the companion galaxy in PJ308-21 system, stretches over about $25\,{\rm kpc}$ and $>1500\,{\rm km\,s^{-1}}$ towards and beyond the quasar host suggesting that the satellite galaxy is tidally stripped by the interaction with the central quasar host galaxy \citep{Decarli+2017,Decarli+2019}. Furthermore, in both PJ231-20 and PJ308-21 systems, Ly$\alpha$ nebular emission have been recently discovered \citep{Farina+2019}, possibly indicating that also PJ231-20 quasar recently underwent a merger event with the close companion galaxy.

Here we present ALMA band 3--6 observations of the PJ231-20 and PJ308-21 systems, in which we sampled the FIR dust continuum and various emission lines probing different ISM phases and conditions; including the FSLs of the singly-ionized atomic nitrogen [NII]$_{\rm 205\mu m}$, and the neutral carbon [CI]$_{\rm 369\mu m}$, tracers of the low-density fully ionized medium (electron density $n_{\rm e} \apprle 1\,{\rm cm^{-3}}$), and the cold dense atomic phase of the ISM ($n_{\rm H}\sim 10^3\,{\rm cm^{-3}}$), respectively. We also present observations of the molecular transitions of CO at intermediate and high-$J$ (7--6, 10--9, 15--14, 16--15), three water vapor (ortho-)H$_2$O rotational transitions ($3_{12}-2_{21}$, $3_{21}-3_{12}$, $3_{03}-2_{12}$) and the hydroxyl molecule OH$_{\rm 163\mu m}$ doublet associated with the warm dense phase of the ISM  ($n_{\rm H}\apprge 10^5\,{\rm cm^{-3}}$, $T>100\,{\rm K}$). Combining the different information and making use of radiative transfer models, we study a variety of dust and gas-phase ISM physical properties in the quasar hosts and companion galaxies, allowing to discriminate line excitation by star formation and AGN processes. This study aims to provide an overview of the complex conditions of the multiphase ISM in galaxies at $z\sim6$.

\setcounter{magicrownumbers}{0}
\begin{table*}
\caption{Information on targeted objects and characteristics of ALMA observations for each frequency setup.}            
\label{tbl:obs_data}      
\centering      
\resizebox{\hsize}{!}{
\begin{tabular}{ l c c c c | c c} 
\toprule\toprule
{\rm Object ID}					&\multicolumn{4}{c|}{\large \bf PJ231-20}										&\multicolumn{2}{c}{\large \bf PJ308-21}\\
			&\multicolumn{2}{c}{\bf QSO}				&\multicolumn{2}{c|}{\bf companion $^{(\star)}$}			&\multicolumn{1}{c}{\bf QSO}				&\multicolumn{1}{c}{\bf companion}\\
\cmidrule(lr){1-7}
R.A. (J2000.0)	&\multicolumn{2}{c}{15h 26m 37.84s}	&\multicolumn{2}{c|}{15h 26m 37.87s}	&\multicolumn{1}{c}{20h 32m 10.00s}	&\multicolumn{1}{c}{20h 32m 10.17s}\\
DEC. (J2000.0)	&\multicolumn{2}{c}{$-20^{\circ}\,50'\,0.8''$}&\multicolumn{2}{c|}{$-20^{\circ}\,50'\,2.3''$}&\multicolumn{1}{c}{$-21^{\circ}\,14'\,2.4''$}&\multicolumn{1}{c}{$-21^{\circ}\,14'\,2.7''$}\\
redshift $^{(1)}$&\multicolumn{2}{c}{$6.58651\pm0.00017$}		&\multicolumn{2}{c|}{$6.5900\pm0.0008$}			&\multicolumn{1}{c}{$6.2342\pm0.0010$}			&\multicolumn{1}{c}{$6.2485\pm0.0005$}
\\
\bottomrule
\toprule
{\bf Frequency setup} $^{(2)}$								&{\rm A}			&{\rm B}		&{\rm C} 		&{\rm D}		&{\rm A}		&{\rm D}\\		
\cmidrule(lr){1-7}								  																	
Band													& 3				& 4				& 5				& 6				& 3				& 6\\
Central frequency (GHz) $^{(3)}$							&99.755			&146.456			&199.350			&234.397			&104.901			&245.870\\			
Time on source (min)									&33.82			&$19.70$			&$28.27$			&8.72			&41.97  			&$39.83$\\
Array configuration										&C43-4			&C43-2, C43-3		&C43-2			&C43-2			&C43-4			&C43-2\\ 			
No. antennas											&$43$			&$48$                 	&$48	$			&$45$			&48				&$44	$\\			
Baselines (m)											&$15-1547$		&$15-500$		&$15-360$		&$15-483$		&$15-783$		&$15-455$\\		
Beam (${\rm arcsec^2}$)	$^{(4)}$								&$0''.99\times0''.89$	&$1''.49\times1''.22$	&$1''.76\times1''.14$	&$1''.10\times1''.01$	&$1''.32\times1''.13$	&$1''.19\times0''.97$\\	
RMS cont. (${\rm mJy\,beam^{-1}}$)							&$0.011$			&$0.019$			&$0.029$        		&$0.042$			&$0.009$			&$0.015$\\			
$\left\langle{\rm RMS}\right\rangle$ cube (${\rm mJy\,beam^{-1}}$) $^{(5)}$	&$0.24$			&$0.27$			&$0.47$			&$0.48$			&$0.19$			&$0.20$\\			
Observation date										&2019 Oct 6		&2018 Apr 23		&2018 Jun 4		&2018 Aug 21		&2019 Oct 14		&2018 Aug 12\\		
\bottomrule                 
\end{tabular}
}
\tablefoot{$^{(1)}$Redshift of ${\rm [CII]_{158\mu m}}$ line from \citet{Decarli+2017}. $^{(2)}$Setup A refers to ALMA Cycle 7 data, while setup C, B and D refer to Cycle 5 data. $^{(3)}$Central frequency of the entire frequency setup. $^{(4)}$Beam sizes are those from the natural maps. $^{(5)}$Averaged RMS sensitivity over the entire bandwidth of the cubes (channel width of $40\,{\rm km\,s^{-1}}$). $^{(\star)}$We note that \citet{Neeleman+2019} found two companions in the PJ231-20 ALMA field located at south and south-est of the QSO. However, our observations do not allow us to spatially resolved both the sources. In this work we assume that the bulk of the emission from companion galaxies, arises from the source located at the south of the central PJ231-20 QSO, that is reported in \citet{Decarli+2017} and denoted with 'C' in \citet{Neeleman+2019} work.}%
\end{table*}

The paper is organized as follows: in Sect.~\ref{sect:data_reduction} we present the ALMA observations and we describe the reduction of the data. In Sect.~\ref{sect:data_presentation} we outline the analysis conducted on the calibrated data and we report FIR line emission and continuum measurements in the sources of the PJ231-20 and PJ308-21 systems. In Sect.~\ref{sect:cloudy_modeling} we describe the set-up and the outputs of our CLOUDY models. In Sect.~\ref{ssect:dust} we present and discuss the results obtained from the dust continuum. Then, we dedicate Sect.~\ref{sect:ionized_medium} to the tracers of the ionized medium and the results obtained from them, while in Sect.~\ref{sect:atomic_medium} we focused on the atomic medium and we put constraints on its physical properties. Afterward, in Sect.~\ref{ssect:molecular_medium} we study the molecular phase of the ISM through the analysis of the CO, H$_2$O and OH lines. In Sect.~\ref{sect:mass_contributions} we estimate the various mass budgets of the different gas components. Finally, in the last Sect.~\ref{sect:conclusions} we summarize our results and we draw the conclusions.

Throughout the paper we assume a standard $\Lambda\rm{CDM}$ cosmology with $H_0=69.3\,\si{km\,s^{-1}Mpc^{-1}}$, $\Omega_{m}=0.287$, $\Omega_{\Lambda}=1-\Omega_{m}$ from \citet{Hinshaw+2013}.

\setcounter{magicrownumbers}{0}
\begin{table}
\caption{Covered emission lines in each frequency setup with their rest frequency.}         
\label{tbl:obs_line}      
\centering      
\resizebox{\hsize}{!}{
\begin{tabular}{ l c c }    
\toprule\toprule
{\bf Covered emission lines} & {\bf Rest frequency (GHz) $^{(1)}$} & {\bf setup ID $^{(2)}$} \\
\cmidrule(lr){1-3}
CO($7-6$) & 806.652 &  \multirow{2}{*}{\Large A}\\ 
${\rm [CI]_{369\mu m}}$ & 809.342 & \\
\cmidrule(lr){1-3}
CO($10-9$) & 1151.985 &  \multirow{3}{*}{\Large B}\\
${\rm H_{2}O}\;3_{12}-2_{21}$ & 1153.127 &\\ 
${\rm H_{2}O}\;3_{21}-3_{12}$ & 1162.912 &\\ 
\cmidrule(lr){1-3}
${\rm [NII]_{\rm 205 \mu m}}$ & 1461.131 & \Large C\\
\cmidrule(lr){1-3}
${\rm H_{2}O}\;3_{03}-2_{12}$ & 1716.770 &\multirow{5}{*}{\Large D}\\
CO($15-14$) & 1726.603 & \\ 
OH $^2\Pi_{1/2}$,$\frac{3}{2}^+\!-\!\frac{1}{2}^-$ $^{(\star)}$ & 1834.744 &\\
OH $^2\Pi_{1/2}$,$\frac{3}{2}^-\!-\!\frac{1}{2}^+$ $^{(\star)}$ & 1837.800 &\\
CO($16-15$) & 1841.346 &\\
\bottomrule                 
\end{tabular}
}
\tablefoot{$^{(1)}$Rest frequency of the lines are taken from the "Splatalogue database for astronomical spectroscopy" available at \url{https://www.cv.nrao.edu/php/splat}, see references therein.  $^{(2)}$Setup IDs are listed in Table~\ref{tbl:obs_data}. $^{(\star)}$The $\Lambda-$doubling transitions of the Hydroxyl molecule listed here are actually
two triplets produced by the hyper-fine structure splitting of the upper and lower level, that are not spectral resolved in the observations. Here we report the average rest frequency of the triplets. The six transitions are OH $^2\Pi_{1/2}$,$\frac{3}{2}^+\!-\!\frac{1}{2}^-$,$1\!-\!1$; OH $^2\Pi_{1/2}$,$\frac{3}{2}^+\!-\!\frac{1}{2}^-$,$2\!-\!1$; OH $^2\Pi_{1/2}$,$\frac{3}{2}^+\!-\!\frac{1}{2}^-$,$1\!-\!0$ and OH $^2\Pi_{1/2}$,$\frac{3}{2}^-\!-\!\frac{1}{2}^+$,$1\!-\!1$;  OH $^2\Pi_{1/2}$,$\frac{3}{2}^-\!-\!\frac{1}{2}^+$,$2\!-\!1$; OH $^2\Pi_{1/2}$,$\frac{3}{2}^-\!-\!\frac{1}{2}^+$,$1\!-\!0$.}
\end{table}

\section{Observations and data reduction}
\label{sect:data_reduction}
We present the ALMA Cycle 5 datasets of quasar PJ231-20, quasar PJ308-21 (hereafter PJ231-20 QSO and PJ308-21 QSO, respectively) and their companion galaxies (program ID: 2017.1.00139.S, PI: R. Decarli) located at redshift $z\sim6.59$ and $z\sim6.24$, respectively. The observations were executed during the period April-August 2018 using 43-48 ALMA 12m-antennas in compact configurations (C43-2, C43-3). The program comprised three frequency settings in each of ALMA bands 4, 5 and 6 respectively, that cover nine atomic fine-structure and molecular lines; [NII]$_{205 \rm \mu m}$, high-$J$ CO(10--9, 15--14, 16--15), H$_2$O ($3_{03}-2_{12}$, $3_{12}-2_{21}$, $3_{21}-3_{12}$), OH doublet, together with dust continuum emission. However, for PJ308-21, only the band 6 setup encompassing CO(15--14, 16--15), H$_2$O $3_{03}-2_{12}$ and the OH$_{\rm 163\mu m}$ doublet was executed. 

We also present follow-up ALMA Cycle 7 observations of the same objects (program ID: 2019.1.00147.S, PI: R. Decarli) in which we sampled the atomic FSL [CI]$_{369 \rm \mu m}$, and the molecular mid-$J$ CO(7--6) transition together with dust continuum emission in ALMA band 3. These observations were executed in October 2019 using 43-48 ALMA 12m-antennas in the C43-4 configuration. All the observations were designed not to spatially resolve the emission of each source, but with high enough angular resolution to resolve the companions from the quasars. This allowed us to maximize the signal-to-noise ratio (S/N) within the beam. The angular resolution ranges between $\sim1''.00$ and $\sim1''.80$. In Table~\ref{tbl:obs_data} we summarize the ALMA observations and in Table~\ref{tbl:obs_line} we report the covered emission lines in each frequency setup together with their rest frequencies. 

We performed data reduction using the default calibration pipeline with the Common Astronomy Software Applications (CASA, v.5.1.1-5 and v.5.6.1-8 for Cycle 5 and Cycle 7 data respectively) package \citep{McMullin+2007}. Calibrated visibilities were imaged using the CASA task \texttt{tclean} with "natural" weighting scheme in order to maximize the sensitivity of the resulting maps. Dirty images were cleaned employing two circular masks superimposed to include continuum emission of both the central quasar and the companion. In the cleaning run, we stopped the procedure when the peak flux in the residual image within the mask dropped close to the root-mean-square (RMS) noise of the image in regions with no source emission (i.e., $\sim$ 1$\sigma$ cleaning threshold). For each setup, we produce two image-frequency cubes for the lower-side band (LSB) and the upper-side band (USB) respectively, with a channel width of $40\,{\rm km\,s^{-1}}$. We obtained continuum images by combining the line-free channels from all spectral windows in multifrequency synthesis mode. The line-free channels were determined by inspecting the visibilities in all the frequency sidebands. In the case where no line was detected, we selected line-free channels by adopting a line width of $300\,{\rm km\,s^{-1}}$ and source redshift based on the [CII]$_{\rm 158\mu m}$ line observations published by \citet{Decarli+2017}.  We used the CASA task \texttt{uvcontsub} to fit the continuum visibilities in the line-free channels and we obtained continuum-subtracted cubes with $40\,{\rm km\,s^{-1}}$ of channel width. We fitted the continuum emission in the LSB and USB, separately, (maximum bandwidth of $\sim4\,{\rm GHz}$) as a constant with frequency. This gave us two continuum-subtracted cubes for each frequency setups. In order to achieve Nyquist-Shannon sampling, we set the cube pixel sizes in the image plane to $0.1\si{\arcsecond}$, equal to $\sim 1/7$ of the FWHM of the minor axis of the ALMA synthesized beam. Therefore, we obtained cubes and continuum images with a pixel size of $0.1\si{\arcsecond}$. Self-calibration was attempted but showed no additional improvement for any of the observations, so we did not use it for what follows. Finally, both continuum images and line cubes were corrected for the primary beam response. In this work we supplement our analysis making use of the ALMA Cycle 3 [CII]$_{158\rm \mu m}$ observations of PJ231-20 and PJ308-21 presented in \citealt{Decarli+2017} (program ID: 2015.1.01115.S), that we analyzed in a consistent way.

\setcounter{magicrownumbers}{0}
\begin{table*}[!htbp]
\caption{Measurements and derived quantities from the ALMA spectra of the QSOs and their companion galaxies.}  
\label{tbl:line_data}      
\centering   
\resizebox{0.8\hsize}{!}{
\begin{tabular}{ l c c c c | c c c c}
\toprule\toprule
{\Large\bf PJ231-20}
			&\multicolumn{4}{c}{\bf QSO}				&\multicolumn{4}{c}{\bf companion}	\\
\vspace{-2mm}\\
Emission line & $z_{\rm line}$ & FWHM & $S\Delta\varv$ & $L_{\rm line}$ & $z_{\rm line}$ & FWHM  & $S\Delta\varv$ & $L_{\rm line}$ \\
&   &(${\rm km\,s^{-1}}$)  &(${\rm Jy\,km\,s^{-1}}$) &$(10^9L_{\astrosun})$  &    &(${\rm km\,s^{-1}}$) &(${\rm Jy\,km\,s^{-1}}$) &$(10^9L_{\astrosun})$\\
\cmidrule(lr){1-9}
${\rm [NII]_{\rm 205 \mu m}}$ 		&$6.5848^{+0.0012}_{-0.0012}$	&$276^{+199}_{-97}$	&$0.21^{+0.09}_{-0.07}$ 		&$0.18^{+0.08}_{-0.06}$ 	&$6.591^{+0.002}_{-0.002}$ 	&$229^{+252}_{-97}$ 	&$0.10^{+0.07}_{-0.06}$ 	&$0.08^{+0.06}_{-0.05}$\\
${\rm [CII]_{\rm 158 \mu m}}$ 		&$6.5864^{+0.0003}_{-0.0003}$	&$372^{+23}_{-21}$	&$2.54^{+0.13}_{-0.13}$ 		&$2.87^{+0.15}_{-0.15}$ 	&$6.5897^{+0.0003}_{-0.0003}$ 	&$492^{+41}_{-39}$ 	&$2.36^{+0.16}_{-0.16}$ 	&$2.68^{+0.18}_{-0.18}$\\
${\rm [CI]_{369\mu m}}$ 			&$6.5873^{+0.0014}_{-0.0014}$	&$243^{+97}_{-70}$		&$0.11^{+0.04}_{-0.04}$ 		&$0.05^{+0.02}_{-0.02}$ &-- &-- &$<0.08$ &$<0.04$ \\ 
\cmidrule(lr){1-9}
CO($7-6$) 					&$6.5863^{+0.0007}_{-0.0007}$	&$335^{+39}_{-34}$ 		&$0.46^{+0.04}_{-0.04}$ 		&$0.22^{+0.02}_{-0.02}$ 	&$6.5870^{+0.0020}_{-0.0014}$ 	&$614^{+242}_{-184}$	&$0.24^{+0.07}_{-0.06}$	&$0.12^{+0.03}_{-0.03}$ \\ 
CO($10-9$) ${^{(\star)}}$					&$6.5878^{+0.0005}_{-0.0005}$ 	&$293^{+53}_{-43}$		&$0.40^{+0.06}_{-0.06}$		&$0.28^{+0.04}_{-0.05}$ 	&$6.5873^{+0.0005}_{-0.0010}$ 	&$183^{+160}_{-67}$ 	&$0.11^{+0.04}_{-0.03}$ 	&$0.08^{+0.03}_{-0.02}$ \\
CO($15-14$) 					&$6.5851^{+0.0017}_{-0.0013}$	&$454^{+81}_{-128}$	&$0.33^{+0.10}_{-0.09}$		&$0.34^{+0.10}_{-0.10}$ 	&-- &-- &<0.16 &<0.16 \\ 
CO($16-15$) 					&$6.5857^{+0.0003}_{-0.0003}$	&$83^{+72}_{-24}$		&$0.15^{+0.05}_{-0.04}$		&$0.16^{+0.06}_{-0.04}$ 	&-- &-- &<0.16 &<0.17  \\
\cmidrule(lr){1-9}
${\rm H_{2}O}\;3_{12}-2_{21}$ ${^{(\star)}}$ 	&$6.5874^{+0.0005}_{-0.0010}$	&$191^{+69}_{-49}$		&$0.18^{+0.06}_{-0.05}$		&$0.12^{+0.04}_{-0.03}$ &-- &-- &<0.09 &<0.06 \\ 
${\rm H_{2}O}\;3_{21}-3_{12}$ 	&$6.5859^{+0.0010}_{-0.0010}$	&$396^{+98}_{-75}$		&$0.33^{+0.07}_{-0.06}$		&$0.23^{+0.05}_{-0.04}$ &-- &-- &<0.09 &<0.06 \\ 
${\rm H_{2}O}\;3_{03}-2_{12}$ 	&$6.5853^{+0.0007}_{-0.0007}$	&$187^{+96}_{-66}$		&$0.18^{+0.07}_{-0.06}$		&$0.18^{+0.07}_{-0.06}$ &-- &-- &<0.16 &<0.16 \\
\cmidrule(lr){1-9}
OH $^2\Pi_{1/2}$,$\frac{3}{2}^+\!-\!\frac{1}{2}^-$			&$6.5863^{+0.0006}_{-0.0006}$ 	&$322^{+70}_{-59}$ 		&$0.45^{+0.09}_{-0.08}$		&$0.49^{+0.10}_{-0.09}$ &$6.5879^{+0.0019}_{-0.0019}$ 	&$296^{+108}_{-118}$ 	&$0.19^{+0.09}_{-0.09}$ 	&$0.21^{+0.10}_{-0.10}$  \\
OH $^2\Pi_{1/2}$,$\frac{3}{2}^-\!-\!\frac{1}{2}^+$ 			&$6.5851^{+0.0006}_{-0.0009}$	&$230^{+91}_{-66}$		&$0.34^{+0.09}_{-0.09}$		&$0.37^{+0.10}_{-0.10}$ &$6.589^{+0.002}_{-0.002}$ 	&$414^{+113}_{-185}$	&$0.27^{+0.12}_{-0.12}$ 	&$0.30^{+0.13}_{-0.13}$\\
\midrule
{Dust continuum} $^{(1)}$ &\multicolumn{4}{c|}{$F_{\it \nu}$ (mJy)} &\multicolumn{4}{c}{$F_{\it \nu}$ (mJy)}  \\
\cmidrule(lr){1-9}
93.85 GHz &\multicolumn{4}{c|}{$0.22\pm0.01$} &\multicolumn{4}{c}{$<0.033$}\\
105.72 GHz &\multicolumn{4}{c|}{$0.31\pm0.02$} &\multicolumn{4}{c}{$0.07\pm0.02$}\\
140.34 GHz &\multicolumn{4}{c|}{$0.72\pm0.02$} &\multicolumn{4}{c}{$0.19\pm0.05$}\\
152.57 GHz &\multicolumn{4}{c|}{$0.86\pm0.03$} &\multicolumn{4}{c}{$0.32\pm0.03$}\\
193.41 GHz &\multicolumn{4}{c|}{$1.91\pm0.04$} &\multicolumn{4}{c}{$0.64\pm0.04$}\\
205.31 GHz &\multicolumn{4}{c|}{$2.19\pm0.03$} &\multicolumn{4}{c}{$0.64\pm0.03$}\\
226.79 GHz &\multicolumn{4}{c|}{$2.88\pm0.05$} &\multicolumn{4}{c}{$0.84\pm0.05$}\\
234.13 GHz &\multicolumn{4}{c|}{$3.29\pm0.05$} &\multicolumn{4}{c}{$0.93\pm0.05$}\\
241.94 GHz &\multicolumn{4}{c|}{$3.46\pm0.06$} &\multicolumn{4}{c}{$0.93\pm0.05$}\\
250.08 GHz &\multicolumn{4}{c|}{$3.94\pm0.06$} &\multicolumn{4}{c}{$1.14\pm0.07$}\\
\bottomrule
\toprule
{\Large \bf PJ308-21}
			&\multicolumn{4}{c}{\bf QSO}				&\multicolumn{4}{c}{\bf companion}	\\
\vspace{-2mm}\\
${\rm [CII]_{\rm 158 \mu m}}$ 		&$6.2354^{+0.0003}_{-0.0002}$	&$533^{+47}_{-43}$	&$1.65^{+0.13}_{-0.12}$ 		&$1.73^{+0.14}_{-0.13}$ 	&$6.2495^{+0.0005}_{-0.0006}$ 	&$239^{+46}_{-40}$ 	&$0.43^{+0.07}_{-0.07}$ 	&$0.46^{+0.08}_{-0.07}$\\
${\rm [CI]_{369\mu m}}$ &-- &-- &<0.06 &<0.03 &-- &-- &<0.06 &<0.03 \\ 
\cmidrule(lr){1-9}
CO($7-6$) &-- &-- &<0.06 &<0.03 &-- &-- &<0.06 &<0.03 \\ 
CO($15-14$) 					&$6.2364^{+0.0020}_{-0.0018}$ 	&$498^{+49}_{-97}$ 	&$0.13^{+0.04}_{-0.04}$ 	&$0.12^{+0.04}_{-0.04}$ &-- &-- &<0.07 &<0.06 \\ 
CO($16-15$) &-- &-- &<0.07 &<0.07 &-- &-- &<0.07 &<0.07  \\
\cmidrule(lr){1-9}
${\rm H_{2}O}\;3_{03}-2_{12}$ 	&$6.234^{+0.003}_{-0.002}$ 	&$264^{+134}_{-120}$ 	&$0.03^{+0.03}_{-0.02}$ 	&$0.03^{+0.03}_{-0.02}$ &-- &-- &<0.07 &<0.06 \\
\cmidrule(lr){1-9}
OH $^2\Pi_{1/2}$,$\frac{3}{2}^+\!-\!\frac{1}{2}^-$			&$6.2348^{+0.0030}_{-0.0017}$ 	&$262^{+145}_{-137}$ 	&$0.04^{+0.03}_{-0.02}$ 	&$0.04^{+0.03}_{-0.02}$ &-- &-- &<0.07 &<0.07  \\
OH $^2\Pi_{1/2}$,$\frac{3}{2}^-\!-\!\frac{1}{2}^+$ 			&$6.234^{+0.002}_{-0.002}$ 	&$400^{+113}_{-165}$ 	&$0.10^{+0.05}_{-0.05}$ 	&$0.10^{+0.05}_{-0.05}$ &-- &-- &<0.07 &<0.07  \\
\midrule
{Dust continuum} {$^{(1)}$} &\multicolumn{4}{c|}{$F_{\it \nu}$ (mJy)} &\multicolumn{4}{c}{$F_{\it \nu}$ (mJy)}  \\
\cmidrule(lr){1-9}
98.95 GHz &\multicolumn{4}{c|}{$0.03\pm0.01$}  &\multicolumn{4}{c}{<0.02}\\
110.87 GHz &\multicolumn{4}{c|}{$0.04\pm0.01$} &\multicolumn{4}{c}{<0.02}\\
237.99 GHz &\multicolumn{4}{c|}{$0.45\pm0.02$} &\multicolumn{4}{c}{$0.08\pm0.02$}\\
246.94 GHz &\multicolumn{4}{c|}{$0.55\pm0.04$} &\multicolumn{4}{c}{<0.12}\\
253.61 GHz &\multicolumn{4}{c|}{$0.59\pm0.02$} &\multicolumn{4}{c}{$0.06\pm0.02$}\\
262.89 GHz &\multicolumn{4}{c|}{$0.59\pm0.06$} &\multicolumn{4}{c}{<0.09}\\
\bottomrule                 
\end{tabular}
}
\tablefoot{In those cases in which the line emission is not detected, upper limits ($3\sigma$) on continuum, line flux and luminosity are reported. $^{(1)}$Here we report the central frequency of the LSB and USB for each frequency setup in which we measured the continuum fluxes. ${^{(\star)}}$The CO(10-9) and H$_2$O $3_{1,2}-2_{2,1}$ line are partially blended.}
\end{table*}

\section{Line and continuum measurements}
\label{sect:data_presentation}
To obtain information on the different ISM phases and dust, we need to measure lines and continuum properties. The observed emission in the ALMA cubes is integrated over the beam and projected along the line of sight. If the source emission is not resolved in the observation, the spectrum measured in the brightest pixel corresponds to the integrated spectrum of the source. By inspecting the continuum and the velocity-integrated line maps we verified that the flux peak (measured in $\rm{Jy\,beam^{-1}}$ and ${\rm Jy\,beam^{-1}\,km\,s^{-1}}$, respectively) is consistent with the integrated flux over an aperture containing the observed emission for each of the objects, even in the case with high S/N and high angular resolution (such as the CO(7--6) line in PJ231-20 QSO in setup A). Therefore, the targeted sources are (at best) marginally resolved in the observations. This conclusion, at least for PJ231-20 system, is supported by the measurements of the size of continuum and [CII]$_{\rm 158\mu m}$ line emitting region obtained from ALMA high angular resolution observations \citep{Neeleman+2019}. This allowed us to safely perform a single-pixel analysis of the data. Indeed, the uncertainties introduced ignoring the aperture correction are comparable to the typical uncertainties ($>10\%$) ascribed to the line and continuum flux measurements taking into account ALMA visibility calibration errors ($\sim20\%$). Furthermore, we take advantage of the single-pixel analysis both to maximize the S/N in the extracted spectra and to simultaneously minimize biases on the flux measurements possibly arising from the blending of quasar and companion galaxy in the frequency setups at low angular resolution. Finally, we assessed that the different beam sizes of the various data cubes in the different ALMA bands do not significantly bias the flux estimates. To do so, we tapered the highest spatial resolution data to the lowest one by using the option \texttt{restoringbeam} within the CASA task \texttt{tclean}, and we checked that the flux measurements obtained from single-pixel analysis in the tapered data are consistent within the uncertainties with those estimated form the data cubes obtained by imaging all the calibrated visibilities\footnote{In particular, we obtained an additional data cube by imaging the calibrated visibilities of ALMA setup A of PJ231-20 system (which has the smallest restoring beam from the original data reduction) by using a "tapering" scheme resulting in a restoring beam of $1''.7\times1''.2$ (similar to that of the setup C, which has the lower angular resolution). Instead, in the case of PJ308-21 data, the beam sizes are comparable.}. 

Hence, we identified the brightest pixel in the continuum map of each ALMA band and we extracted the (beam-)integrated spectra of the quasars and the companion galaxies from the full data cubes (i.e., with continuum present inside) in the selected pixel. We then fit each spectrum as a constant representing the continuum\footnote{\rm For a typical ${\rm S/N}\sim60-100$ over a bandwidth of ${\sim4\,{\rm GHz}}$ in ALMA band 3-6, the continuum emission is well-described by a zeroth-order polynomial within the uncertainties.}, and a Gaussian for the line. %
In order to explore the parameter space, we used the Markov chain Monte Carlo (MCMC) Ensemble sampler \texttt{emcee} \citep{Foreman+2013} and we obtained the posterior probability distributions of the free parameters. Finally, we computed the best values and uncertainties from the 50th, 16th and 84th percentile of the distributions, respectively. In addition, we derived line luminosities as \citep[see, e.g.][]{CarilliWalter2013},
\begin{align}
\label{eq:lum}
&L_{\rm line}\,[L_{\asun}]= 1.04\times10^{-3} S {\Delta \varv}\,\nu_{\rm obs}D_L^2\,,\\
\label{eq:lum'}
&L'_{\rm line}\,[{\rm K\,km\,s^{-1}pc^2}]=3.25\times10^7 S {\Delta \varv}\frac{D_L^2}{(1+z)^3\nu_{\rm obs}^2}\,,
\end{align} 
where $S {\Delta \varv}$ is the line velocity-integrated flux in ${\rm Jy\,beam^{-1}\,km\,s^{-1}}$, $\nu_{\rm obs}$ is the observed central frequency of the line in GHz, and $D_L$ is the luminosity distance in Mpc. The relation between Eq.~\ref{eq:lum} and ~\ref{eq:lum'} is $L_{\rm line}=3\times 10^{-11}\nu_{\rm rest}^3L'_{\rm line}$.

\begin{figure*}[!htbp]
	\resizebox{\hsize}{!}{%
	\centering
	\includegraphics[width=\hsize]{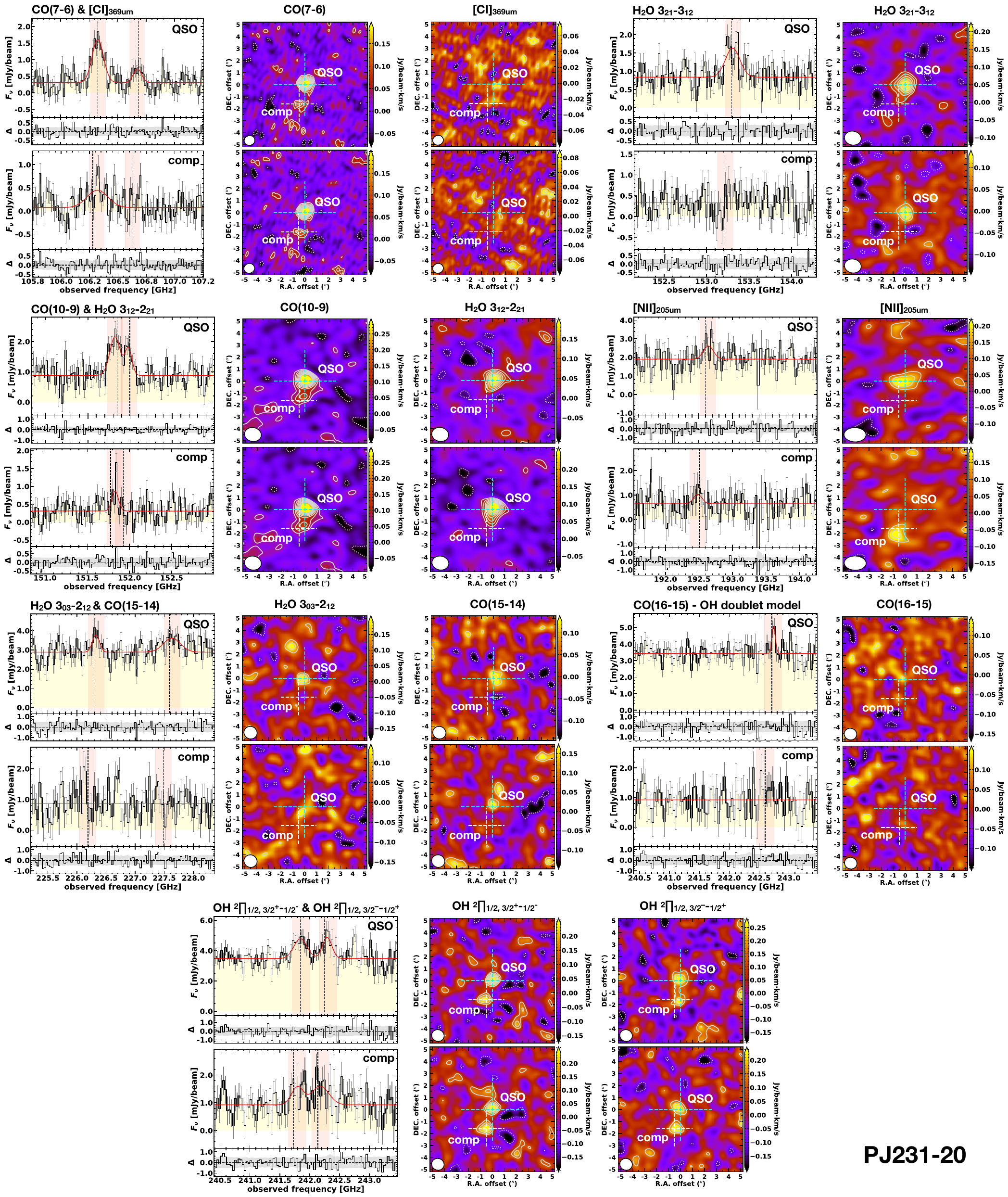}
	}
      	\caption{Detections and non-detections in PJ231-20 system. Here the observed integrated spectra for the quasar and its companion galaxy are shown in a frequency range around the targeted transition. In the panels showing the spectra, we show the best-fit model (red line), the fit residuals (at the bottom of each panel), the expected frequency of the line (vertical black dashed line) and the channels used to obtain the continuum-subtracted line velocity-integrated maps (over $360\,{\rm km s^{-1}}$) that are shown to the right of each spectrum (red shaded area). In the case that the line is not detected, the best-fit model is a constant polynomial fitting the continuum emission only. In the spectra covering the CO(16--15) line, we performed the fit after subtracting the best-fit model for the OH doublet (bottom panels). The shown spectra were extracted at the position of the quasar and companion galaxy. The positions of the sources are indicated in the line velocity-integrated maps with a cyan and a white cross, respectively. The solid contours in the maps show the $[2,3,4,5,6,7,8]\times\sigma$ levels, while dashed contours indicate the $-2\sigma$ level.
	    }
         \label{fig:pj231_spectramaps}
\end{figure*}

\begin{figure*}[!htbp]
	\resizebox{\hsize}{!}{%
	\centering
	\includegraphics[width=\hsize]{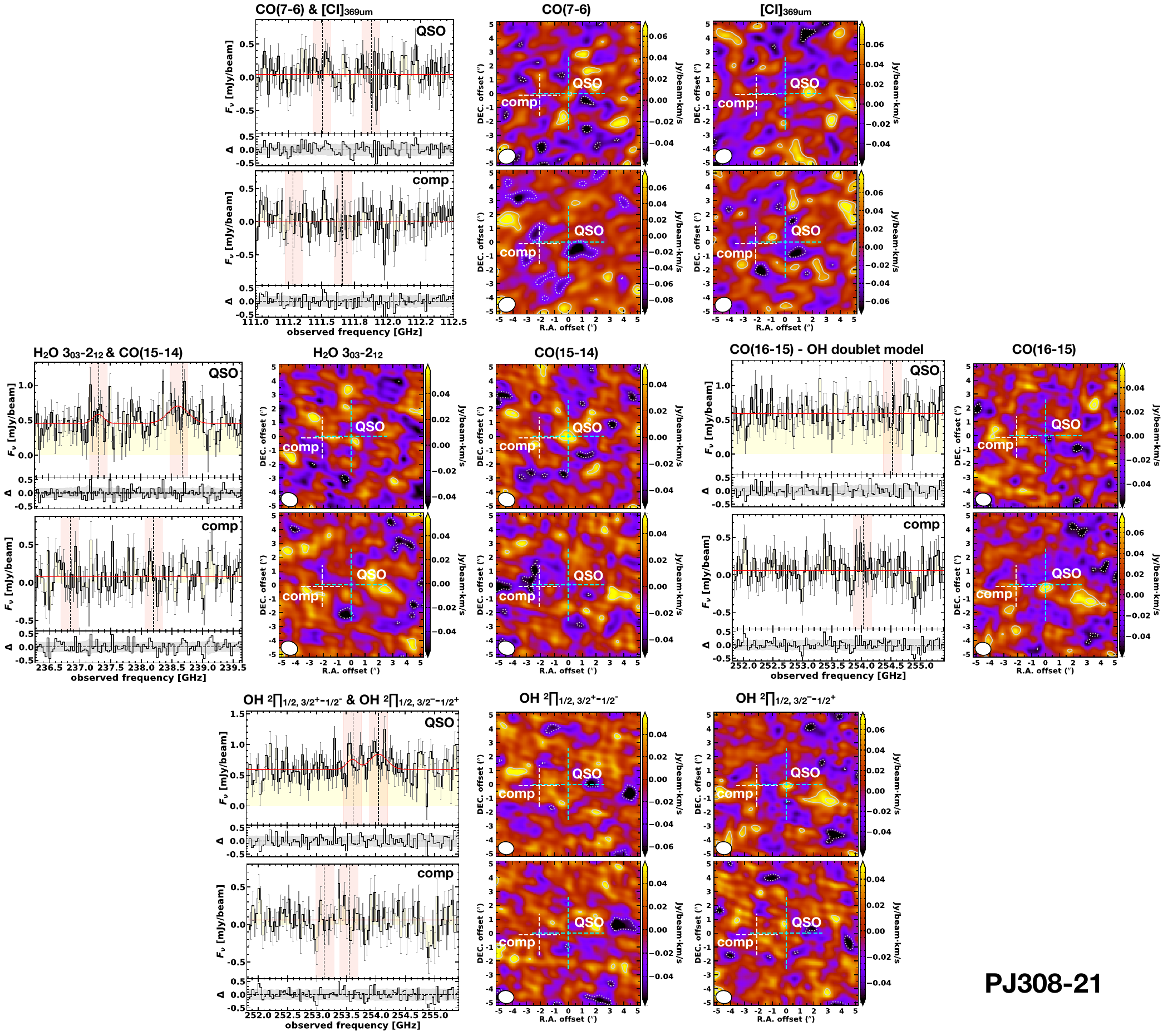}
	}
      	\caption{Detections and non-detections in the PJ308-21 system. See Fig.~\ref{fig:pj231_spectramaps} for a description of the various panels.}
         \label{fig:pj308_spectramaps}
\end{figure*}

\begin{figure}
	\centering
	\includegraphics[width=\hsize]{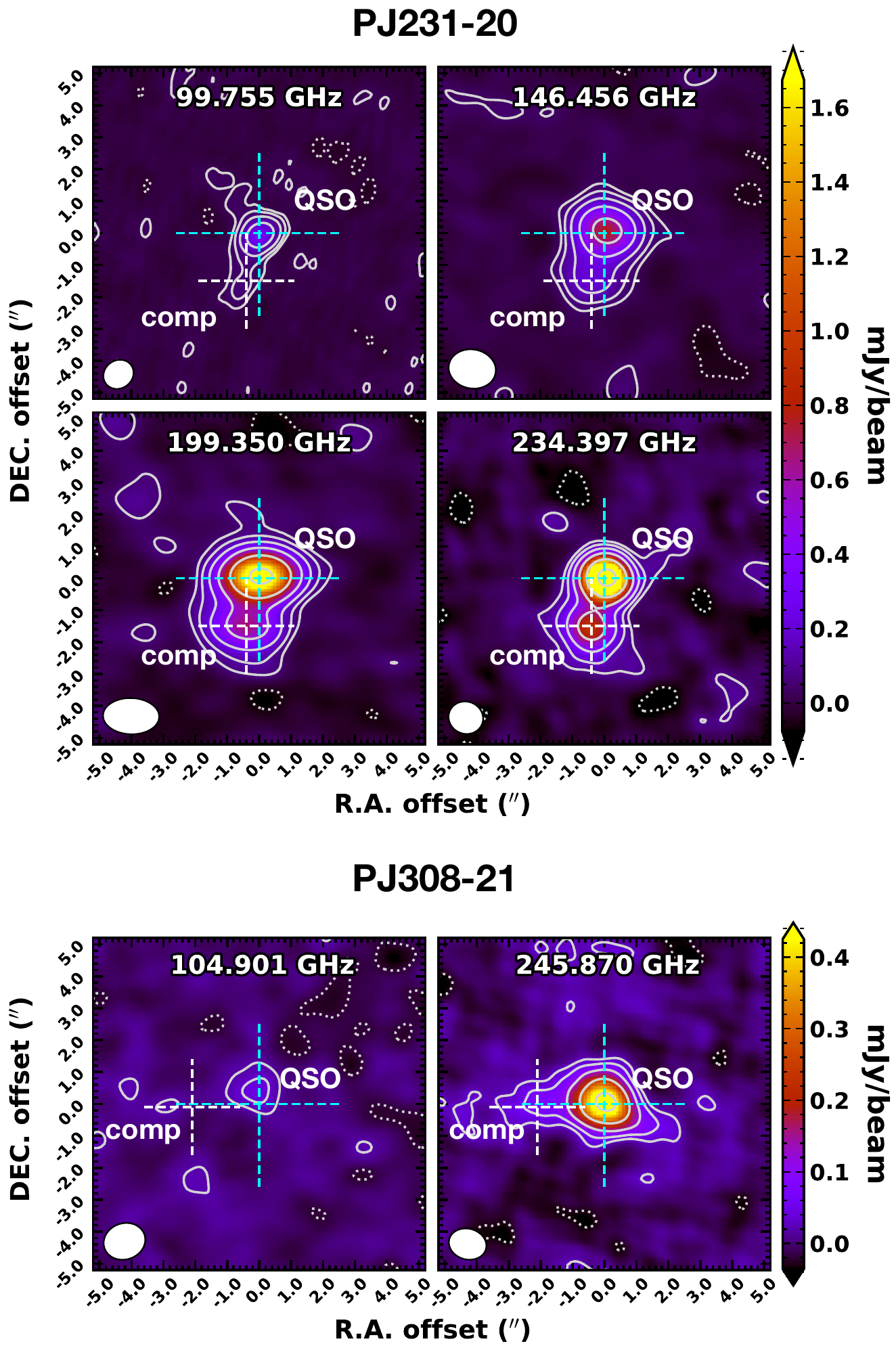}
      	\caption{Continuum maps in the four frequency setups of PJ231-20 (\textit{upper panels}), and the two setups of PJ308-21 system (\textit{lower panels}). The central frequencies of each setup are reported at the top of each panel. The white ellipses indicate the ALMA FWHM synthesized beam. The positions of the quasars and the companions from which we extracted the spectra from line cubes, are marked by cyan and white crosses respectively. The solid contours indicate $[2,4,8,16,32,64]\times\sigma$, while dashed contours are the $-2\sigma$ value. }
         \label{fig:cont_maps}
\end{figure}

In Table~\ref{tbl:line_data} we report the measured and derived quantities for the observed lines and continuum emission in each frequency setup for the quasars and their companions. {\rm In order to check the robustness of the spectral fit results against spurious effects of the noise (especially in the very critical cases of low S/N, e.g. the lines in the spectra of PJ308-21 QSO), we performed different spectral fits by accurately varying the imposing priors and starting points in the MCMC fitting procedure. We therefore assessed that the line is detected (although at low significance) when the derived best-fit models in all the cases, point to a self-consistent solution. In addition, we performed all the fits to the data at a lower spectral resolution (${80\,{\rm km\,s^{-1}}}$) and we verified that all derived parameters are consistent within ${\sim1\sigma}$ with those reported in Table~\ref{tbl:line_data}.} In Figs.~\ref{fig:pj231_spectramaps} and ~\ref{fig:pj308_spectramaps} we report the observed spectra of the PJ231-20 and PJ308-21 systems, respectively, together with the line and continuum best-fit models. We also report the corresponding continuum-subtracted line velocity-integrated maps over $360\,{\rm km\,s^{-1}}$. This value maximizes the S/N of the CO(10--9) velocity-integrated map of PJ231-20 QSO. We used this as a reference value in order to obtain all the other line maps. Then, in Fig.~\ref{fig:cont_maps} we show the line-free continuum maps in the various frequency bands. In some cases (see Figs.~\ref{fig:pj231_spectramaps},~\ref{fig:pj308_spectramaps}) the peaks of line velocity-integrated maps of PJ231-20 and PJ308-21 systems, show a shift with respect to the reference position of the sources. However, in all the cases the shift is well inside the beam of the observed map. This effect can be likely ascribable to the continuum subtraction and/or the low S/N of the emission. Therefore, we conclude that the line and the continuum peaks are spatially consistent at the available resolution.

Throughout the text, we report the significance on the measured flux of a line emission in unit of $\sigma = \sqrt{\Delta \varv\,{\rm FWHM}} \times \left\langle{\rm RMS}\right \rangle$, where $\Delta\varv$ is the channel width ($40\,{\rm km\,s^{-1}}$), FWHM is the full width at half maximum of the line best-fit Gaussian, and  the $\left\langle {\rm RMS}\right\rangle$ is the average RMS noise in the line cube (see Table~\ref{tbl:obs_data}). In cases in which line emission is not detected in the observation, we report $3\sigma$ upper limits assuming ${\rm FWHM} = 300\,{\rm km\,s^{-1}}$. Finally, tor the non-detections of continuum emission we report $3\sigma$ upper limit with $\sigma$ equal to RMS noise of the continuum.
\section{CLOUDY models of photo- and X-ray dissociation regions }
\label{sect:cloudy_modeling}
The observed molecular and atomic FSLs are key diagnostics of the physical conditions present in the ISM of galaxies. The observed FSL to dust-continuum flux ratios require a relatively efficient ($0.1 - 1\%$) mechanism to convert the continuum flux into atomic and molecular gaseous line emission. \citet{Tielens+1985a} showed that this condition is satisfied by PDRs in which dust grain photoelectric heating in regions where the far-UV (FUV) radiation ($6 < h\nu < 13.6\,{\rm eV}$) from hot stars (O and B) impinges on molecular clouds. PDRs represent the interface between the molecular and ionized medium (i.e. 'HII regions') at the outskirt of molecular clouds, where FUV radiation plays a significant role in the heating and/or chemistry of the gas phase. \citet{Tielens+1985a, Tielens+1985b} also pointed out that the emission from such regions depends largely on the cloud density ($n_{\rm H}$) and the strength of the FUV radiation field ($G_0$) illuminating the cloud (where $G_0$ is in Habing flux units, i.e. $1.6\times10^{-3}\,{\rm erg\,s^{-1}\,cm^{-2}}$, see \citealt{Habing+1968}). PDRs include all the atomic neutral medium and most of the molecular clouds, therefore the majority of the mass of the ISM is in PDRs. Densities in PDRs range from $n_{\rm H}\approx0.25\,{\rm cm^{-3}}$ to $n_{\rm H}\apprge 10^{7}\,{\rm cm^{-3}}$ and FUV flux from $G_0\apprle 1$ to $G_0\apprge 10^6$ \citep[see e.g.][]{McKee+1977, Hollenbach+1999, Kaufman+1999, Meijerink+2007}.

Molecular gas can also be exposed to X-ray radiation, for example near supernova shocks or radiation emitted from accreting BH in quasars \citep[see e.g.][]{Langer+1978, Krolik+1983, Krolik+1988, Lepp+1983}. X-ray photons are able to penetrate much deeper into the cloud (column density of $N_{\rm H}\apprge 10^{22}\,{\rm cm^{-2}}$) and can effect the chemical and thermal structure of FUV-opaque molecular gas. Therefore, completely analogous to the PDRs, XDRs can be defined as predominantly neutral gas in which X-rays dominate the gas heating and much of the chemistry through X-ray dissociation and ionization (e.g. \citealt{Maloney+1996, Hartquist+1998}; \\ \citealt{Hollenbach+1999}). In this case, the incident radiation field is typically expressed in terms of X-ray flux, $F_{\rm X}({\rm erg\,s^{-1}\,cm^{-2}})$, over the range $1-100\,{\rm keV}$.

Considerable effort has been made in constructing PDR/XDR models in order to constrain the physical properties of the ISM where molecular and atomic transitions take place \citep[e.g.][]{Tielens+1985a, Meixner+1993, Maloney+1996, Spaans+1996, Kaufman+1999, Kaufman+2006, Elitzur+2006, Meijerink+2005, Meijerink+2007, vanderTak+2007, Pound+2008, Bisbas+2012, Ferland+2017}. Each of these codes is optimized to simulate various astrophysical environments and are based on different assumptions of PDR geometry, thermal and chemical balance, and different implementation of radiative transfer through PDR. Here we use CLOUDY radiative-transfer code (version c.17.01, \citet{Ferland+2017}) to characterize the ISM in the quasar and companion galaxy in the PJ231-20 and PJ308-21 systems, as it is highly customizable and allows the simultaneous treatment of all the species studied in this work. %
We modeled the observed line-emitting region as a single plane-parallel semi-infinite cloud impinged by a radiation field in both the PDR and XDR regimes, and we obtained a suite of diagnostics to interpret the observed line ratios in these systems. In the following we describe how the models were set up.

Within {CLOUDY}, we ran a total of $270\times2$ PDR/XDR models assuming a 1D gas slab of constant total hydrogen density in the range $\log(n_{\rm H}/{\rm cm^{-3}}) = [2,6]$ (15 models, $\sim0.29$ dex spacing) varying the strength of the radiation field (18 models), for two values of cloud total hydrogen column density of $N_{\rm H}=10^{23},10^{24}\,{\rm cm^{-2}}$. In the case of PDR models, we defined the spectral energy distribution (SED) of the impinging radiation field as a black body emission with a temperature of $T=5\times10^4{\rm K}$ scaled to obtain Habing fluxes at the gas slab surface in the range $\log G_0 = [1,6]$ ($\sim0.29$ dex spacing). In the case of XDR, we choose a standard AGN template as the incident continuum SED,
\eq{f_{\nu} = \nu^{\alpha_{\rm UV}}\exp\tonda{-h\nu/k_{\rm B}T_{\rm BB}}\exp\tonda{-k_{\rm B}T_{\rm IR}/h\nu}+a\nu^{\alpha_X},}
with parameters set in order to generate the continuum used in a large atlas of model broad line region line intensities \citep{Korista+1997}. We normalized this incident SED in order to have X-ray flux between $1-100\,{\rm keV}$ in the range $\log\,[F_{\rm X}/({\rm erg\,s^{-1}\,cm^{-2}})]=[-2.0,2.0]$ ($\sim0.24$ dex spacing) at the cloud surface. This range fully covers the X-ray radiation field strengths generally considered in standard XDR models \citep[e.g.][]{Maloney+1996, Meijerink+2007}.

For the gas slab we assumed the default interstellar medium metal abundances (\texttt{abundance ISM}) stored in CLOUDY. We adopted dust ISM grains (\texttt{grains ISM}) stored in {CLOUDY} with a size distribution and abundance from \citet{Mathis+1997}. In addition, we included polycyclic aromatic hydrocarbons (PAHs, \texttt{grains PAH}) with distribution from \citet{Abel+2008}, which generally dominate the grain photoelectric heating. We also included the cosmic microwave background (CMB) radiation at $z=6$ that affects the far-IR line luminosity of galaxies at high $z$. We adopted the default {CLOUDY} prescriptions for the cosmic ray ionization rate background (\texttt{cosmic rays background}) which are an important source of heating deep into the molecular cloud. The mean H cosmic ray ionization rate is $2\times10^{-16}\,{\rm s^{-1}}$ \citep{Indriolo+2007}, and the H$_2$ secondary ionization rate is $4.6\times10^{-16}\,{\rm s^{-1}}$ \citep{Glassgold+1974}.  Finally, we added $1.5\,\si{km\,s^{-1}}$ in quadrature to the thermal motions within the cloud, to model the effect of line broadening produced by microturbulence.

In order to simulate the absorption of ionizing radiation due to atomic hydrogen; in both PDR and XDR case, we modified the incident continuum by extinction due to photoelectric absorption by a cold neutral slab with a column density of $10^{24}\,{\rm cm^{-2}}$ (\texttt{extinguish column 24}, see \citealt{Cruddace+1974}). The code computes the radiative transfer through the slab up to total hydrogen column density of $N_{\rm H}=10^{23},10^{24}\,{\rm cm^{-2}}$. We choose this stopping criterion to fully sample the molecular component typically located at $N_{\rm H} >10^{22}\,{\rm cm^{-2}}$ as is observed in giant molecular clouds \citep[e.g.][]{McKee+1977}. However, a higher column density $N_{\rm H}\apprge 5\times10^{23}\,{\rm cm^{-2}}$ is required to properly model H$_2$O and OH emission \citep[see e.g.][]{Goicoechea+2005, Gonzalez-Alfonso+2014, Spinoglio+2005, Liu+2017}.

With our PDR and XDR models we obtained grids of line intensities and emitted continuum emerging from the cloud's surface (both in unit of ${\rm erg\,s^{-1}\,cm^{-2}}$), as function of the density of the medium (total hydrogen density, $n_{\rm H}$ ranges $10^2-10^6\,{\rm cm^{-3}}$) and the strength of the radiation field (parametrized by $G_0 = 10^1-10^6$ or $F_{\rm X}\,({\rm erg\,s^{-1}\,cm^{-2}}) = 10^{-2}-10^2$ in the case of PDR and XDR, respectively) for two values of total hydrogen column density $N_{\rm H}=10^{23}, 10^{24}\,{\rm cm^{-2}}$.

The models adopted in this work are designed to simulate a "typical" cloud in a galaxy under very simplistic assumptions. Undoubtedly, these simple models cannot provide a realistic picture of the ISM in galaxies that are composed of an ensemble of clouds and diffuse medium with a wide ranges of physical and geometrical properties. However, quantities such as line luminosity ratios can be used to mitigate the large uncertainties related to the unknown parameters, thus allowing us to quantitatively compare model predictions with observations.

In the following sections, we present various results obtained from dust FIR continuum and line detections in the targeted sources. Then, by comparing our model results with the observed emission we put constraints on the key parameters characterizing the ISM in the quasar hosts and companion galaxies.

\begin{figure*}[!htbp]
	\centering
	\resizebox{\hsize}{!}{%
	\centering %
	\includegraphics[width=\hsize]{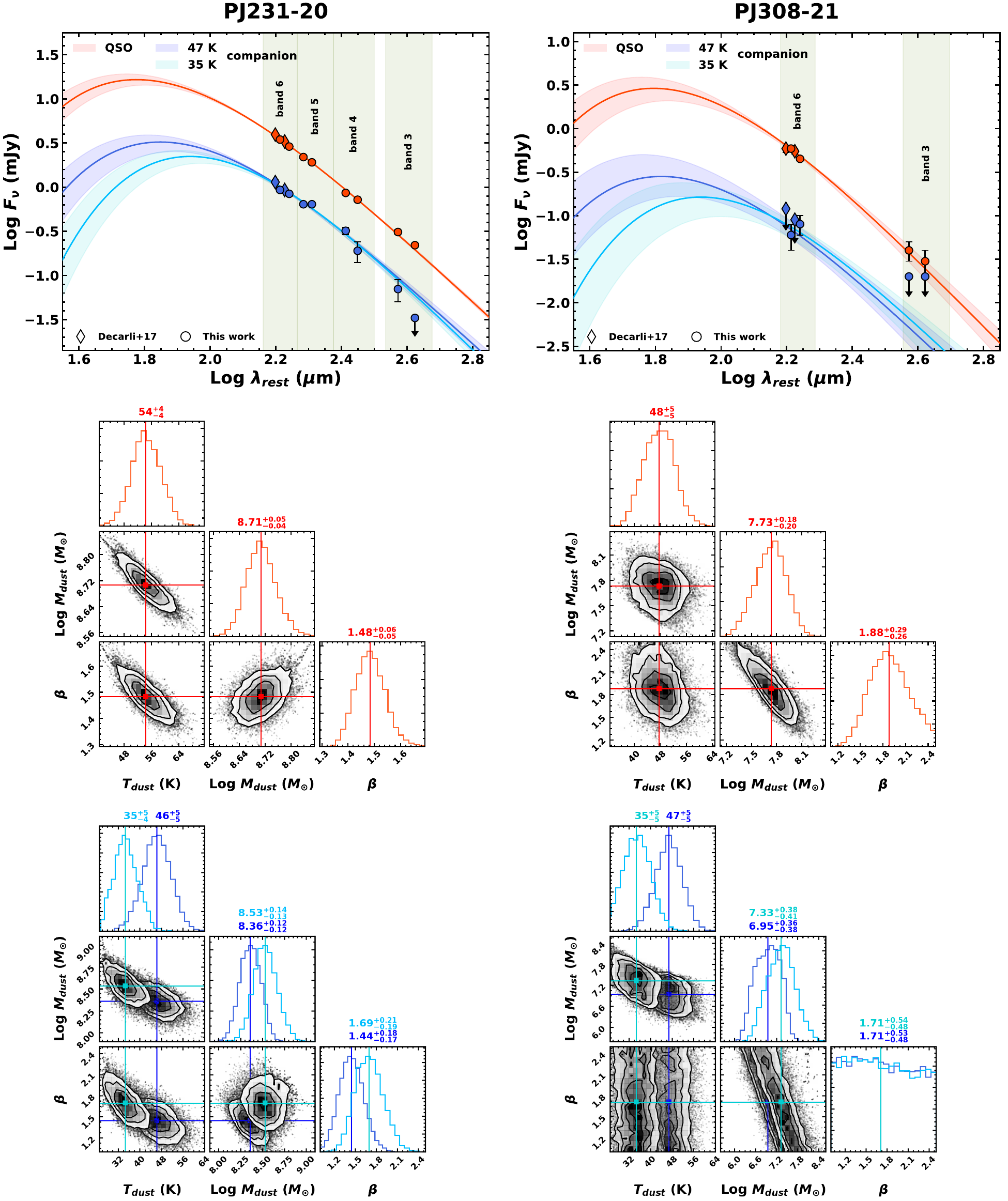}
	}
      	\caption{Models of the dust SEDs of the PJ231-20 (\textit{left panels}) and PJ308-21 (\textit{right panels}) sources. In the upper panels, circles and diamonds indicate, respectively, continuum data obtained in this work and those obtained from the ALMA Cycle 6 [CII]$_{\rm158 \mu m}$ observations \citep{Decarli+2017}. Downward arrows indicate $3\sigma$ upper limits. The best-fit modified black body model is indicated with red solid line for the quasars, while blue and cyan line are the models for companion galaxies assuming $T_{\rm dust}=47\,{\rm K}$ and $35\,{\rm K}$, respectively. The shadowed areas show the $1\sigma$ confidence intervals. Green bands indicate the ALMA frequency bands redshifted to the quasar-companion pair rest-frame ($z_{\rm pair} = [z_{\rm QSO}+z_{\rm comp}]/2$). The lower panels show the posterior probability distributions of the free parameters (red for quasars, blue and cyan for companions). The vertical lines indicate the best-fit values computed as $50{\rm th}$ percentile of the distributions. The best-fit values with their uncertainties are also reported.}
         \label{fig:dust_continuum}
\end{figure*}

\section{Dust properties: FIR continuum}
\label{ssect:dust}
The continuum emission detected in the FIR and (sub-)mm bands, is dominated by emission from dust distributed in the diffuse medium and star-forming regions of a galaxies. The bulk of the dust in star-forming regions is heated by the diffuse rest-frame optical/UV radiation field produced by young stars. The typical dust temperature for quasar hosts lies in the range $T_{\rm dust}\sim40-70\,{\rm K}$ \citep[e.g.][]{Priddey+2001,Beelen+2006,Wang+2008,Leipski+2013,Leipski+2014}, and increases to $\apprge 100-500\,{\rm K}$ in the proximity of the central BH accretion disk \citep{Rowan-Robinson+1995, Charmandaris+2004, Beelen+2006}. Since the dust grains are in local thermodynamical equilibrium (LTE) with the radiation field, they emit thermal radiation. However, dust grains are not ideal black bodies. %
For this reason, dust thermal emission is usually modeled as a modified black body where the grains' emission efficiency is expressed in terms of the dust opacity coefficient, $k_{\rm d}(\nu)\propto \nu^\beta$. 
Therefore, by assuming optically thin dust emission \citep[see e.g.][]{Beelen+2006}, in the Rayleigh-Jeans limit, the continuum flux is proportional to the dust mass, $M_{\rm dust}$, and temperature, $T_{\rm dust}$ \citep[see e.g.,][]{Downes+1992, Dunne+2000}. Based on observational studies at low and high redshift \citep[e.g.][]{Blain+2003, Conley+2011, Rangwala+2011, Casey+2012, Riechers+2013, Faisst+2020}, the optically thin dust approximation may not be true at short wavelengths, however likely holds for our observations since most of the continuum data is at rest-frame $\lambda>200\,{\rm \mu m}$.
In order to estimate $M_{\rm dust}$ and $T_{\rm dust}$ in the quasar host galaxies and their companions, we modeled the continuum data with a modified black body taking into account the effect of the cosmic microwave background radiation \citep[CMB; see][]{daCunha+2013} whose temperature scales as $\propto (1+z)$ and is $T_{\rm CMB}\approx 19\, {\rm K}$ at $z=6$. Indeed, at the redshifts of the sources the CMB is an additional non-negligible source of heating and provides a strong background against which we measure line and continuum emission. %
Therefore, following \citet{daCunha+2013} we define the flux density in the optically thin limit as:
\begin{align}
&F_{\nu/(1+z)}^{\rm intr} = B_\nu[T_{\rm dust}(z)]k_{\rm d}(\nu)(1+z)M_{\rm dust}D_L^{-2}\,,\\
&F_{\nu/(1+z)}^{\rm obs} = \tonda{1 - \frac{B_{\nu}[T_{\rm CMB}(z)]}{B_{\nu}[T_{\rm dust}(z)]}}F_{\nu/(1+z)}^{\rm intr}\,;
\label{eq:BB_model}
\end{align}
where $F_{\nu/(1+z)}^{\rm intr}$ and $F_{\nu/(1+z)}^{\rm obs}$ are the intrinsic and observed flux density against the CMB, respectively; $B_\nu(T)$ is the Planck function $B_\nu(T)  = [2h\nu^3/c^2][\exp(h\nu/k_{\rm B}T)-1]^{-1}$, $k_{\rm d}(\nu)$ is the frequency-dependent dust mass opacity coefficient defined as $k_{\rm d}(\nu) = k_0 (\nu/\nu_0)^\beta$, where $\nu_0 = 352.7$ GHz, and $\beta$ is the dust spectral emissivity index \citep[see e.g.,][]{Dunne+2000}. \citet{daCunha+2013} also show that the observed galaxy dust temperature at certain redshift, $T_{\rm dust}(z)$, is higher than the dust temperature that we would observe at $z=0$, due to the absorption of CMB photons by the dust grains: $T_{\rm dust}(z) = \{(T_{\rm dust}^{z=0})^{4+\beta}+(T_{\rm CMB}^{z=0})^{4+\beta}[(1+z)^{4+\beta}-1]\}^{1/(4+\beta)}$. However, at $z\approx6.5$ and typical dust temperatures of $T_{\rm dust} > 35\,{\rm K}$, this correction is $<2\%$, much less than relative uncertainties that we expect on dust temperature estimate. Therefore, we ignore this correction.

Hence, we used the model defined in Eq.~\ref{eq:BB_model} to fit the quasar and companion continuum data reported in Table~\ref{tbl:line_data}, together with ALMA band 6 data presented in \citet{Decarli+2017}, from which we obtained line and continuum measurements. In order to find the best-fit model, we made use of the \texttt{emcee} code \citep{Foreman+2013}, allowing $M_{\rm dust}$, and $\beta$ to freely vary during the fitting procedure. We limited the parameter space by imposing box-like priors defined as $\log M_{\rm dust}[M_{\astrosun}] > 0$, and $\beta\in(1,2.5)$. In addition, the sampled frequency range of the observations mostly covers the Rayleigh-Jeans tail of the dust thermal emission. The observed flux density in this regime is $\propto T_{\rm dust}M_{\rm dust}$, making these parameters degenerate, therefore we employed a relative tight Gaussian prior on $T_{\rm dust}$ with mean and standard deviation equal to $47\,{\rm K}$ and $5\,{\rm K}$, respectively. These choices are consistent with the typical measurements in quasar host galaxies reported in the literature \citep[e.g.][]{Priddey+2001,Beelen+2006,Wang+2008,Leipski+2013,Leipski+2014} and also validated by the results obtained below for the water vapor lines (see Sect.~\ref{ssect:water_tir_relation}). On the other hand, if the companion galaxies actually do not host a quasar (as suggested by rest-frame UV and X-ray studies, see \citealt{Decarli+2017, Connor+2019, Connor+2020}), we expect that the dust temperature may not be as high as the quasar host galaxies due the lack of the additional dust heating by an AGN. Indeed, the non-detection of water lines in the companions also argue against a high $T_{\rm dust}$ (albeit the detections of OH$_{\rm 163\mu m}$ doublet that likely traces similar regions of those traced by H$_2$O lines, see Sects.~\ref{ssect:water_vapor}, ~\ref{ssect:oh_lines}, ~\ref{ssect:h2o_oh_ratio}). Therefore, in the case of the companion galaxies, we also considered a lower-temperature scenario. To do so, we performed an additional fit centering the Gaussian dust temperature prior on $T_{\rm dust}=35\,{\rm K}$, that is a representative value of the population of SMGs at $z\sim 1-3$ \citep[e.g.][]{Chapman+2005, Kovacs+2006}. In the fits we also take into account the upper limit measurements on dust continuum by inserting a penalty term in the log-likelihood function in the form $\mathcal{P}=-0.5[({\rm model}-{\rm data})/{\rm data}]^2$, if ${\rm model}>{\rm data}$, else $\mathcal{P}=0$. Finally, we assumed the sources' [CII]-based redshifts reported in \citealt{Decarli+2017} (see Table~\ref{tbl:line_data}). In Fig.~\ref{fig:dust_continuum} we show the best-fit models of the continuum emission in PJ231-20 and PJ308-21 systems, respectively, together with the posterior probability distributions of the parameters. From these results, we estimated the total (TIR, $8-1000\,{\rm \mu m}$, \citealt{Sanders+2003}) and FIR ($40-400\,{\rm \mu m}$, \citealt{Helou+1988}) rest-frame luminosities of the sources, by integrating the best-fit models over the corresponding frequency ranges.  
Finally, we inferred the SFR, using the local scaling relation from \citet{Murphy+2011}; ${\rm SFR_{IR}}/(M_{\astrosun}\,{\rm yr^{-1}}) = 1.49\times 10^{-10}L_{\rm TIR}/L_{\astrosun}$, assuming that the IR-luminosity is dominated by star formation and under the hypothesis that the entire Balmer continuum (i.e. $912\,\AA <\lambda<3646\,\AA$) is absorbed and re-irradiated by dust in the optically thin limit. Therefore, any contribution of the central AGN to the IR luminosity will result in an overestimation of the ${\rm SFR_{IR}}$. The initial mass function (IMF) implicitly assumed in this $L_{\rm TIR}$ to ${\rm SFR}$ conversion is a Kroupa IMF \citep{Kroupa2001}, having a slope of $-1.3$ for stellar masses between $0.1-0.5\,M_{\astrosun}$, and $-2.3$ for stellar masses ranging between $0.5$ and $100\,M_{\astrosun}$. In Table~\ref{tbl:dust_data} we report all the derived quantities obtained with the dust continuum modeling.

\setcounter{magicrownumbers}{0}
\begin{table}
\caption{Dust properties in the PJ231-20 and PJ308-21 sources.}      
\label{tbl:dust_data}
\centering      
\resizebox{\hsize}{!}{
\begin{tabular}{l c c c c c c }
\toprule\toprule
								&$T_{\rm dust}$		&Log $M_{\rm dust}$		&$\beta$				&Log $L_{\rm FIR}$  $^{(1)}$		&Log $L_{\rm TIR}$ $^{(2)}$		&${\rm SFR_{IR}}$\\
								&(${\rm K}$)			&($M_{\astrosun}$)			&					&($L_{\astrosun}$)				&($L_{\astrosun}$)				&($M_{\astrosun}\,{\rm yr^{-1}}$)\\
\cmidrule(lr){1-7}
\parbox[t]{2mm}{\multirow{3}{*}{\rotatebox[origin=c]{90}{\bf PJ231-20}}}
\hspace{2mm} \multirow{1}{*}{\bf QSO} 	&54 	&$8.71^{+0.05}_{-0.04}$ 	&$1.48^{+0.06}_{-0.05}$	&$13.13^{+0.06}_{-0.06}$ 		&$13.28^{+0.09}_{-0.09}$			&$2849^{+633}_{-533}$ \\
\cmidrule(lr){2-7}
\hspace{5mm} \multirow{2}{*}{\bf comp.} 	&46 	&$8.36^{+0.12}_{-0.12}$ 	&$1.44^{+0.18}_{-0.17}$ 	&$12.40^{+0.09}_{-0.10}$ 		&$12.47^{+0.12}_{-0.13}$ 		&$442^{+140}_{-112}$\\
								&35 	&$8.53^{+0.14}_{-0.13}$ 	&$1.69^{+0.21}_{-0.19}$ 	&$12.16^{+0.12}_{-0.10}$ 		&$12.18^{+0.14}_{-0.11}$ 			&$224^{+75}_{-55}$\\
\cmidrule(lr){1-7}								
\parbox[t]{2mm}{\multirow{3}{*}{\rotatebox[origin=c]{90}{\bf PJ308-21}}}
\hspace{2mm} {\bf QSO} 				&48 	&$7.73^{+0.18}_{-0.20}$ 	&$1.9^{+0.3}_{-0.3}$		&$12.31^{+0.14}_{-0.13}$			&$12.42^{+0.18}_{-0.17}$			&$392^{+193}_{-128}$ \\
\cmidrule(lr){2-7}
\hspace{5mm} \multirow{2}{*}{\bf comp.} 	&47 	&$7.0^{+0.4}_{-0.4}$ 	&$1.7^{+0.5}_{-0.5}$ 	&$11.28^{+0.25}_{-0.19}$ 			&$11.4^{+0.3}_{-0.2}$ 			&$38^{+28}_{-16}$\\
								&35 	&$7.3^{+0.4}_{-0.4}$ 	&$1.7^{+0.5}_{-0.5}$ 	&$11.0^{+0.2}_{-0.2}$ 			&$11.1^{+0.2}_{-0.3}$			&$19^{+14}_{-8}$\\
\bottomrule                 
\end{tabular}
}
\tablefoot{$^{(1),} {^{(2)}}$Far-infrared and total infrared luminosity obtained by integrating the best-fit modified black body model in the (rest-frame) wavelength range $40-400\,{\rm \mu m}$ \citep{Helou+1988}, and $8-1000\,{\rm \mu m}$ \citep{Sanders+2003}, respectively.}
\end{table}

In the case of the quasar and companion galaxy in PJ231-20 system, we find good constraints both on the dust mass ($\apprle 0.1\,{\rm{dex}}$) and the spectral emissivity index (relative error of $\apprle10\%$). On the other hand, the dust temperature values obtained from the posterior probability distribution are consistent within $1\sigma$ with the imposed prior, confirming that $T_{\rm dust}$ cannot be constrained with the fit. In the case of the quasar and companion in the PJ308-21 system, we found only tentative constraints on $M_{\rm dust}$ and $\beta$ due to the poor coverage of the dust spectral energy distributions (SEDs). The best fit values of $\beta$ in both PJ231-20 and PJ308-21 quasars and companions range between $\sim1.5-1.9$, and are consistent with the values estimated in both low redshift sources \citep[e.g.][]{Conley+2011, Casey+2012} and those at $z>5-6$ \citep[e.g.][]{Riechers+2013, Carniani+2019, Novak+2019, YangJ+2019, Faisst+2020}. We note that, although assuming different dust temperatures of $47\,{\rm K}$ and $35\,{\rm K}$  as prior in the dust SED fit of the companion galaxies, we obtained similar results that are consistent within $1.5\sigma$. As shown in Fig.~\ref{fig:dust_continuum}, this is mainly due to lack of high-frequency continuum measurements, where the peak of dust SED is placed. However, an additional large source of uncertainty in the dust fitting is the optically thin dust approximation. In fact, the shape of the dust SED may be altered by the dust opacity at short wavelength, thus affecting the derived $L_{\rm TIR}$. Furthermore, if a warmer dust component is present, it may significantly contributes to the infrared luminosity.

\section{Ionized medium}
\label{sect:ionized_medium}
\subsection{The [NII]$_{\rm 205\mu m}$ and [CII]$_{\rm 158\mu m}$ fine-structure lines}
\label{ssect:ionised_med}

The ionization potential of the nitrogen atom (${\rm 14.5\,eV}$), is greater than that of neutral hydrogen (${\rm 13.6\,eV}$), therefore the ${\rm N^+}$ emission is due to collisions with electrons in the fully ionized medium. For an electron temperature of 8000~K, the critical density\footnote{The density at which the rate of collisional depopulation of a quantum level equals the spontaneous radiative decay rate.} of [NII]$_{\rm 205\mu m}$ line is $n_{\rm e}^{\rm crit}\approx44{\rm\,cm^{-3}}$ \citep{Oberst+2006}, thus this line traces the low-density diffuse medium ($n_{\rm e}\sim0.1-1\,{\rm cm^{-3}}$) rather than the dense (HII) regions ($n_e\sim0.5-1.0\times10^{3}\,{\rm cm^{-3}}$) where it is rapidly thermalized and its emission is collisionally quenched \citep[e.g.][]{Herrera-Camus+2016}.

The fine-structure transition (${^3P_{3/2}}\!\!\to\!\!{^3P_{1/2}}$) of the C$^+$ ion is one of the major coolants of the cold gas phase ($50\,{\rm K} < T<3000\,{\rm K}$) of the ISM in star-forming galaxies, and is easily detected in ALMA band 6 at $z\apprge 6$. Singly-ionized carbon has a lower ionization potential than HI ($11.3\,{\rm eV}$), therefore [CII]$_{\rm 158\mu m}$ arises both from the neutral medium, including atomic clouds and PDRs on the surface of molecular clouds exposed to stellar UV radiation in star-forming regions; and from low-density ionized gas. In the ionized medium, the C$^+$ ion is excited by collisions with electrons and its $158\,{\rm\mu m}$ transition has a critical density of $n_{\rm e}^{\rm crit}\approx46\,{\rm cm^{-3}}$ \citep{Oberst+2006}. On the other hand, in the neutral medium, most of the [CII]$_{\rm 158\mu m}$ emission is expected to arise from the dense PDRs where C$^+$ is collisionally excited by neutral and molecular hydrogen ($n_{\rm H}^{\rm crit}\simeq3.0\times10^3\,{\rm cm^{-3}}$, $n_{\rm H_2}^{crit}\simeq6.1\times10^3\,{\rm cm^{-3}}$, see \citealt{Goldsmith+2012}). 

The [NII]$_{\rm 205\mu m}$ line provides us with important clues to interpret the observed [CII]$_{\rm158\mu m}$ emission \citep{Decarli+2017}. We detected the [NII]$_{\rm 205\mu m}$ (${^3P_1}\!\!\to\!\!{^3P_0}$) line in the PJ231-20 system with a significance of $\sim 4\sigma$ and $\sim2\sigma$ for the quasar and the companion galaxy respectively (see Figs.~\ref{fig:pj231_spectramaps},~\ref{fig:pj308_spectramaps}). The line measurements are reported in Table~\ref{tbl:line_data}. 

\subsection{Fraction of [CII]$_{\rm 158\mu m}$ emission from PDRs}
\label{ssect:cii_fraction}
Since [CII]$_{\rm 158\mu m}$ and [NII]$_{\rm 205\mu m}$ have similar ionization potentials and critical densities, in the optically thin case, the line ratio in the ionized medium depends essentially only on their relative abundance. Therefore, the observed [NII]$_{\rm 205\mu m}$ line emission reveals the fraction of [CII]$_{\rm 158\mu m}$ arising from neutral gas. Following \citet{Diaz-Santos+2017}, we estimated the fraction of [CII]$_{\rm 158\mu m}$ emission arising from PDRs, by assuming a typical $\cii^{\rm ion}_{\rm 158\mu m}/\nii_{\rm 205\mu m}\approx 3$ observed in HII regions \citep{Oberst+2006}:
\eq{f(\cii^{\rm PDR}) = \frac{\cii^{\rm PDR}_{\rm 158\mu m}}{\cii_{\rm 158\mu m}} \approx  1 - 3\frac{\nii_{\rm 205\mu m}}{\cii_{\rm 158\mu m}}.}

\begin{figure}
	\centering
   	\includegraphics[width=\hsize]{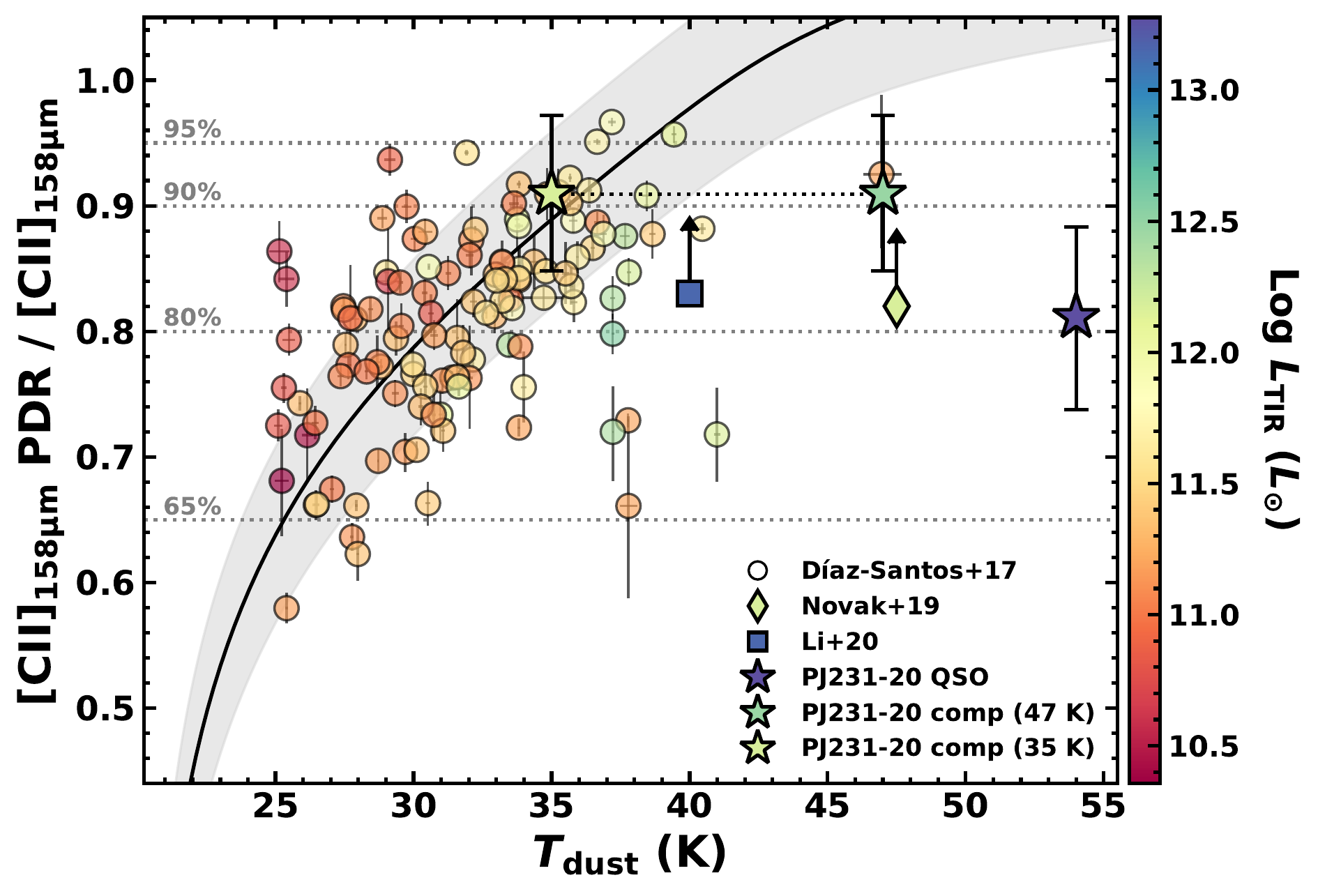}
      	\caption{Fraction of \cii$_{158\mu m}$ arising from PDR as a function of dust temperature, color coded by the logarithm of the total infrared luminosity. The PJ231-20 QSO and companion galaxy are marked by star symbols. For the companion we report both the cases with assumed dust temperature of $47\,{\rm K}$ and $35\,{\rm K} $, these points are connected by a dashed line. The diamond marks the $3\sigma$ upper limit on quasar ULAS J1342+0928 at $z=7.54$ \citep{Novak+2019}, while the square is the upper limit on SDSS J2310+1855 at $z=6.00$ \citep{Li+2020b,Li+2020}. The latter limits are computed by use the [NII]$_{\rm 122\mu m}$/[NII]$_{\rm 205\mu m}$ line luminosity ratio to estimate the electron density and then the fraction of [CII]$_{\rm 158\mu m}$ from the ionized gas from \citet{Oberst+2006}. $T_{\rm dust}$ and $L_{\rm TIR}$ for SDSS J2310+1855 are derived via a dust SED fitting by \citet{Shao+2019}, while for ULAS J1342+0928, $L_{\rm TIR}$ is estimated in \citet{Novak+2019} by assuming $T_{\rm dust}=47\,{\rm K}$ (here slightly shifted for better visualization). Circles represent the LIRG sample presented in \citet{Diaz-Santos+2017}. The black solid line is the best fit model $f({\rm [CII]_{158\mu m}^{\rm PDR}}) = 0.82(\pm0.01)+0.41(\pm0.04)\log(S_{63}/S_{158})$, where $S_{63}$ and $S_{158}$ are the continuum flux density at $63\,{\rm \mu m}$ and $158\,{\rm \mu m}$, respectively. The gray shaded area is the $1\sigma$ confidence interval. The approximate relation between the flux density ratio and $T_{\rm dust}$ is provided by \citet{Diaz-Santos+2017}, $T_{\rm dust} = 20.24 + 14.54(S_{63}/S_{158}) - 3.75(S_{63}/S_{158})^2 + 0.46(S_{63}/S_{158})^3$.}
         \label{fig:cii_pdr}
\end{figure}

We note that, while the [CII]/[NII] ratio for HII regions assumes Solar metallicities, the high dust content seen in $z\apprge 6$ quasar host galaxies suggests that the metallicities are indeed high (see e.g. \citealt{Novak+2019}, furthermore this is also supported --although indirectly-- by studies on the metallicity of the broad line region in $z\sim6$ quasars, see e.g. \citealt{Pentericci+2002, Jiang+2007, Onoue+2020}). Combining our [CII]$_{\rm 158\mu m}$ line luminosity measurements with those of the [NII]$_{\rm 205\mu m}$ line (see Table~\ref{tbl:line_data}), we conclude that PDRs account for $\apprge 80\%$ of the [CII]$_{\rm 158\mu m}$ emission in both the PJ231-20 QSO and its companion galaxy. This result is consistent with previous studies on local starburst galaxies showing that no more that $\sim30\%$ of the [CII]$_{\rm 158\mu m}$ emission is emitted by the diffuse ionized medium \citep[e.g.][]{Carral+1994, Lord+1996, Colbert+1999}. Studies of the  [CII]/[NII] ratio in FIR bright galaxies at higher redshifts ($z>4-5$) have also concluded that only a small fraction of [CII]$_{\rm 158\mu m}$ emission arises from the ionized gas-phase ISM \citep[see e.g.][]{Decarli+2014, Pavesi+2016}.

In Figure~\ref{fig:cii_pdr} we compare $f(\cii^{\rm PDR})$ and the dust temperature of the PJ231-20 QSO and its companion with a sample of local luminous infrared galaxies (LIRGs) reported in \citet{Diaz-Santos+2017}, and with the value of two high-$z$ quasars: the SDSS J2310+1855 at $z=6.00$ \citep{Li+2020b,Li+2020} and the most distant quasar know so far, ULAS J1348+0928 at $z=7.54$ \citep{Novak+2019}. \citet{Diaz-Santos+2017} find that galaxies with warmer $T_{\rm dust}$ show an higher $f(\cii^{\rm PDR})$; therefore the fraction of [CII]$_{\rm 158\mu m}$ emitted from PDR increases in highly star-forming systems. The authors also rule out a significant role for AGN in increasing $f(\cii^{\rm PDR})$ in galaxies. A possible interpretation is that HII regions are more enshrouded by dust in high $T_{\rm dust}$ galaxies than those of more evolved systems with lower $T_{\rm dust}$ in which stellar feedback processes have already cleared out most of the dust from the star-forming regions \citep[][and references therein]{Diaz-Santos+2017}. That is, the fractions of [CII]$_{\rm 158\mu m}$ arising from the ionized medium is associated with low-density 'fossil' HII regions and diffuse ionized gas not associated with hot young stars. While the high dust temperatures ($47 - 54\,{\rm K}$) of the high-$z$ galaxies presented in this work lie outside the range of the LIRG sample, in the case of the PJ231-20 companion, the assumption of $T_{\rm dust}=35\,{\rm K}$ place this object in agreement with the trend of local LIRGs shown in Fig.~\ref{fig:cii_pdr}. 

\subsection{IR line deficits}
\label{ssect:line_deficit}
The ratio of the luminosity of ISM cooling lines, such as $L_{\rm [CII]{158\mu m}}$, $L_{\rm [NII]{205\mu m}}$ to the total IR luminosity ($L_{\rm TIR}$), measures the ratio of the cooling of the gas to that of the dust. These ratios decrease by $\sim1-2$ orders of magnitude with increasing dust temperature and IR luminosity in local galaxies \citep[e.g.][]{Malhotra+1997,Gracia-Carpio+2011,Farrah+2013, Diaz-Santos+2013,Diaz-Santos+2017,Zhao+2013,Zhao+2016,Herrera-Camus+2018} and those at high-redshift \citep[e.g.][]{Decarli+2012, Decarli+2014, Banados+2015, Novak+2019, Rybak+2019}. The underlying cause of the so-called 'line deficits' are still debated. The physical arguments most often proposed include changes in ionization parameter producing high dust to gas opacity ratio, optically thick dust screening part of the line emission, and progressive thermalization of the lines with low critical densities. However, a possible explanation is that HII regions are dustier in IR-bright galaxies than in low-luminosity systems \citep[e.g.][]{Luhman+2003, Gonzalez-Alfonso+2008, Abel+2009, Gracia-Carpio+2011, Farrah+2013, Riechers+2014, Diaz-Santos+2017, Herrera-Camus+2018b}. In this scenario, a higher fraction of UV photons produced by young stars are absorbed by dust, thus suppressing ionizing photons that would otherwise be absorbed by the neutral medium, thereby decreasing the photoionization heating efficiency the net effect is to decrease the line emission relative to the IR luminosity. On the other hand, if [CII]$_{\rm 158\mu m}^{\rm ion}$ and [NII]$_{\rm 205\mu m}$ emission arise from different ISM phases (i.e., low-density regions and diffuse ionized gas not associated with hot young stars) than the IR continuum emission, then the line deficit could be driven by a boosting of IR luminosity rather than a deficit in the observed line flux. However, consistently with what was discussed in Sect.~\ref{ssect:cii_fraction}, a fraction of this excess of energy can be transferred to the surrounding neutral/molecular medium increasing the [CII]$_{\rm 158\mu m}^{\rm PDR}$ emission and therefore decreasing the deficit.

\begin{figure}
	\centering
   	\includegraphics[width=0.75\hsize]{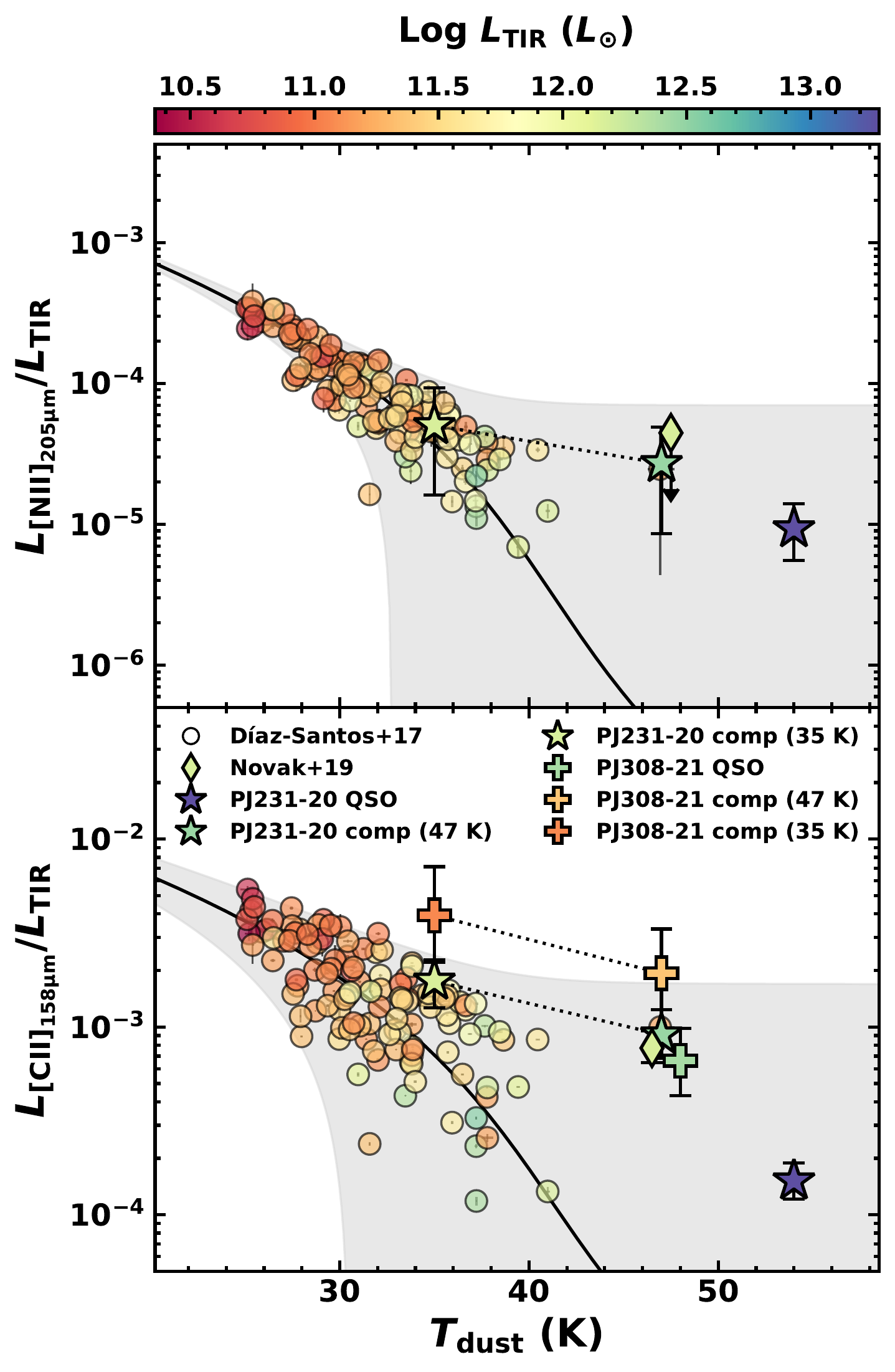}
      	\caption{[NII]$_{\rm 205\mu m}$ and [CII]$_{\rm 158\mu m}$ line deficits (top and bottom panel, respectively) with respect to total infrared luminosity ($L_{\rm TIR}(8-1000\,{\rm \mu m})$) for the PJ231-20 and PJ308-21 quasars and companion galaxies, and the LIRG sample of \citet{Diaz-Santos+2017} as a function of dust temperature and color coded by $\log L_{\rm TIR}$. We also report the measurements obtained for the $z=7.54$ quasar ULAS J1342+0928 \citep{Novak+2019}. The sources in PJ231-20 systems are indicated by stars, while those in PJ308-21 system are marked with crosses. For the companion galaxies we report both the cases with assumed dust temperature of $47\,{\rm K}$ and $35\,{\rm K}$, these points are connected by a dashed line. The solid lines are the best fit to the LIRG sample performed in \citet{Diaz-Santos+2017} using the model $L_{\rm line}/L_{\rm FIR} = \epsilon_0 \exp[(-S_{63}/S_{158})/\delta]$ with $(\epsilon_0,\delta)=((1.26\pm0.16)\times10^{-3},0.50\pm0.04)$ for the [NII]$_{\rm 205\mu m}$ line, and $(\epsilon_0,\delta)=((14.0\pm0.9)\times10^{-3},0.68\pm0.04)$ for [CII]$_{\rm 158\mu m}$ line. We scaled the best fits by a constant factor due to the different FIR luminosity definition employed by \citet{Diaz-Santos+2017}. The best-fit scaling values are $0.56$ for [NII]$_{\rm 205\mu m}$ line, and $0.44$ for [CII]$_{\rm 158\mu m}$ line. Grey shaded area represents the $1\sigma$ scatter of the relation. The \citet{Novak+2019} points were slightly shifted along the dust temperature axis (assumed to be $47\,{\rm K}$) for a better visualization of the data.
		}
         \label{fig:line_deficit}
\end{figure}

In Fig.~\ref{fig:line_deficit} we compare the [CII]$_{\rm 158\mu m}$ and [NII]$_{\rm 205\mu m}$ line deficits in the quasars and companion galaxies presented in this work, with the local LIRG sample of \citet{Diaz-Santos+2017} and the $z=7.54$ J1342+0928 quasar \citep{Novak+2019}. Our findings are consistent within the scatter with the overall trends observed in LIRGs. In particular, the line deficit of the companion galaxy in PJ231-20 system, in the low dust temperature case ($T_{\rm dust} = 35\,{\rm K}$), is typical of the bulk of local LIRG population. Noticeably, while we find that PJ231-20 QSO host and companion galaxies show [NII]$_{\rm 205\mu m}$ line deficits that are consistent within the errors, the deficit in [CII]$_{\rm158\mu m}$ is significantly more pronounced in the quasar than in the companion galaxy. However, we do not believe that this is evidence that the presence of a powerful AGN affects the [CII]$_{\rm158\mu m}$ line deficit (unless the presence of an AGN biases our $L_{\rm TIR}$ measurement): this scenario is excluded from the study of local ULIRGs sample  \citep[e.g.][]{Diaz-Santos+2013, Diaz-Santos+2017, Farrah+2013} showing no strong dependence of line deficit and [CII]$_{\rm 158\mu m}^{\rm PDR}$ fraction with the AGN contribution. In this context, not even higher ionization lines like [OIII]$_{\rm 88\mu m}$ seem to be significantly influenced by AGN \citep[see e.g.][]{Walter+2018}. Indeed, in the PJ308-21 system, the quasar shows similar [CII]$_{\rm 158\mu m}$ deficit to that of the companion galaxy, and both sources exhibit less extreme deficits than in the PJ231-20 QSO and its companion, thus suggesting that the observed line deficits is not trivially affected by the presence of an AGN. The observed scatter in the $L_{\rm [CII]{\rm 158\mu m}}/L_{\rm TIR}$ ratio could be due in fact to the variation of the PDR contribution to the total [CII]$_{\rm 158\mu m}$ with increasing $T_{\rm dust}$ \citep[][see Fig.~\ref{fig:cii_pdr}]{Diaz-Santos+2017}.

\section{Atomic medium}
\label{sect:atomic_medium}
\subsection{The [CI]$_{\rm 369\mu m}$ fine-structure line}
\label{ssect:atomic_medium}
We detect the atomic fine-structure line [CI]$_{\rm 369\mu m}$ (${^3P_2}\!\!\to\!\!{^3P_1}$) in the PJ231-20 QSO with a significance of $\sim 4.5\sigma$ while we derived upper limits for the companion and for objects in the PJ308-21 system. This line has a critical density of $1.2\times10^3\,{\rm cm^{-3}}$ \citep[see e.g.][and references therein]{CarilliWalter2013}, hence it traces the cold dense atomic neutral medium. As the ionization potential of neutral carbon is close to the dissociation energy of the CO molecule it is expected to exist in a narrow range of physical conditions. The [CI] emission was once thought to emerge from a thin C$^+$/C/CO transition layer in the molecular clouds \citep{Tielens+1985a, Tielens+1985b, Hollenbach+1999,Kaufman+1999}. However, several lines of evidence highlight a close connection between [CI] emission and CO rotational transitions, suggesting that the [CI] line traces the bulk of molecular gas in galaxies \citep[see e.g.,][]{Ikeda+2002, Papadopoulos+2004, Walter+2011, Alaghband-Zadeh+2013, Israel+2015, Bothwell+2017, Valentino+2018, Valentino+2020}. This is also predicted by hydrodynamic simulations \citep[e.g.][]{Tomassetti+2014} in which CO and [CI] are found to co-exist throughout the bulk of the cold molecular component. A viable explanation is provided by clumpy, inhomogeneous PDR models \citep[e.g.,][]{Meixner+1993,Spaans+1996} in which the surface layers of [CI] are distributed across clumpy molecular clouds. 

\begin{figure}
	\centering
   	\includegraphics[width=\hsize]{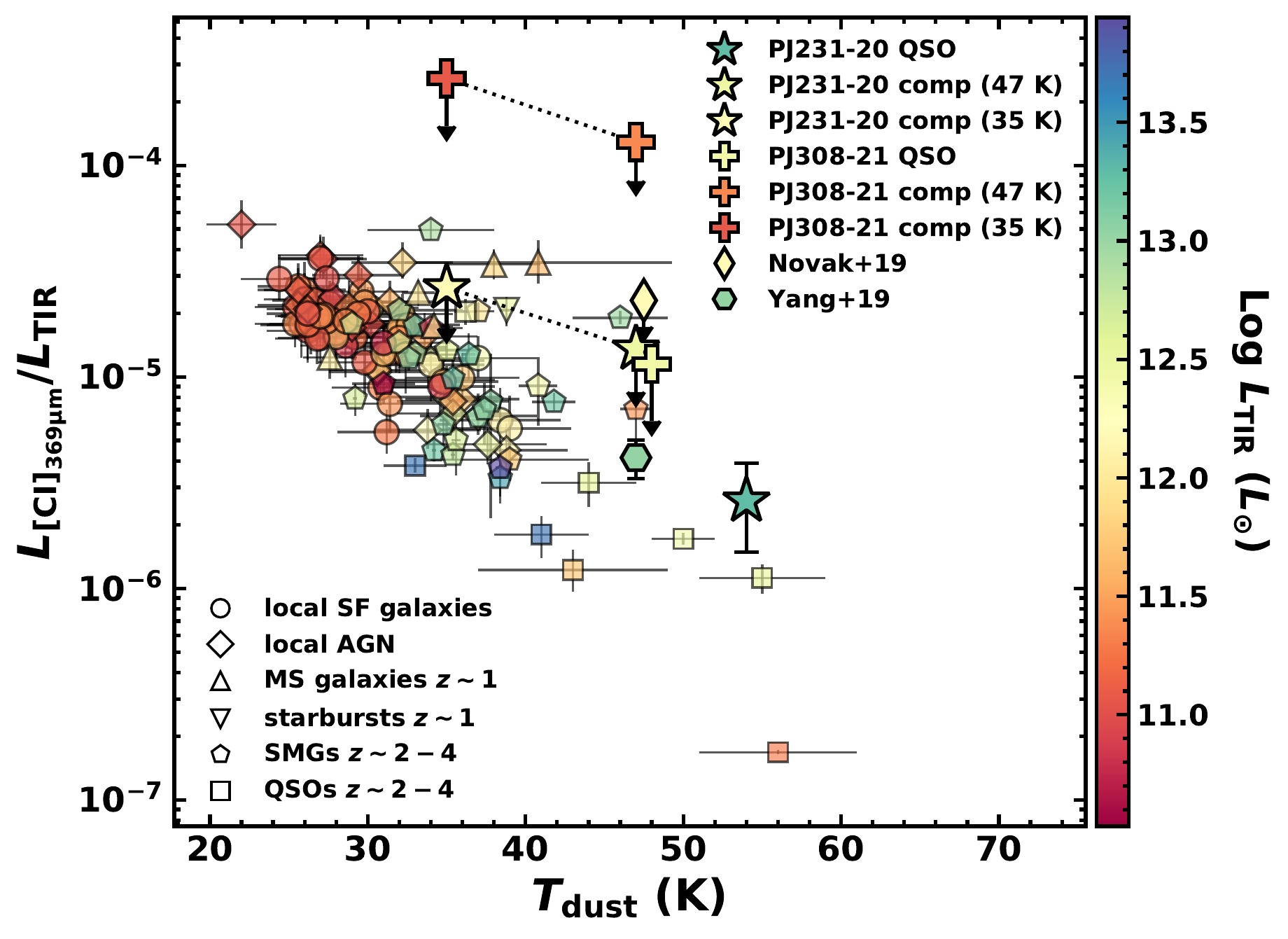}
      	\caption{Comparison of $L_{\rm [CI]{369\mu m}}/L_{\rm TIR}$ and dust temperature ($T_{\rm dust}$) observed in the PJ231-20 and PJ308-21 systems, with various local and high-$z$ galaxies. Data points are color-coded by $\log L_{\rm TIR}$. The reported sample is taken from \citet{Valentino+2020} (and references therein) and comprises local star-forming (SF) galaxies (circles) and AGN (diamonds); main-sequence (MS) galaxies (upright triangles) and starburst galaxies (upside-down triangles) at $z\sim 1$; and $z\sim2-4$ SMGs (pentagons) and QSOs (squares). The PJ231-20 QSO and its companion are indicated with stars, while the PJ308-21 sources are marked with crosses. For the companions we report both the cases with assumed dust temperature of $47\,{\rm K}$ and $35\,{\rm K}$, these points are connected by a dashed line. The upper limit in our sources correspond to $3\sigma$. We also report measurements in the quasar ULAS J1342+0928 at $z=7.54$ \citep[see][]{Novak+2019}, and the quasar J0439+1634 at $z\simeq6.5$ \citep{YangJ+2019}.} 
         \label{fig:ci_fir}
\end{figure}
\begin{figure*}[!htbp]
	\centering
	\resizebox{0.85\hsize}{!}{
	\includegraphics[width=\hsize]{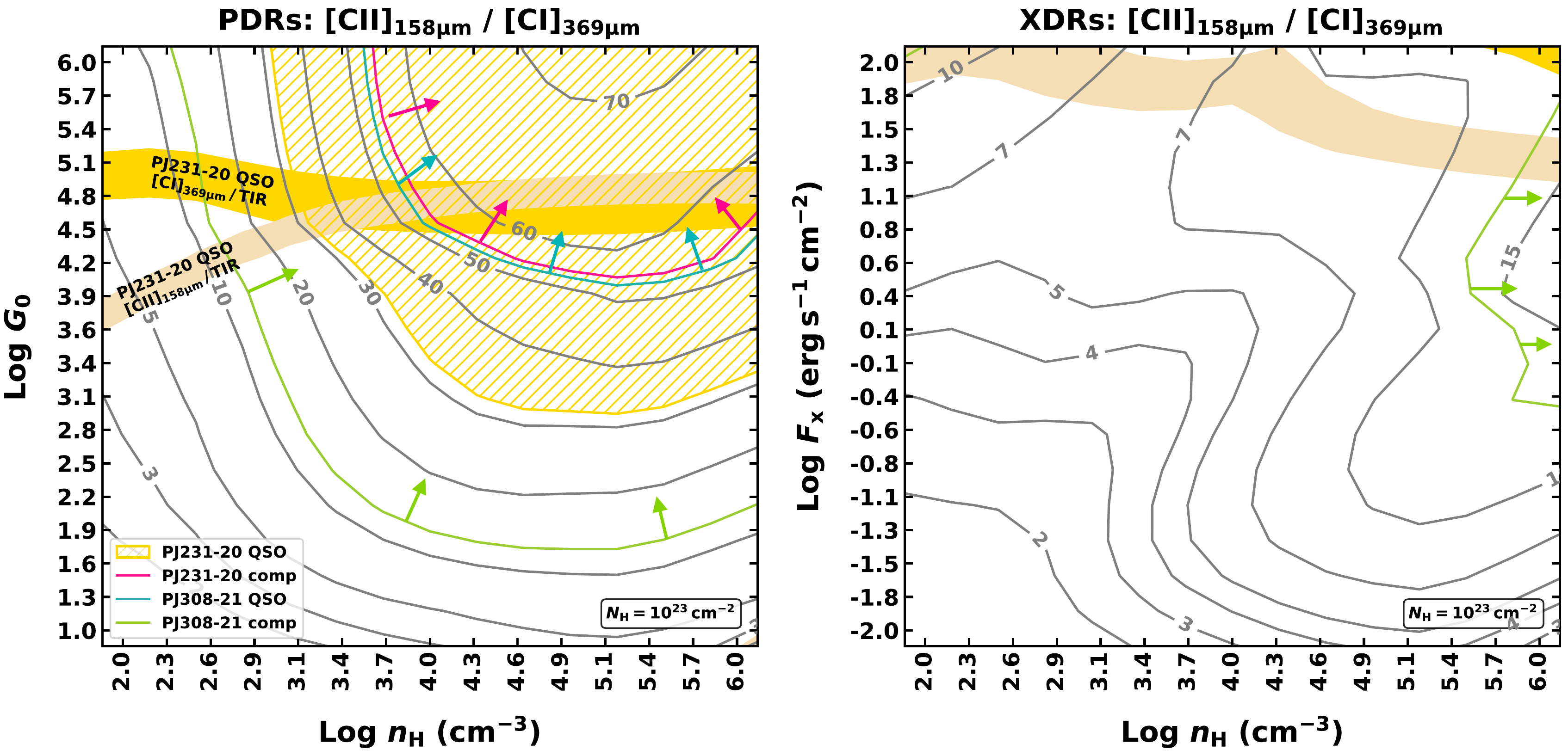}}
      	\caption{Grids of [CII]$_{\rm 158\mu m}$/[CI]$_{\rm 369\mu m}$ intensity line ratio as function of the strength of radiation field ($G_0$ or $F_{\rm X}$) and total hydrogen density ($n_{\rm H}$), in the PDR case ({\it left panel}) and XDR case ({\it right panel}). The adopted total hydrogen column density is $N_{\rm H}=10^{23}\,{\rm cm^{-2}}$. Model values are indicated by gray contours (linear scale), while constraints on observed line ratios are superimposed. Gold dashed area indicates the [CII]$_{\rm 158\mu m}$/[CI]$_{\rm 369\mu m}$ luminosity ratio measured in the PJ231-20 QSO within its uncertainties, while $3\sigma$ lower limits on the other sources are reported (see legend at the bottom left of the left panel). [CII]$_{\rm 158\mu m}$/[CI]$_{\rm 369\mu m}$ ratios in PJ231-20 QSO and companion are scaled according to the estimated fraction of [CII]$_{\rm 158\mu m}$ arising from PDRs. The XDR case does not reproduce the observed ratios in any source except for a small range of values corresponding to the lower limit measured in PJ308-21 QSO. We also report the [CI]$_{\rm 369\mu m}$/TIR and [CII]$_{\rm 158\mu m}$/TIR luminosity ratios estimates in PJ231-20 QSO as gold and light brown filled areas, respectively, providing additional constraints on the radiation field flux.}
         \label{fig:cii_ci_ratio}
\end{figure*}

\subsection{The [CI]$_{\rm 369\mu m}/{\rm TIR}$ ratio}
In Fig.~\ref{fig:ci_fir} we show $L_{\rm [CI]{369\mu m}}/L_{\rm TIR}$ ratio as a function of dust temperature for a variety of galaxies and AGN in the local and high-$z$ Universe \citep[see][and references therein for full details]{Valentino+2020}. The observed average ratio is $L_{\rm [CI]{369\mu m}}/L_{\rm TIR}\sim 10^{-5}$, but there is a negative correlation with $T_{\rm dust}$. However, contrary to the [NII]$_{\rm 205\mu m}$ and [CII]$_{\rm 158\mu m}$ line deficits (see Fig.~\ref{fig:line_deficit}), this trend is not surprising. Indeed, [CI] in PDR models are predicted to arise from regions with gas kinetic temperature comparable to dust temperature $T_{\rm dust}$ \citep[e.g.][]{Tielens+1985a} with the IR luminosity rises with a high power of $T_{\rm dust}$ ($L_{\rm TIR}\propto T_{\rm dust}^{4+\beta}$), while the line emission into the cloud is only proportional to the gas temperature. Then, the trend shown in Fig.~\ref{fig:ci_fir} is also driven by a redshift dependence. Indeed, \citet{Valentino+2018, Valentino+2020} found that, on average, the observed $L_{\rm [CI]{369\mu m}}/L_{\rm TIR}$ ratios are lower in high-$z$ SMGs than in local star-forming and main sequence (MS) galaxies (i.e. those defining a relative tight distribution in SFR and stellar mass $M_\star$, see e.g. \citealt{Noeske+2007, Elbaz+2011}). Hence, the most plausible underlying reason of this trend is a potential evolution in the dust temperature with redshift enhancing $L_{\rm TIR}$ \citep[e.g.][]{Schreiber+2018, Liang+2019, Ma+2019, Faisst+2020,Riechers+2020}.  
Furthermore, the IR luminosity measured in local AGN and high-$z$ quasars (that also are expected to have higher $T_{\rm dust}$ than star-forming galaxy) could be affected by a substantial contribution from a dusty torus, boosting the observed $L_{\rm TIR}$, thus driving the observed decrease of $L_{\rm [CI]{369\mu m}}/L_{\rm TIR}$ with $T_{\rm dust}$. However, $L_{\rm [CI]{369\mu m}}$ and $L_{\rm TIR}$ trace the neutral gas and the SFR, respectively, thus their ratio is a proxy of the gas depletion timescale $\tau_{\rm dep} = M_{\rm gas}/{\rm SFR}$, representing the efficiency of the star formation. Therefore, the observed trend of $L_{\rm [CI]{369\mu m}}/L_{\rm TIR}$ could be interpreted as due to a decrease of $\tau_{\rm dep}$ (or equivalently higher star-formation efficiency) in high-$z$ SMGs and quasar hosts than in MS and local star-forming galaxies. This result is consistent with several CO-based studies \citep[e.g.][]{Daddi+2010, Genzel+2015, Tacconi+2018, Birkin+2020}.

We obtained $3\sigma$ upper limits of [CI]$_{\rm 369\mu m}$ line for all sources in the systems studied in this paper, except the PJ231-20 QSO for which we measured $L_{\rm [CI]{369\mu m}}/L_{\rm TIR} =2.6^{+1.4}_{-1.1}\times 10^{-6}$. This value together with the  assumed $T_{\rm dust}$, locate this source in the parameter ranges observed in other high-$z$ quasars. In particular a similar $L_{\rm [CI]{369\mu m}}/L_{\rm TIR}$ ratio ($4.2^{+0.8}_{-0.8}\times10^{-6}$) has been measured by \citet{YangJ+2019} in the $z\simeq 6.5$ quasar \object{J0439+1634}. This result, suggests high star-formation efficiency in these sources relative to their local counterparts. For the other sources studied in this work, we obtained $L_{\rm [CI]{369\mu m}}/L_{\rm TIR} < 10^{-4}-10^{-5}$ limits that are $\sim 1-2$ order of magnitude lower than $L_{\rm [CII]{158\mu m}}/L_{\rm TIR}$. We also estimated a $L_{\rm [CI]{369\mu m}}/L_{\rm [CII]{158\mu m}} \sim 2\%$ in PJ231-20 QSO and $<2-6\%$ in all the other sources. This values are similar to that measured in the quasar host galaxy J1148+5251 at $z=6.42$ by \citet{Riechers+2009}. These results support the assumption that [CI]$_{\rm 369 \mu m}$ is a less important coolant than the dust continuum and [CII]$_{\rm 158 \mu m}$ at such high redshifts. In Sect.~\ref{sect:mass_contributions} we further exploit the [CI]$_{\rm 369\mu m}$ detections in quasar hosts and companions to put constraints on the amount of neutral carbon and molecular gas in the ISM.

\subsection{[CII]$_{\rm 158\mu m}$/[CI]$_{\rm 369\mu m}$ ratio}
\label{ssect:cii_ci}
The [CII]$_{\rm 158\mu m}$/[CI]$_{\rm 369\mu m}$ ratio is extremely useful in discerning between PDR and XDR models, as has been used in other high-$z$ studies \citep{Venemans+2017b, Venemans+2017c, Novak+2019}. In Fig.~\ref{fig:cii_ci_ratio} we show the [CII]/[CI] intensity ratio obtained from our {CLOUDY} grids with $N_{\rm H}=10^{23}\,{\rm cm^{-2}}$. In the PDR case, the [CII]/[CI] ratio ranges between $\apprle 2$ and $\apprge 70$, while in the XDR case it does not exceed a value of $\sim 15$. These ranges are consistent with those that are found in other models in the literature \citep[e.g.][]{Kaufman+1999, Meijerink+2007}. The different values of the [CII]/[CI] ratio predicted in PDR and XDR models can be explained by the lower CO/C abundance ratio produced in the XDR regime \citep[see e.g.][]{Maloney+1996, Meijerink+2005}. Indeed, while in PDRs there is a sharp C$^+$/C/CO transition layer at a certain depth in the clouds (depending on both density and FUV field strength), X-ray photons can penetrate much deeper into the cloud affecting its whole  structure. As a result, in the XDR regime, both C and C$^+$ are present throughout most of the cloud and their column density increases much more gradually as a function of cloud depth, than in the PDR case.

Additional constraints on the radiation field in galaxies are provided by [CI]$_{\rm 369\mu m}/{\rm TIR}$ and [CII]$_{\rm 158\mu m}/{\rm TIR}$ luminosity ratios. Indeed, in the PDR case, for high FUV fluxes ($G_0\apprge 10^2$) these ratios are expected to decease as $G_0$ increases, with almost no dependence on ISM density. In fact, by increasing $G_0$, the C$^+$/C/CO transition layer, where [CI] transitions take place, is pushed deeper into the cloud while the C column density remains substantially unaffected. The gas heating efficiency due to photoelectric effect on dust grains reaches its maximum values at $G_{0}\sim 10-100$, and becomes less efficient at higher fluxes, while the dust heating per UV photon remains at the same level \citep[e.g.][]{Kaufman+1999}. In this case, the [CII]$_{\rm 158\mu m}$ line luminosity increases logarithmically while IR-luminosity is linearly proportional to $G_0$. As a result, both [CI]$_{\rm 369\mu m}$/TIR and [CII]$_{\rm 158\mu m}$/TIR luminosity ratios, depends mostly on $G_0$ for high UV fluxes \citep[see also e.g.][]{Tielens+1985a, Kaufman+1999, Gerin+2000}.

The observed luminosity ratios, corrected for the fraction of [CII]$_{\rm 158\mu m}$ actually arising from PDRs (see Sect.~\ref{ssect:cii_fraction}) are $L_{\rm [CII]{158\mu m}}/L_{\rm [CI]{\rm 369\mu m}}=46^{+30}_{-14}$, for the PJ231-20 QSO,
and $3\sigma$ limits of $> 55, 53, 13$ in the PJ231-20 companion, the PJ308-21 QSO, and the PJ308-21 companion galaxy respectively. Thus for all the sources, the PDR scenario is favored. In the case of the PJ231-20 QSO, this diagnostic points to a density $n_{\rm H}>10^3\,{\rm cm^{-3}}$ and a FUV radiation field strength of $> 10^3\,G_0$. Similar constraints are inferred for all the other sources (see Fig.~\ref{fig:cii_ci_ratio}). In addition, by combining the estimated values of $L_{\rm [CI]{369\mu m}}/L_{\rm TIR}$ and $L_{\rm [CII]{158\mu m}}/L_{\rm TIR}$ ratios with the [CII]$_{\rm 158\mu m}$/[CI]$_{\rm 369\mu m}$ CLOUDY models (see Fig.~\ref{fig:cii_ci_ratio}), we constrain FUV flux to $G_0 \simeq 3\times10^4-10^5$ in the PJ231-20 QSO. A similar value is found for the J0439+1634 quasar at $z\simeq6.5$ by \citet{YangJ+2019}.

\section{Molecular medium}
\label{ssect:molecular_medium}
\subsection{CO rotational lines}
\label{ssect: co_rotational_lines}
Molecular clouds consist almost entirely of molecular hydrogen (H$_2$). Unfortunately, H$_2$ has a zero electric dipole moment, and high vibrational energy levels, thus it is a poor radiator in the physical conditions of the cold ISM. The most abundant molecule after H$_2$ is CO; it has a weak permanent dipole moment and its rotational levels are primarily populated by collisions with H$_2$. CO emission has been widely detected in normal galaxies in the local Universe (see e.g. \citealt{Saintonge+2017,Saintonge+2018}; and \citealt{Tacconi+2020} for a review) and in the highest redshift quasars and SMGs \citep[e.g.][]{Riechers+2013, Strandet+2017, Venemans+2017a, Novak+2019, Li+2020}, and it provides key information to characterize the ISM. Low-$J$ ($J_{\rm up} < 6$) CO lines have low critical density ($n_{\rm crit}\sim10^3\,{\rm cm^{-3}}$) and excitation temperature, therefore they are associated with cold molecular gas in PDRs. In contrast, high-$J$ ($J_{\rm up} > 7$) transitions have high critical density ($n_{\rm crit}\sim10^5\,{\rm cm^{-3}}$) and mainly trace the high density and/or high temperature gas, and their excitation is typically explained by the presence of an intense X-ray radiation field impinging the molecular clouds in the central region of galaxies (XDRs), or by shocks produced by AGN-driven outflows \citep[e.g.][]{Meijerink+2013, Gallerani+2014, Mingozzi+2018, Carniani+2019}. Therefore, by combining multiple CO lines, one can study the CO spectral line energy distribution (SLED) that is a powerful tool to disentangle the contribution of star formation and AGN (PDRs vs. XDRs) to the CO line emission and to constrain the properties of different gas phases such as the density of the medium and the intensity of the radiation field \citep[e.g.][]{Riechers+2009, Riechers+2013, Stefan+2015, Yang+2017, Canameras+2018, Carniani+2019}.

We detect multiple CO lines (7--6, 10--9, 15--14, 16--15) in the PJ231-20 QSO host with various degrees of significance, while we detect only CO(7--6) and CO(10--9) in the companion galaxies. We also only have a marginal CO(15--14) detection in the PJ308-21 QSO. The measurements are reported in Table~\ref{tbl:line_data}. 

\subsection{CO(1--0)--normalized SLEDs}
\label{ssect:co_sleds}
The CO SLED shows the relative luminosity of various CO transitions in an astrophysical source. It is used to gauge the underlying physical conditions (density, intensity and source of the radiation field) of the CO-emitting gas. The intensity of the observed lines is usually normalized to the ground transition, CO(1--0), which however is not observable with ALMA in the targeted sources. Nevertheless, we estimate its intensity following empirical relations between the CO(1--0) and the IR luminosity \citep[see e.g.][for a review]{CarilliWalter2013}. Alternatively, the dust masses estimated via the Rayleigh-Jeans dust continuum (Sect.~\ref{ssect:dust}) can be converted into a CO(1--0) luminosity via a gas-to-dust mass ratio, $\delta_{\rm gdr}$, and a CO-to-H$_2$ (light-to-mass) conversion factor, $\alpha_{\rm CO}$ \citep[see][for a review]{Bolatto+2013}: $L'_{\rm CO(1-0)} = M_{\rm dust} \, \delta_{\rm gdr}/\alpha_{\rm CO}$.

Therefore, by starting from the dust masses derived in Sect.~\ref{ssect:dust} we first estimated the molecular gas mass $M_{\rm H_2}$ of our sources, by adopting $\delta_{\rm gdr}= 100$ (e.g. \citealt{Genzel+2015}, see Sect.~\ref{sect:mass_contributions}) then, by employing the typical value of $\alpha_{\rm CO} = 0.8\,({\rm K\, km\,s^{-1}\,pc^2})^{-1}$ \citep{Downes+1998}, adopted in the literature for SMGs and quasar hosts (see \citealt{CarilliWalter2013} for a comprehensive discussion) and assuming that $75\%$ of the dust-derived gas mass is in molecular form \citep[e.g.][]{Riechers+2013,Wang+2016, Venemans+2017a}, we find $L_{\rm CO(1-0)} \sim 2.4 \times 10^6\,L_{\astrosun}$ for PJ231-20 QSO, and $L_{\rm CO(1-0)} \sim 2.6 \times 10^5\,L_{\astrosun}$ for PJ308-21 QSO. On the other hand, in the case of companions galaxies,  assuming $T_{\rm dust}=47\,{\rm K}$ we find $L_{\rm CO(1-0)} \sim1.1 \times 10^6\,L_{\astrosun}$ and
$\sim0.4 \times 10^5\,L_{\astrosun}$ for the source in PJ231-20 and PJ308-21 system, respectively. These values, increase by a factor of $\sim2\times$, assuming a dust temperature of $35\,{\rm K}$. However, the latter are consistent within the uncertainties with the values obtained adopting a higher dust temperature.

\begin{figure}
	\centering
   	\includegraphics[width=\hsize]{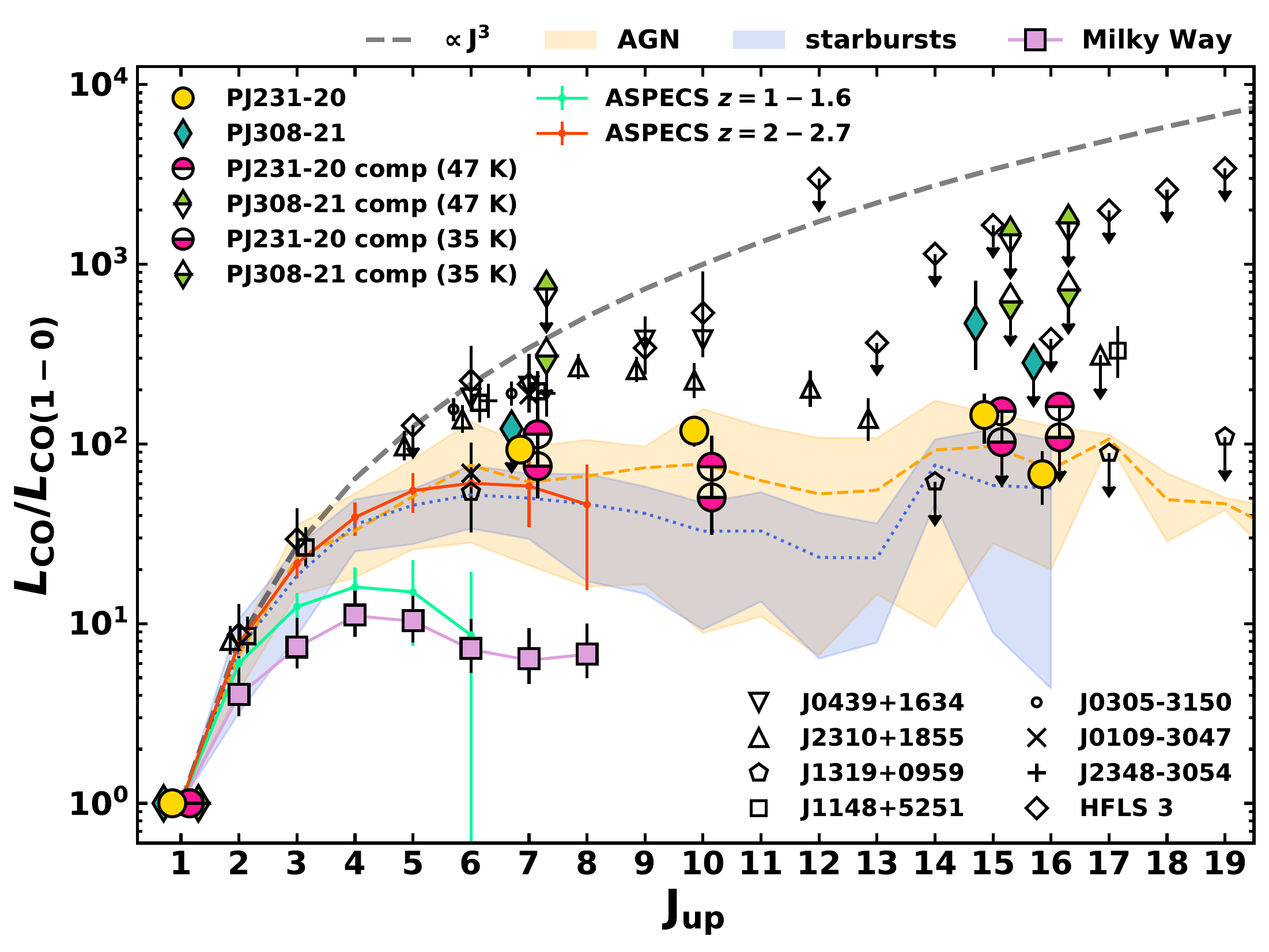}
      	\caption{Average CO(1--0)--normalized SLEDs (in $L_{\astrosun}$). Colored circles and diamonds are measurements obtained in this work for PJ231-20 and PJ308-21 systems. For companion galaxies we show both SLEDs obtained assuming $T_{\rm dust}=47\,{\rm K}$ (half top filled symbols) and $35\,{\rm K}$ (half bottom filled symbols). Empty symbols refers to the CO ladders of other high-$z$ quasars and one SMG (HFLS 3) from the literature \citep[see][and references therein]{Riechers+2013, Venemans+2017a, Carniani+2019, YangJ+2019, Li+2020}. Average CO SLEDs of local AGN and starburst galaxies from the literature \citep{Mashian+2015, Rosenberg+2015} are reported by the orange dashed and blue dotted lines, respectively; shaded areas are the confidence limits (computed as 16th and 85th percentile of the distribution, respectively). Violet squares are the Milky Way Galactic Center CO-normalized ladders from \citet{Fixsen+1999}. Average CO SLEDs of starburst galaxies at $z\sim1-1.6$ and $z\sim2-2.6$ from ASPECS, are shown as green and red lines, respectively. The gray dashed line is the expected thermalized CO SLED in the Rayleigh-Jeans limit.}
         \label{fig:co_sleds_comp}
\end{figure}

In Fig.~\ref{fig:co_sleds_comp} we compare the CO measurements of our sources, normalizing by CO(1--0) with the average CO(1--0)--normalized SLEDs of low-$z$ starburst galaxies and AGN, respectively \citep{Mashian+2015, Rosenberg+2015}, the Milky Way Galactic Center \citep{Fixsen+1999} and various measurements of $z>6$ sources retrieved from the literature. High-$z$ sources shown include the quasars \object{J2310+1855} at $z=6.00$ \citep{Carniani+2019, Shao+2019, Li+2020}; \object{J1319+0959} at $z=6.12$, \citep{Wang+2013, Carniani+2019}; \object{J1148+5251} at $z=6.4$, \citep{Walter+2003, Riechers+2009, Stefan+2015, Gallerani+2014}; J0439+1634 at $z=6.52$ \citep{YangJ+2019}, and \object{J0305-3150}, \object{J0109-3047}, \object{J2348-3054} at $z= 6.61,6.79,6.90$ respectively \citep{Venemans+2017a}\footnote{See Table 2 in \citet{Carniani+2019} and Table 1 in \citet{Venemans+2017a} for the complete collection of measurements.}, and the $z=6.34$ SMG HFLS 3 \citep{Riechers+2013}. For those quasars for which $L_{\rm CO(1-0)}$ was not available in the literature, we derive $L_{\rm CO(1-0)}$ from dust masses provided in \citet{Carniani+2019}, using the same assumptions described above for the PJ231-20 and PJ308-21 systems. In addition, we report the CO SLED models obtained from the ASPECS ALMA large program for $z\sim 1-2$ starburst galaxies \citep{Boogaard+2020}.

Despite the large uncertainties we can identify an overall trend for the low-$z$ starburst and AGN CO SLEDs. Indeed, starburst CO SLEDs typically reach a peak at $J_{\rm up} \simeq 6-7$ showing a steady decline afterwards. The average CO SLED of local AGN is more excited at high-$J$  ($J_{\rm up}\apprge9-10$) than are that of the starbursts, consistent with the scenario in which high-$J$ CO transitions are associated with a highly excited medium due to the strong radiation field or shocks produced by the AGN activity.

Here, the PJ231-20 QSO CO SLED is similar to those of local AGN, in particular the observed CO(10--9)/CO(1--0) luminosity ratio help us to discriminate between AGN and starbursts regime. On the other hand, the CO SLED of PJ231-20 companion galaxy is consistent within $\sim 1\sigma$ with both the starbursts and AGN population, even taking into account the upper limits on CO(15--14)/CO(1--0) and CO(16--15)/CO(1--0) luminosity ratios. On the other hand, if we assume a dust temperature of $35\,{\rm K}$, the CO SLED appears similar to the average CO SLED of starburst galaxies. However, \citet{Mazzucchelli+2019} do not detect the PJ231-20 companion in the rest-frame optical/UV wavelengths, revealing that this source is highly dust-enshrouded with a SED similar to Arp 220, that is both highly star-forming and highly dust-obscured ULIRG in the local Universe. In addition, \citealt{Connor+2020} do not detect the PJ231-20 companion even in X-ray observations, thus ruling out the presence of an AGN in this source, at least with luminosity similar to those of optically selected quasars. These previous studies suggest that star formation should dominate the radiation field (and therefore the CO excitation) in this source. While little information is available for the PJ308-21 companion, in the case of the quasar host we observe a high CO(15--14)/CO(1--0) luminosity ratio, possibly revealing the contribution of the AGN exciting this high-$J$ CO ladder \citep[see e.g.][]{Schleicher+2010, Gallerani+2014, Carniani+2019}. Finally, the observed CO(7--6)/CO(1--0) luminosity ratios in both PJ231-20 quasar host and the companion, and the upper limit available for PJ308-21 QSO, are lower than other $z>6$ quasars \citep[e.g.][]{Venemans+2017a}, while are consistent within uncertainties with average CO SLED model of the $z\apprge 2$ starburst galaxies in the ASPECS field. Interestingly, the latter appears fairly similar to the average CO SLED of local starburst galaxies. However, we note that the CO luminosity ratios in high-$z$ sources suffer from large uncertainties. In particular, the estimate of $L_{\rm CO(1-0)}$ is significantly affected by the assumed gas-to-dust ratio and the uncertainties on dust mass obtained from the dust SED (see e.g. Sect.~\ref{ssect:dust}). 

\subsection{CO SLED models}  
\label{ssect:co_models}
\begin{figure*}[!htbp]
	\centering
	\resizebox{\hsize}{!}{
	\centering
	\includegraphics[width=\hsize]{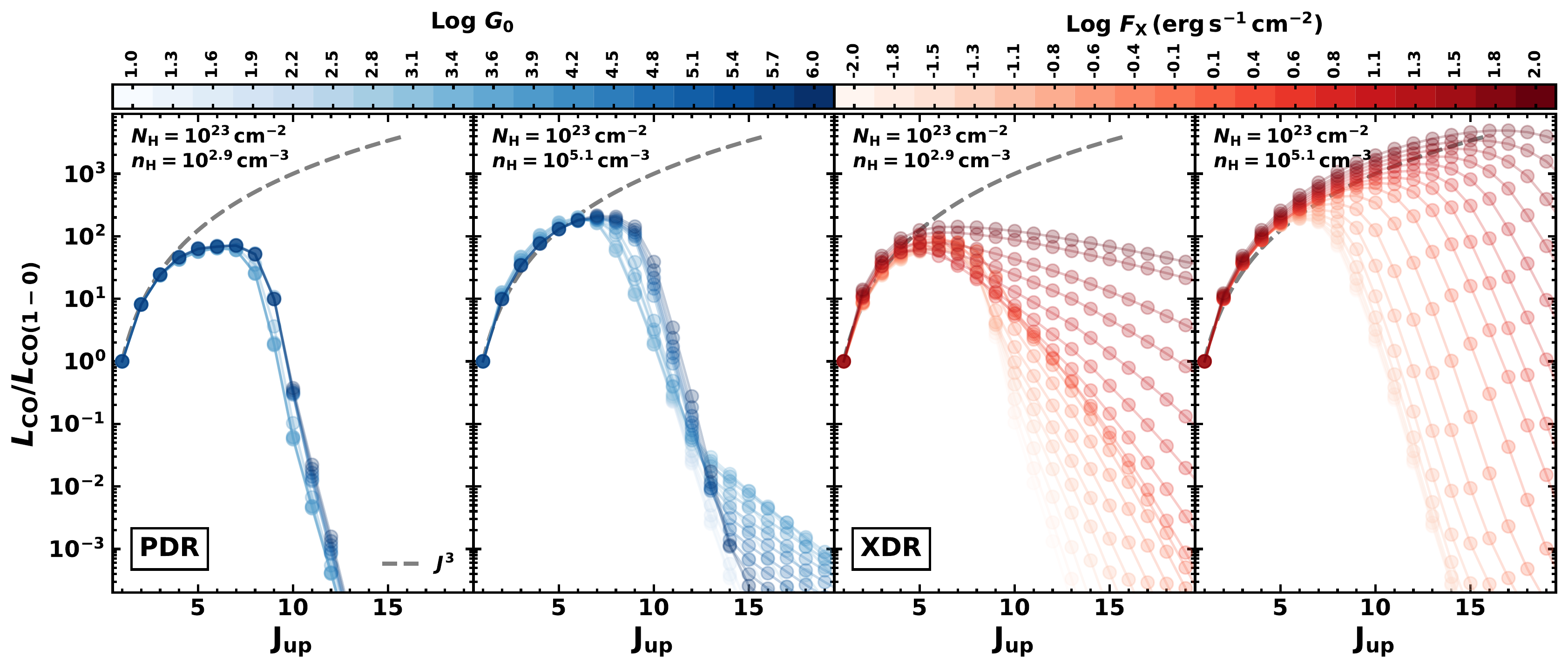}}
      	\caption{CLOUDY CO(1--0)-normalized SLED models in the PDR ({\it left panels}) and XDR regime ({\it right panels}) varying the intensity of the incident radiation field over the entire parameter space and for two reference values of total hydrogen density ($n_{\rm H}=10^{2.9},\,10^{5.1}\,{\rm cm^{-3}}$), and column density $N_{\rm H}=10^{23}\,{\rm cm^{-2}}$. PDR cases are shown in blue while XDRs are in red, both are color-coded according to the values of the radiation field flux. The gray dashed line indicates the $J^3$ curve, that is the expected trend of the CO SLED in LTE (in the optically thick and high temperature -- or low frequency -- limit).  }
         \label{fig:co_sled_models}
\end{figure*}

From the {CLOUDY} grids, we retrieved the CO line intensities and we normalized them to the CO(1--0) line intensity, thus obtaining the normalized CO SLEDs as function of $n_{\rm H}$ and radiation field flux ($G_0$ or $F_X$ in the case of PDR and XDR models, respectively). In Fig.~\ref{fig:co_sled_models} we show the CO SLED models in the PDR and XDR case for two values of cloud gas density, $n_{\rm H} \simeq 0.7\times10^{3}$ and $\simeq1\times10^{5}\,{\rm cm^{-3}}$, with varying radiation field strength at fixed hydrogen column density $N_{\rm H}=10^{23}\,{\rm cm^{-2}}$. The results show that in the PDR case the CO SLED is almost independent of $G_0$, contrary to the XDR case. These behaviors are the direct consequence of a completely different heating mechanism driven by absorption of UV and X-ray photons. Indeed, X-rays have very low cross sections and therefore can penetrate much deeper into the cloud (at column densities $>10^{22}\,{\rm cm^{-2}}$), furthermore, X-ray photons have high heating efficiency compared to UV photons, thus keeping molecular clouds at high temperature even at high column density \citep{Maloney+1996, Lepp+1996, Meijerink+2005}. As expected, in both the PDR and XDR cases, the peak of the CO SLED rises and shifts to higher $J$ values as the density increases. Indeed, the population of high-$J$ CO levels (set by the competition of collisional excitation and radiative de-excitation) increases with increasing cloud density due to the higher critical densities of such lines, and progressively thermalizes, saturating at a  certain value. In the optically thick and high temperature (or low-frequency) limit, the luminosities (in $L_{\astrosun}$ unit) of CO rotational lines are approximately proportional to $J^3$ \citep[see e.g.][]{Obreschkow+2009, daCunha+2013, Narayanan+2014}. At higher-$J$ transitions the PDR CO SLED drops considerably and flattens on a low level due to the small amount of hot gas in the outer layer of the molecular cloud. On the other hand, in XDRs, a much larger fraction of the gas is at higher temperature, thus the CO SLED does not drop as dramatically. For this reason, observations of high-$J$ CO lines should reveal the presence of an additional source of heating, such for example X-rays \citep[see e.g.][]{Schleicher+2010}. We note that our CO SLED model predictions are in agreement with other PDR/XDR CLOUDY models reported in literature \citep[e.g.][]{Vallini+2018, Vallini+2019}. 

\begin{figure}[]
	\centering
	\resizebox{\hsize}{!}{
	\centering
	\includegraphics[width=\hsize]{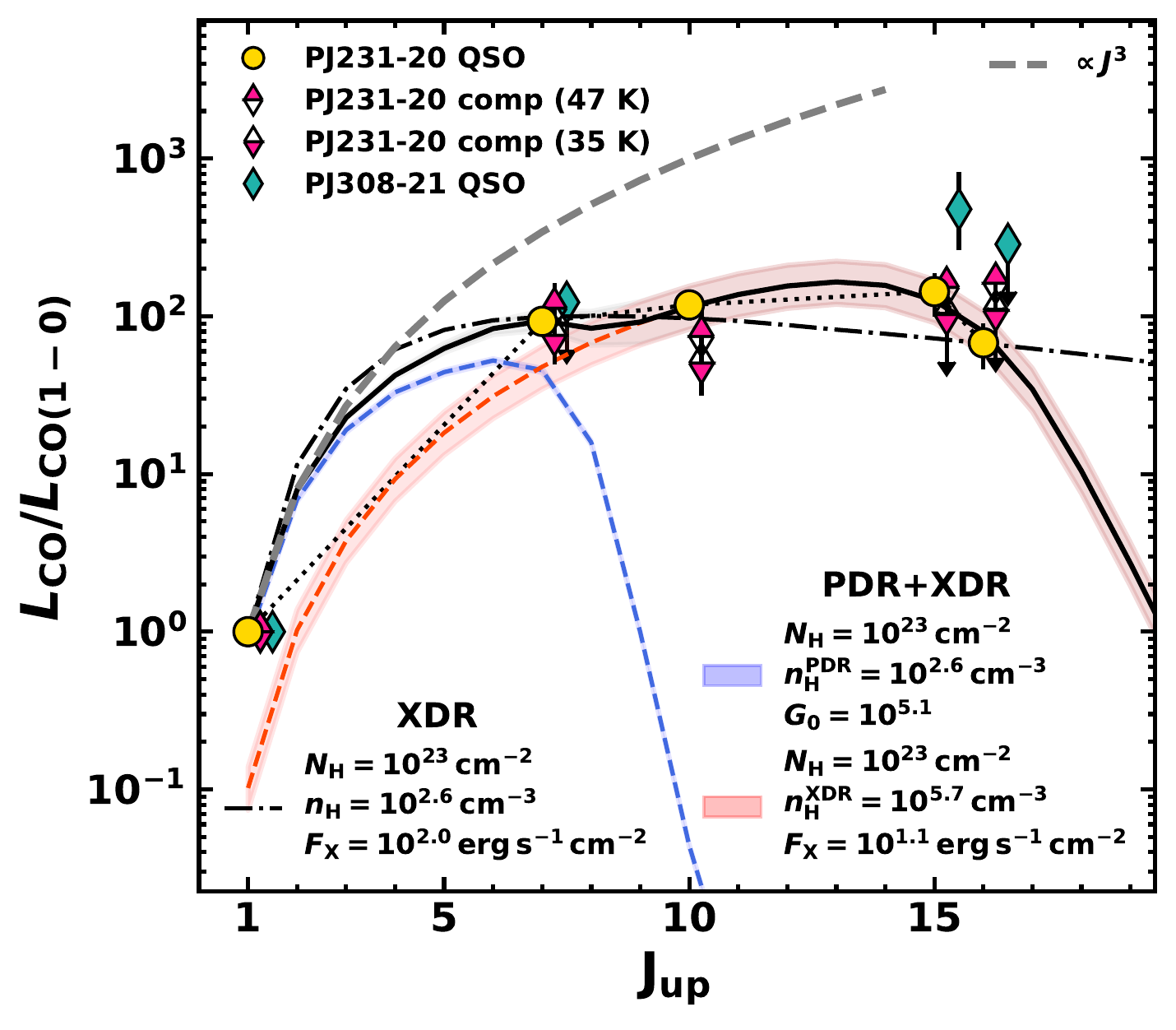}}
      	\caption{CO(1--0)-normalized SLED fits of the PJ231-20 QSO. Gold circles indicate the CO luminosity ratios measured in the quasar. The dot-dashed black line is the best-fit model using a single XDR model. The composite (PDR+XDR) model described in the text is represented with solid black line. The blue and red dashed line show the PDR and XDR component respectively, with the shadowed area indicating the $1\sigma$ uncertainties on normalization values. Best-fit parameters are reported in the legends at the bottom of the figure. The gray dashed line is the theoretical trend ($\propto J^3$) of the CO SLED in LTE, optically thick and high-temperature (or low frequency) limit. The red and blue diamonds (slightly shifted horizontally for clarity) are the CO(1--0)-normalized SLED in the PJ231-20 companion galaxy and the PJ308-21 system, respectively. For the PJ231-20 companion we report both the CO SLEDs obtained assuming $T_{\rm dust}=47\,{\rm K}$ (half top filled magenta diamonds) and $35\,{\rm K}$ (half bottom filled magenta diamonds, see Sects.~\ref{ssect:co_sleds}).}
         \label{fig:co_sled_fit}
\end{figure}

For the PJ231-20 QSO, the mere detection (although at modest significance) of very high-$J$ CO transitions, with estimated $L_{\rm CO(15-14)}/L_{\rm CO(1-0)}$ and $L_{\rm CO(16-15)}/L_{\rm CO(1-0)}$ ratios of $\sim 100$, immediately points to a significant XDR contribution. In Fig.~\ref{fig:co_sled_fit} we show our best-fit CLOUDY models of the CO SLED in PJ231-20 QSO obtained by minimizing $\chi^2$. The high-$J$ normalized CO fluxes ($J_{\rm up}=15, 16$) in the PJ231-20 QSO cannot be reproduced by a single PDR model. We obtained excellent agreement by using a composite (PDR+XDR) model in the form $W_{\rm PDR}F_{\rm CO}^{\rm PDR}(n_{\rm H}^{\rm PDR},G_0)+W_{\rm XDR}F_{\rm CO}^{\rm XDR}(n_{\rm H}^{\rm XDR},F_{\rm X})$, adopting $N_{\rm H}=10^{23}\,{\rm cm^{-3}}$. Here $W_{\rm PDR}$ and $W_{\rm XDR}$ are the relative contributions of the PDR ($F_{\rm CO}^{\rm PDR}$) and XDR component ($F_{\rm CO}^{\rm XDR}$) to the CO(1--0) emission, respectively. The best-fit model has $W_{\rm PDR}=0.90\pm0.04$ and $W_{\rm XDR}=0.10\pm0.04$, total hydrogen density $n_{\rm H}^{\rm PDR}\sim 4\times10^2\,{\rm cm^{-3}}$, $n_{\rm H}^{\rm XDR}\sim5\times10^{5}\,{\rm cm^{-3}}$, and UV/X-ray radiation field strength of $G_0\sim1\times10^{5}$ and $F_{\rm X}\sim10\,{\rm erg\,s^{-1}\,cm^{-2}}$, respectively. Since we consider the same column densities for PDR and XDR models, $W_{\rm PDR}$ and $W_{\rm XDR}$ are directly connected with the CO mass budget in the ISM. Therefore, this result indicates that PDRs account for $\sim90\%$ of the molecular mass in the quasar host. However, while XDRs account for a small fraction of the molecular mass, they dominate the CO emission at $J_{\rm up}\ge 10$. In the PJ231-20 QSO, we measured $L_{\rm CO(7-6)}/L_{\rm CO(1-0)}=93_{-12}^{+13}$. This ratio is $\sim 30\%$ with respect to the value expected for CO thermalized emission in the optically thick and Rayleigh-Jeans limit (described by $J^3$ curve in Fig.~\ref{fig:co_sled_fit}). 

However, although the composite model allows us to accurately reproduce the observed CO SLED in the PJ231-20 QSO, the number of free parameters is greater than data points, thus producing degeneracies in best-fit parameters. In Fig.~\ref{fig:co_sled_fit}, we also show the best-fit model obtained employing a single XDR component. In this case, the best-fit model is roughly in agreement with the observed CO SLED in the PJ231-20 QSO, with a value $n_{\rm H} \sim 4\times10^{2}\,{\rm cm^{-3}}$, but it points to an extreme value of the X-ray radiation field intensity of $F_{\rm X} \sim 10^2\,{\rm erg\,s^{-1}\,cm^{-2}}$, corresponding to the upper boundary of the parameter space. Therefore, this result should be taken into account with caution. 

In Fig.~\ref{fig:co_sled_fit} we also report the observed CO SLED in the PJ231-20 companion galaxy and the PJ308-21 QSO. Despite detecting $L_{\rm CO(7-6)}/L_{\rm CO(1-0)}$ and $L_{\rm CO(10-9)}/L_{\rm CO(1-0)}$ in the PJ231-20 companion, we cannot safely infer a significant XDR contribution in this source due to the lack of high-$J$ CO detections. On the other hand, the high $L_{\rm CO(15-14)}/L_{\rm CO(1-0)}\sim 3\times 10^2$ measured in the PJ308-21 QSO (although with low S/N), can only be reproduced with our models by adopting a strong XDR component.

Overall, our CO SLED modeling suggest that a significant contribution to CO emission arises from clouds exposed to X-ray radiation, at least in quasar hosts. This result appears to be in contrast with our ${\rm [CII]_{158\mu m}}/{\rm [CI]_{369\mu m}}$ modeling discussed in Sect.~\ref{ssect:cii_ci}. However, from our composite PDR+XDR model, we found that PDRs dominate the molecular mass budget and thus dominate the [CII]$_{\rm 158\mu m}$ and [CI]$_{\rm 369\mu m}$ emission. In this scenario, the high-$J$ CO lines arise mainly from molecular medium in the central region of quasar host galaxy, while emission from lower $J$ CO lines and FSLs trace the ISM on larger scales. Indeed, the good  correlation between [CI] and $J\le 7$ CO lines, suggests that these lines trace the spatially extended gas reservoir associated with vigorous star formation rather than denser and more concentrated gas traced by high-$J$ CO lines (see e.g. \citealt{Gerin+2000, Engel+2010, Ivison+2011, Jiao+2017} for studies on nearby sources, and e.g. \citealt{Weiss+2005, Tacconi+2008, Bothwell+2010, Bothwell+2013, Riechers+2011, Walter+2011, Yang+2017, Valentino+2018, Valentino+2020}, for intermediate and high-$z$ works).

Finally, we stress that the uncertainties reported in our CO(1--0)-normalized SLEDs are statistical errors, ignoring any systematic uncertainties associated with the assumption of a gas-to-dust ratio by which we estimated the CO(1--0) luminosity in this work; therefore our results should be considered tentative. In order to better constrain the relative contribution of PDRs and XDRs in these sources we need observations of other CO lines over a wide range of values of $J$. We also point out that the presented analysis does not rule out the possibility that the other source of heating such as cosmic rays, shocks or mechanical heating from turbulence may contribute significantly to the mid-/high-$J$ CO excitation \citep[see e.g.][]{Hollenbach+1989, Flower+2010, Mingozzi+2018, Godard+2019, Vallini+2019}

\subsection{H$_2$O rotational lines}
\label{ssect:water_vapor}
Water vapor emission is a tracer of the molecular warm dense phase of the interstellar medium \citep[$n_{\rm H_{2}}\apprge 10^5 - 10^6 {\rm cm^{-3}}$, $T\sim 50-100\,{\rm K}$, see][]{Liu+2017}, where UV/X-ray radiation from newly formed star or powerful AGN raises the dust temperature above the ice evaporation temperature \citep[e.g.][]{Cernicharo+2006}. In such regions, the H$_2$O molecule can be released into the gas phase by photodesorption from dust grains \citep{Hollenbach+2009}, or by sputtering of grains in shocked-heated regions \citep[e.g.][]{Bergin+2003,Gonzalez-Alfonso+2012, Gonzalez-Alfonso+2013} where H$_2$O becomes the third most abundant species and the strongest molecular emitter after the high-$J$ ($J_{\rm up}>7-8$) CO transitions. The H$_2$O molecule can also be formed in the gas phase through ion-neutral chemistry and neutral-neutral endothermic reactions \citep{Graff+1987, Wagner+1987, Hollenbach+2009}. Lower level transitions ($E_{\rm upper} < 200\,{\rm K}$) arise in collisionally excited gas with kinetic temperature of $\sim100\,{\rm K}$ and clump densities of the order of $3\times10^6 {\rm cm^{-3}}$ where they may play an important role as a cooling agent \citep[e.g.][]{Neufeld+1993,Neufeld+1995,Bradford+2011}. On the other hand, due to the high critical density for collisional excitation of higher levels ($n_{\rm crit} \apprge 10^7 - 10^8\,{\rm cm^{-3}}$, e.g. \citealt{Faure+2007,Faure+2008,Daniel+2012}), water excitation require instead radiative excitation by intense far-IR radiation field from warm dust \citep{Weiss+2010,Gonzalez-Alfonso+2014} and the intensity of the lines is found to be nearly linear proportional with the total infrared luminosity of the galaxies \citep{Omont+2013, Yang+2013, Yang+2016, Liu+2017}. As both collisions and IR pumping are responsible for populating the water energy levels, H$_2$O lines not only probe the physical conditions of the warm and dense molecular gas-phase ISM but also provide important clues about the dust IR radiation density. 

Previous ground-based observations of water lines in nearby galaxies were limited by telluric atmospheric absorption, and consequently have been restricted to radio maser transitions and a few transitions in IR-luminous galaxies \citep[e.g.][]{Combes+1997,Menten+2008}. On the other hand, observational campaigns using space telescopes (e.g., Infrared Space Observatory (ISO), {\it Herschel}/PACS/SPIRE) have been successful in detecting H$_2$O features in local galaxies \citep[mainly in absorption][]{Fischer+1999,Fischer+2010,Gonzalez-Alfonso+2004,Gonzalez-Alfonso+2008, Gonzalez-Alfonso+2010,Gonzalez-Alfonso+2012,vanderWerf+2010,Weiss+2010, Rangwala+2011,Spinoglio+2012,Pereira-Santaella+2013}. Furthermore, various H$_2$O emission lines have been detected at higher redshifts ($z>1-2$) in starburst galaxies and Hyper/Ultra Luminous Infrared Galaxies \citep[Hy/ULIRGs, e.g.][]{Bradford+2011, Lis+2011, Omont+2011, Omont+2013, vanderWerf+2011, Combes+2012, Lupu+2012, Bothwell+2013, Yang+2013, Yang+2016, Yang+2019, Yang+2020} and also in a $z>6$ SMG \citep{Riechers+2013} and a lensed $z>6$ quasar host galaxy \citep{YangJ+2019}.

\begin{figure}
	\centering
	\resizebox{\hsize}{!}{
	\centering
   	\includegraphics[width=\hsize]{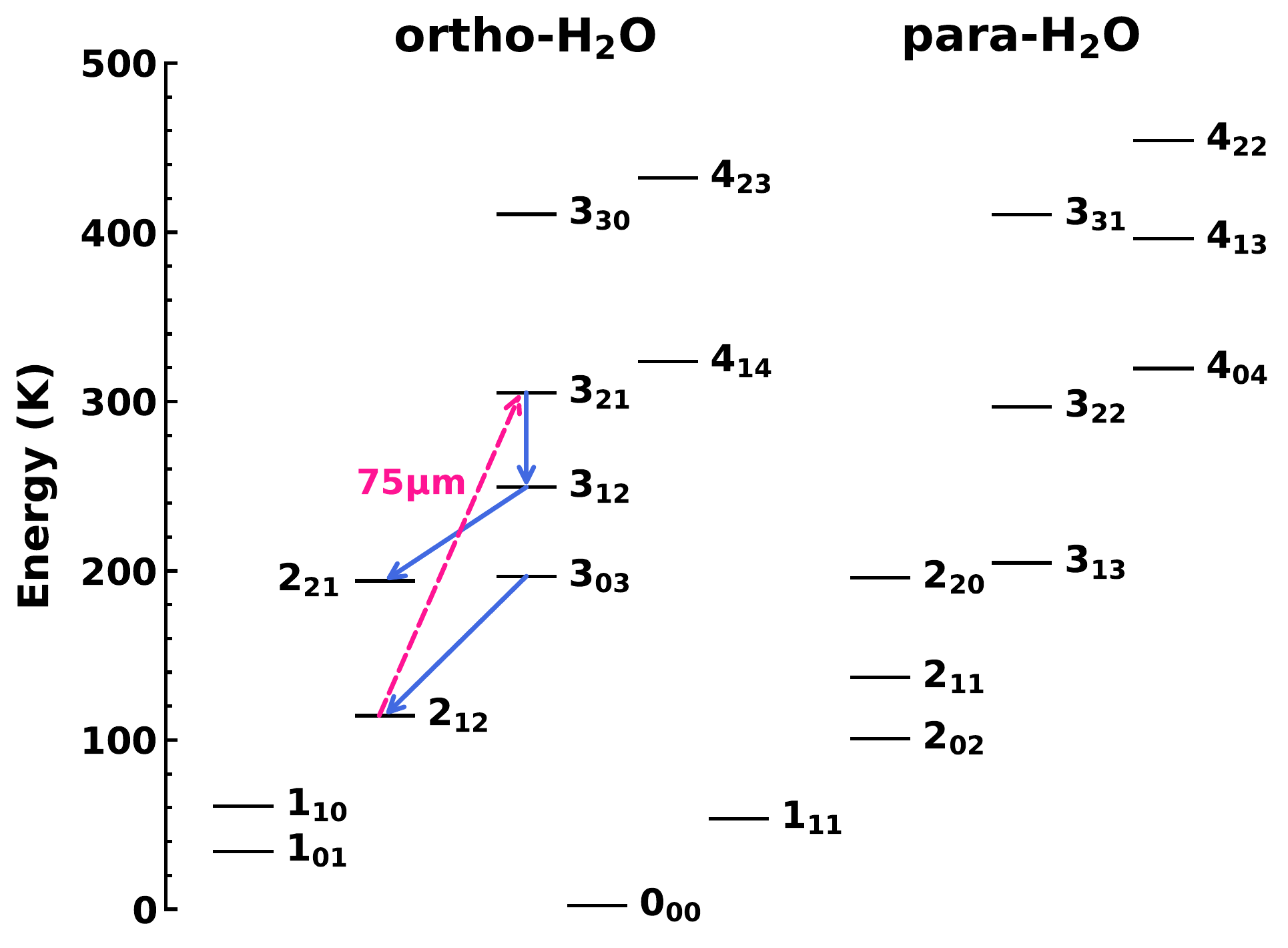}
	}
      	\caption{Part of the H$_2$O energy level diagram. Para- and ortho-H$_2$O ladders are shown together with the detected H$_2$O line transitions (blue arrows). The magenta arrow indicates the IR radiative-pumped transition at $75\,{\rm \mu m}$ that populates the $3_{21}$ level \citep[see][]{Gonzalez-Alfonso+2014,Liu+2017}.}
         \label{fig:water_levels}
\end{figure}

In this work, we present sub-mm H$_2$O emission in our two $z>6$ non-lensed quasar host galaxies. In particular, we report three H$_2$O line detections in PJ231-20 QSO ($3_{12}-2_{21}$; $3_{21}-3_{12}$; $3_{03}-2_{12}$), and a tentative detection of H$_2$O $3_{03}-2_{12}$ in PJ308-21 QSO, while we only have upper limits on water lines from the companion galaxies. The water vapor energy levels, and the targeted (ortho-)H$_2$O transitions are reported in Fig.~\ref{fig:water_levels}. The H$_2$O $3_{12}-2_{21}$ and $3_{21}-3_{12}$ lines are likely mainly produced by the cascade process in response to absorption of IR photons at $75\,{\rm \mu m}$, while the H$_2$O  $3_{03}-2_{12}$ line is more sensitive to collisions and is enhanced in systems with higher gas kinetic temperature \citep[e.g.][]{vanderWerf+2011,Gonzalez-Alfonso+2014,Liu+2017}.

\begin{figure}
	\centering
   	\includegraphics[width=\hsize]{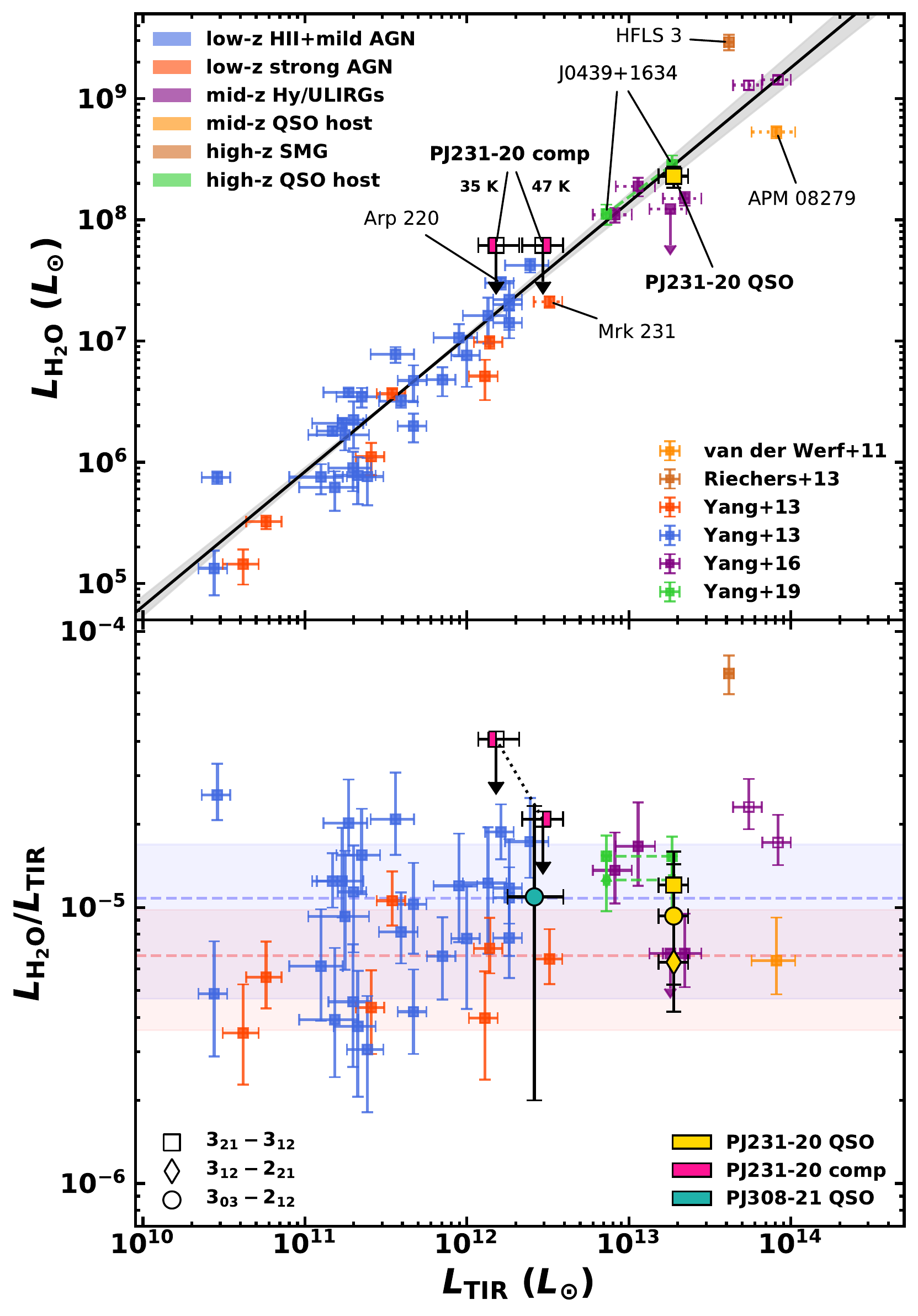}
      	\caption{{\it Upper panel:} The $L_{\rm H_2O(3_{21}-3_{12})} - L_{\rm TIR}$ relation. Black solid line is the best power-law fit from \citet{Yang+2013} considering the low-$z$ HII+mild- and strong AGN sample (blue and red squares respectively). The gray area is the 1$\sigma$ confidence limit. The gold and magenta squares indicate the PJ231-20 quasar host and its companion, respectively. In the case of companion, we report both the cases assuming $T_{\rm dust}=47\,{\rm K}$ (half left filled magenta square) and $35\,{\rm K}$ (half right magenta square). These points are connected by a dotted black line. Purple squares are the mid-$z$ Hy/ULIRGs of \citet{Yang+2016}. We also report measurements of APM 08279+5255 \citep{vanderWerf+2011}, the $z\sim6.3$ SMG HFLS 3 \citep{Riechers+2013}, and the $z\sim6.5$ J0439+1634 quasar \citep{YangJ+2019}. Empty squares indicate measurements not corrected for gravitational lensing. {\it Lower panel:} $L_{\rm H_2O}/L_{\rm TIR}$ versus total infrared luminosity. Symbols are the same as in the upper panel. Together with  $L_{\rm H_2O}/L_{\rm TIR}$ for the $3_{21}-3_{12}$ transition, we also report those for $3_{12}-2_{21}$ and $3_{03}-2_{12}$ for PJ231-20 QSO, the companion galaxy, PJ308-21 QSO and the quasar J0439+1634, as indicated in the legend.}
	         \label{fig:lh2o_ltir}
\end{figure}

\subsection{The $L_{\rm H_2O}- L_{\rm TIR}$ relation}
\label{ssect:water_tir_relation}
In the upper panel of Fig.~\ref{fig:lh2o_ltir} we compare the H$_2$O $3_{21}-3_{12}$ line with the total IR luminosity obtained through the fit of the dust SED of the PJ231-20 QSO and its companion (see Table~\ref{tbl:dust_data}), with a sample of nearby ULIRGs presented in \citet{Yang+2013}. For this comparison, we also included H$_2$O detections of $z\sim 2.5-3.5$ Hy/ULIRGs from \citet{Yang+2016}, APM 08279+5255 at $z\simeq3.9$ \citep{vanderWerf+2011}, the $z=6.34$ SMG HFLS 3 \citep{Riechers+2013}, and J0439+1634, a $z\simeq6.5$ quasar host galaxy \citep{YangJ+2019}. Our data are consistent within the uncertainties with the almost linear relation found by \citet{Yang+2013,Yang+2016}, $L_{\rm H_2O}\propto L_{\rm TIR}^{1.1-1.2}$. This correlation appears as the straightforward consequence of the IR-pumping mechanism responsible for the population of the upper level ($3_{21}$) of the H$_2$O molecules \citep{Yang+2013}, that, after absorption of far-IR photons, cascade via the lines we observe in an approximately constant fraction. However, \citet{Liu+2017} emphasized that the medium-excitation H$_2$O transitions (such as $3_{21}-3_{12}$) probe the physical regions of galaxies in which a large fraction of FIR emission is generated. Therefore, the observed $L_{\rm H_2O}- L_{\rm TIR}$ correlation could be largely driven by the sizes of FIR- and water vapor-emitting regions. The H$_2$O detections presented in this work together with those of \citet{Riechers+2013, YangJ+2019}, extend the redshift range explored in the previous works by \citet{Yang+2013,Yang+2016}, confirming that this relation seems to hold also at very high-$z$. 

\citet{Yang+2013} found slightly different values of $L_{\rm H_2O}/ L_{\rm TIR}$ ratios for the $3_{21}-3_{12}$ transition in local AGN and star-forming-dominated galaxies with possible mild AGN contribution. However, apart from individual studies (e.g. \citealt{Gonzalez-Alfonso+2010} for Mrk 231, and \citealt{vanderWerf+2011} for APM 08279) showing that AGN could be the main power source exciting the H$_2$O lines in this objects, it is still not clear how a strong AGN could affect the H$_2$O emission. 
One possibility is that the buried AGN gives a larger TIR-to-$75{\rm \mu m}$ luminosity ratio than in starburst galaxies due to the different shape of the IR SED of these sources \citep[e.g.][]{Kirkpatrick+2015, Yang+2016}. In addition, high X-ray fluxes can photodissociate H$_2$O molecules. On the other hand, high dust continuum opacity at $100\,{\rm \mu m}$ ($\tau_{100}$) and/or large velocity dispersion can enhance the H$_2$O luminosity \citep[e.g. Arp220,][]{Gonzalez-Alfonso+2014,Yang+2016}.

For the $3_{21}-3_{12}$ transition toward PJ231-20 QSO, we find $L_{\rm H_2O}/ L_{\rm TIR}\sim 1.2\times10^{-5}$, comparable with the ratio measured in the $z\sim6.5$ quasar J0439+1634 \citep{YangJ+2019}. This value is higher than the average ratio measured by \citet{Yang+2013} in local AGN-dominated sources ($7\pm3\times10^{-6}$) and is consistent with that measured in star-forming galaxies ($1.1\pm0.6\times 10^{-5}$), suggesting a significant contribution of star formation to the IR radiation exciting the H$_2$O line. In Fig.~\ref{fig:lh2o_ltir} we also report the $L_{\rm H_2O}/ L_{\rm TIR}$ ratios for $3_{12}-2_{21}$ and $3_{03}-2_{12}$ transitions detected in the PJ231-20 QSO and the tentative H$_2$O $3_{03}-2_{12}$ detection in the PJ308-21 QSO. These values are in the range measured for the $3_{21}-3_{12}$ transition in local ULIRGs.

\subsection{H$_2$O($3_{21}-3_{12}$)--normalized SLED }
\label{ssect:h2o_sled}
\begin{figure}
	\centering
   	\includegraphics[width=\hsize]{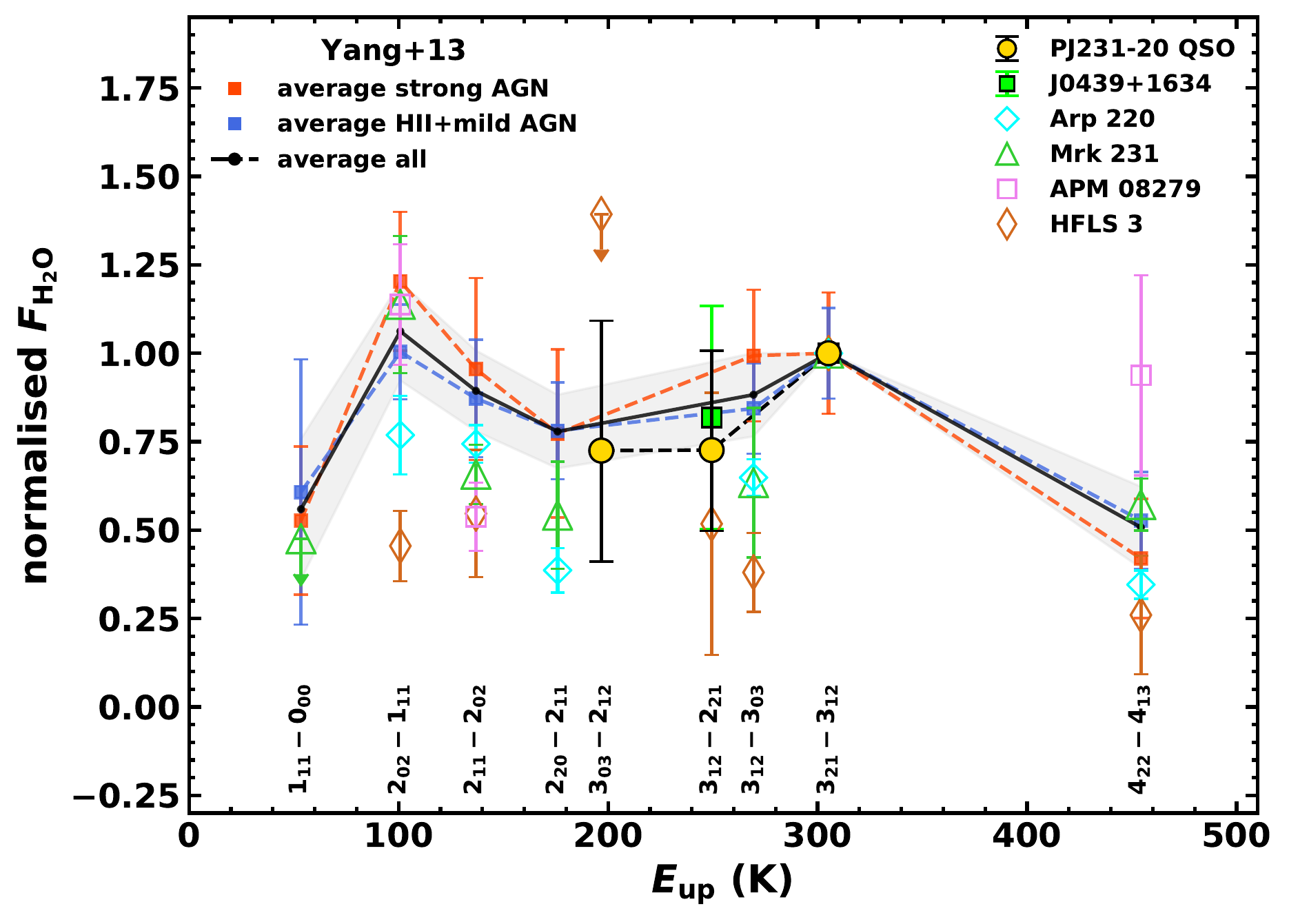}
      	\caption{H$_2$O($3_{21}-3_{12}$)--normalized H$_2$O intensities (in ${\rm Jy\,km\,s^{-1}}$) as a function of the excitation temperature of the upper level. Gold circles are the H$_2$O ratios measured in the PJ231-20 QSO. The black solid line reports the average H$_2$O SLED of the whole sample of \citet{Yang+2013}, with gray shadowed area representing the $1\sigma$ uncertainty. The red and blue dashed lines indicate the average values of strong-AGN- and HII+mild-AGN-dominated galaxies, respectively, as classified by \citet{Yang+2013}. We also report data retrieved from the literature for the two nearby sources Arp 220 \citep{Rangwala+2011} and the AGN-dominated Mrk 231 \citep{Gonzalez-Alfonso+2010}. For comparison we also show the H$_2$O flux ratios observed in high-$z$ sources such as APM 08279+5255 \citep{vanderWerf+2011}, the lensed $z>6$ quasar J0439+1634 \citep{YangJ+2019}, and the SMG HFLS3 \citep{Riechers+2013}. The energy of upper levels of $2_{20}-2_{11}$ and $3_{12}-3_{03}$ transitions, were shifted  for clarity to $-20$ and $+20\,{\rm K}$, respectively.} 
         \label{fig:h2osled_comparing}
\end{figure}

By combining the three water vapor lines detected in PJ231-20 QSO, in Fig.~\ref{fig:h2osled_comparing} we trace the line flux ratios, that is the normalized H$_2$O SLED, and we compare with the results on local and high-$z$ sources found in the literature. The SLEDs were normalized to H$_2$O $3_{21}-3_{12}$, a medium-excitation line that is predicted to be easily radiatively pumped in the warm dense medium \citep[see][]{Liu+2017}. Therefore the ratios with the other transitions directly reflect the effect of IR-pumping line excitation.

In Fig.~\ref{fig:h2osled_comparing}, the average H$_2$O SLED of the local ULIRG sample by \citet{Yang+2013}, shows two peaks at the energy of H$_2$O $2_{02}-1_{11}$ and $3_{21}-3_{12}$ transitions, indicating high IR pumping efficiency at $75\,{\rm \mu m}$. This is consistent with the models of \citet{Liu+2017} showing that these two lines are efficiently pumped in the warm ISM component at $T_{\rm dust}\apprge 40\,{\rm K}$. The bulk H$_2$O emission observed in star-forming galaxies of \citet{Yang+2013} is explained by \citet{Gonzalez-Alfonso+2014} by model with $T_{\rm dust}=55-75\,{\rm K}$, $\tau_{100}\sim 0.1$ and H$_2$O column density $N_{\rm H_2O}\sim (0.5-2)\times 10^{17}\,{\rm cm^{-2}}$, including a significant contribution from both collisional excitation and line radiative pumping. Despite significant uncertainties, the observed H$_2$O line ratios in the PJ231-20 QSO show a clear peak at $3_{21}-3_{12}$ transition. In particular the line ratio $3_{12}-2_{21}/3_{21}-3_{12}$  is consistent with the value observed in $z\sim6.3$ SMG HFLS 3 \citep{Riechers+2013} and in the $z\sim6.5$ quasar J0439+1634 \citep{YangJ+2019} whose H$_2$O SLEDs also peak at $3_{21}-3_{12}$. The stronger intensity of H$_2$O $3_{21}-3_{12}$ measured in the PJ231-20 QSO than the other transitions lower in the cascade, in conjunction with the high critical density of the observed transitions ($\apprge10^8\,{\rm cm^{-3}}$), suggests that the contribution from collisional excitation of the upper energy levels of the observed water vapor transitions is minor. Other factors that could contribute to the PJ231-20 QSO H$_2$O SLED's peak at $3_{21}-3_{12}$ are the shape of the IR dust SED that peaks close to $75\,{\rm \mu m}$, thus allowing high pumping efficiency, together with the large FWHM ($\sim 400\,{\rm\, km\,s^{-1}}$) of the H$_2$O $3_{21}-3_{12}$ line, which increases the cross-sections for absorbing pumping photons \citep[e.g.][]{Gonzalez-Alfonso+2014}.

In the next section we retrieve quantitative information on the ISM associated with the H$_2$O emission by modeling the water vapor SLED using our {CLOUDY} models.
\begin{figure*}
	\centering
	\resizebox{\hsize}{!}{
	\centering
	\includegraphics[width=\hsize]{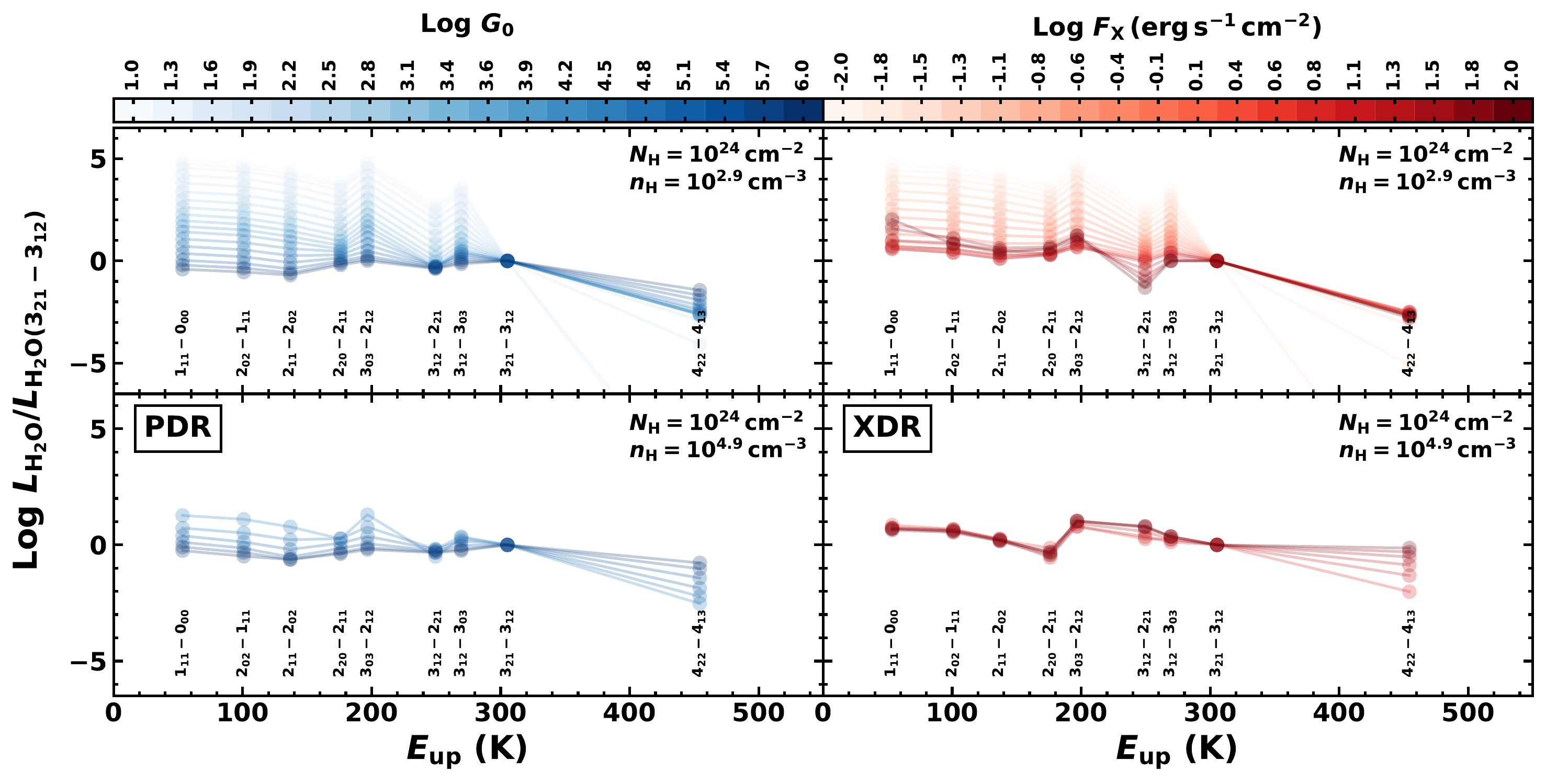}}
      	\caption{CLOUDY H$_2$O($3_{21}-3_{12}$)--normalized SLED models in the PDR ({\it left panels}) and XDR regime ({\it right panels}) varying the intensity of the incident radiation field over the entire parameter space and for two reference values of total hydrogen density ($n_{\rm H}=10^{2.3},\,10^{4.6}\,{\rm cm^{-3}}$), and column density $N_{\rm H}=10^{24}\,{\rm cm^{-2}}$. PDR cases are shown in blue while XDRs are red; both are color-coded according to the values of the radiation field flux. The energy of upper levels of the $2_{20}-2_{11}$ and $3_{12}-3_{03}$ transitions, were shifted for clarity to $-20$ and $+20$ K, respectively. In order to exclude unphysical models, here we report H$_2$O SLEDs in which H$_2$O $3_{12}-2_{21}$ intensity is $>10^{-2}\times$ that of the corresponding CO(10--9) line.}
         \label{fig:h2o_sled_models}
\end{figure*}

\subsection{H$_2$O SLED models}
\label{ssect:h2o_models}
Our CLOUDY outputs predict emergent intensities of a large number of water vapor transitions. In Fig.~\ref{fig:h2o_sled_models} we report the H$_2$O SLED normalized to H$_2$O $3_{21}-3_{12}$ line in the PDR and XDR regimes. We do so for two representative values of total hydrogen density $n_{\rm H}\simeq 2\times 10^2\,{\rm cm^{-3}}$ and $\simeq 4\times 10^4\,{\rm cm^{-3}}$ with varying radiation field strength with fixed total hydrogen column density $N_{\rm H}=10^{24}\,{\rm cm^{-2}}$.

In both PDR and XDR cases the H$_2$O SLED flattens as the intensity of the radiation field increases. In particular, $L_{\rm H_2O}/L_{\rm H_2O(3_{21}-3_{12})}$ line ratios for transitions with energy upper level $E_{\rm up} < 300\,{\rm K}$, approach similar values varying within a factor of $\sim 4$. Since the total hydrogen density of the cloud models spans a range that is well below the typical critical density of H$_2$O lines ($\sim 10^8-10^9{\rm cm^{-3}}$), this could be explained in terms of a progressively more efficient IR pumping of the high-lying H$_2$O energy levels as the UV/X-ray photon flux increases, that subsequently cascade radiatively toward low energy levels. This implies that these H$_2$O lines tend to statistical equilibrium, in agreement with the analysis of, e.g. \citet{Gonzalez-Alfonso+2014,Liu+2017}. Furthermore, as also mentioned in Sect.~\ref{ssect:water_vapor}, at gas kinetic temperature $\apprge300\,{\rm K}$, water molecules can form in the gas phase via the neutral-neutral endothermic reaction ${\rm OH + H_2\,\to\,H_2O + H}$. Therefore, an increase of the total hydrogen density ($n_{\rm H} > 10^4\,{\rm cm^{-3}}$) inside the cloud requires higher radiation field flux ($G_0>2\times10^4$, $F_{\rm X}>0.05 - 0.1\,{\rm erg\,s^{-1}\,cm^{-2}}$) to maintain gas temperature above the reaction activation barrier in a large volume of the cloud in order to produce a significant H$_2$O column density $N_{\rm H_2O}> 10^{15}\,{\rm cm^{-2}}$ \citep[see also, e.g.][]{Neufeld+1995, Neufeld+2002, Cernicharo+2006, Meijerink+2011}. On the other hand, in XDRs, strong X-ray radiation can photodissociate a large fraction of the H$_2$O molecules, thus quenching their line emission. However, since the H$_2$O $3_{21}-3_{12}$ transition is one of the most sensitive to the pumping by IR photons \citep[e.g.][]{Liu+2017}, low UV/X-ray fluxes cannot efficiently populate its upper level, producing unphysical results in the predicted $L_{\rm H_2O}/L_{\rm H_2O(3_{21}-3_{12})}$. In Fig.~\ref{fig:h2o_sled_models}, to exclude such non-physical results that clearly at odds with observations, we report those H$_2$O SLEDs for which the H$_2$O $3_{12}-2_{21}$ predicted intensity is at least $10^{-2}\times$ that of CO(10--9) line, since these two lines have similar frequency and are observed to have similar fluxes (see Fig.~\ref{fig:pj231_spectramaps} of this work, but see also e.g. \citealt{Riechers+2013, YangJ+2019}).

\begin{figure}[!htbp]
	\centering
	\resizebox{\hsize}{!}{
	\centering
	\includegraphics[width=\hsize]{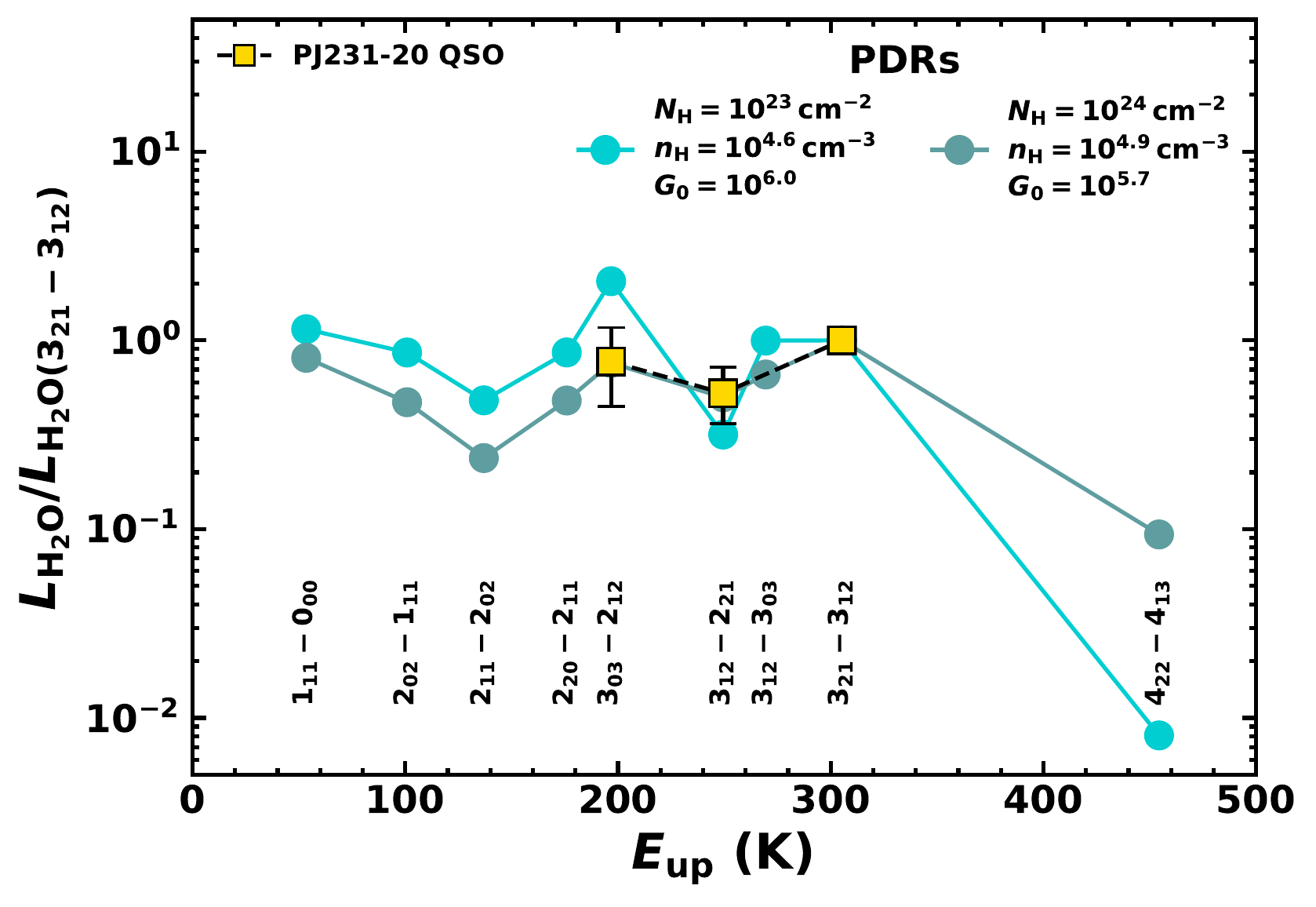}}
      	\caption{Best-fit models of the H$_2$O($3_{21}-3_{12}$)--normalized SLED in PJ231-20 QSO. Gold squares are the observed line ratios while light and dark blue circles are the PDR best-fit models with $N_{\rm H}=10^{23}$ and $10^{24}\,{\rm cm^{-2}}$, respectively. The best-fit parameters are indicated in the upper part of the panel.}
         \label{fig:h2o_sled_fit_pj231}
\end{figure}

By using the H$_2$O line ratios predicted by our CLOUDY models, we performed the H$_2$O SLED fit of PJ231-20 QSO. We report the results in Fig.~\ref{fig:h2o_sled_fit_pj231}. The observed H$_2$O line ratios of PJ231-20 QSO cannot be satisfyingly reproduced by any models with $N_{\rm H}=10^{23}\,{\rm cm^{-2}}$, while the agreement between data and best-fit model significantly improves employing higher column density PDR model with $N_{\rm H}=10^{24}\,{\rm cm^{-2}}$. The resulting best-fit model points to a high density medium with $n_{\rm H}\sim 0.8\times10^5\,{\rm cm^{-3}}$, exposed to a strong FUV radiation field with $G_{0}\sim 5\times 10^5$. The mere requirement of a high cloud column density implies that the observed water vapor transitions arise deep in the cloud where the water vapor column density reaches $N_{\rm H_2O}\apprge 3\times 10^{17}\,{\rm cm^{-2}}$. Furthermore, since the H$_2$O line ratios observed in PJ231-20 QSO are similar, and given that the three H$_2$O transitions $3_{21}-3_{12}$; $3_{12}-2_{21}$; $3_{03}-2_{12}$ have similar frequencies and are connected by a cascade process, our result suggests that these lines are near statistical equilibrium. However, the low number of data points makes the result uncertain, in particular the cloud density is poorly constrained with the fit. On the other hand, the high density value found for the PJ231-20 QSO H$_2$O lines, is in the range typically inferred from H$_2$O SLED modeling in local sources \citep[see e.g.][]{Gonzalez-Alfonso+2014,Liu+2017} and high-$z$ quasar \citep[e.g.][]{vanderWerf+2011}

\subsection{OH transitions}
\label{ssect:oh_lines}
The hydroxyl radical (OH) is a key intermediary molecule that forms in warm PDRs exposed to FUV radiation via the endothermic reaction between atomic oxygen and vibrationally-excited H$^*_2$. Further reaction with H$_2$ leads to the formation of H$_2$O. On the other hand, water vapor can photodissociate in gas unshielded to FUV radiation producing OH. OH can also form in shocked gas where high gas temperatures can overcome the activation barrier of the O $+$ H$_2$ reaction, or in quiescent obscured regions irradiated by cosmic rays or X-rays via ion-neutral reactions starting from H$^+_3$ \citep{Hollenbach+1997, Agundez+2010}. The OH molecule has a large dipole moment ($1.668$ Debye), therefore radiative rates of its rotational transitions are generally high compared to those of CO. The critical densities of OH transitions are of the order of $10^9-10^{10}\,{\rm cm^{-3}}$ \citep{Destombes+1977, Dewangan+1987}, implying that OH excitation cannot be driven by collisions, indeed radiative pumping by IR photons is typically invoked to explain its emission. The electronic ground state has two fine-structure levels, $^2\Pi_{3/2}$ (the ground state) and $^2\Pi_{1/2}$, each of which has two $\Lambda$-doubling levels (further split in two hyper-fine sublevels) . Furthermore, the OH molecule can also have rotational angular momentum ($J$), resulting in a ladder of rotational states with increasing $J$. The OH ground state $^2\Pi_{3/2}$ ($J=3/2$) can absorb IR radiation near $119\,{\rm \mu m}$, $79\,{\rm \mu m}$, $53\,{\rm \mu m}$ and $35\,{\rm \mu m}$, followed by spontaneous radiative decay which repopulates the four ground sublevels, including the FIR OH cross-ladder transitions whose radiative rates are $\sim 100\times$ weaker than the rotational transitions. OH transitions, either in absorption or emission, were detected in local Seyfert galaxies, including M82 \citep{Colbert+1999}, NGC 253 \citep{Fischer+1999, Bradford+1999}, IRAS 20100-4156 and 3Zw35 \citep{Kegel+1999}, Arp 220 \citep{Gonzalez-Alfonso+2004}, NGC 1068 \citep{Smith+2004,Spinoglio+2005}, Mrk 231 \citep{Gonzalez-Alfonso+2008}, and NGC 4418 \citep{Gonzalez-Alfonso+2014}. Systematic searches for OH emission have revealed that some OH lines exhibit P-Cygni profiles, suggesting massive molecular outflows in samples of local (U)LIRGs and quasars \citep{Fischer+2010, Gonzalez-Alfonso+2010, Gonzalez-Alfonso+2012, Gonzalez-Alfonso+2013,Gonzalez-Alfonso+2017, Sturm+2011,Spoon+2013,Veilleux+2013, Herrera-Camus+2020}. OH emission has also been detected at $z\apprge6$ in the SMG HFLS 3 \citep{Riechers+2013}, and in the quasar J1319+0950 \citep{Herrera-Camus+2020}, demonstrating the feasibility in detecting such lines at very high-$z$ with ground-based observations.  

Here we present detections of $\Lambda$-doublet OH$_{\rm 163 \mu m}$ in the PJ231-20 QSO and companion galaxy together with a tentative detection in the PJ308-21 QSO. So far, this doublet, when detected in other galaxies, has been mostly observed in emission. These lines are likely produced by a fluorescent-like mechanism through absorption of IR photons at $35\,{\rm \mu m}$ and $75\,{\rm \mu m}$ emitted by warm dust, and thus they trace the warm moderate-dense molecular medium ($T\sim 150-300\,{\rm K}$, $n_{H_2}\apprle 10^4-10^{5}\,{\rm cm^{-3}}$) in star-forming regions \citep[e.g.][]{Goicoechea+2002, Goicoechea+2005}. The observed OH line fluxes in our systems are comparable with those of mid-$J$ CO, and greater than H$_2$O lines, showing that OH can be a powerful tracer of the molecular medium at $z\sim6$. We find similar $L_{\rm OH}/L_{\rm TIR}$ ratios $\sim 2-4\times10^{-5}$ in both PJ231-20 and PJ308-21 quasars, while we estimated a larger value $\sim 10^{-4}$ in the PJ231-20 companion galaxy. A similar value is found in the SMG AzTEC-3 and HFSL 3 at $z=5.3$ and $z=6.34$, respectively \citep{Riechers+2013, Riechers+2014}. 

\subsection{H$_2$O/OH$_{\rm 163\mu m}$ ratio}
\label{ssect:h2o_oh_ratio}
To obtain further constraints on the molecular medium in the targeted systems, we inspected the line intensity ratios between the observed H$_2$O lines and the OH$_{\rm 163\mu m}$ doublet. Indeed, formation of the OH molecule can occur through the neutral-neutral reaction ${\rm H_2 + O \to OH + H}$ that can successively react to form H$_2$O, or by destruction of water \citep[see e.g.][]{Elitzur+1978, Poelman+2005, Meijerink+2011}. Furthermore, both H$_2$O and the OH molecule have very high critical densities $\apprge10^8\,{\rm cm^{-3}}$, and IR pumping possibly has a dominant role in their excitation. Therefore, the H$_2$O lines and OH$_{\rm 163\mu m}$ are expected to arise from regions with similar physical properties. 

\begin{figure}
	\centering
	\resizebox{0.9\hsize}{!}{
	\centering
	\includegraphics[width=\hsize]{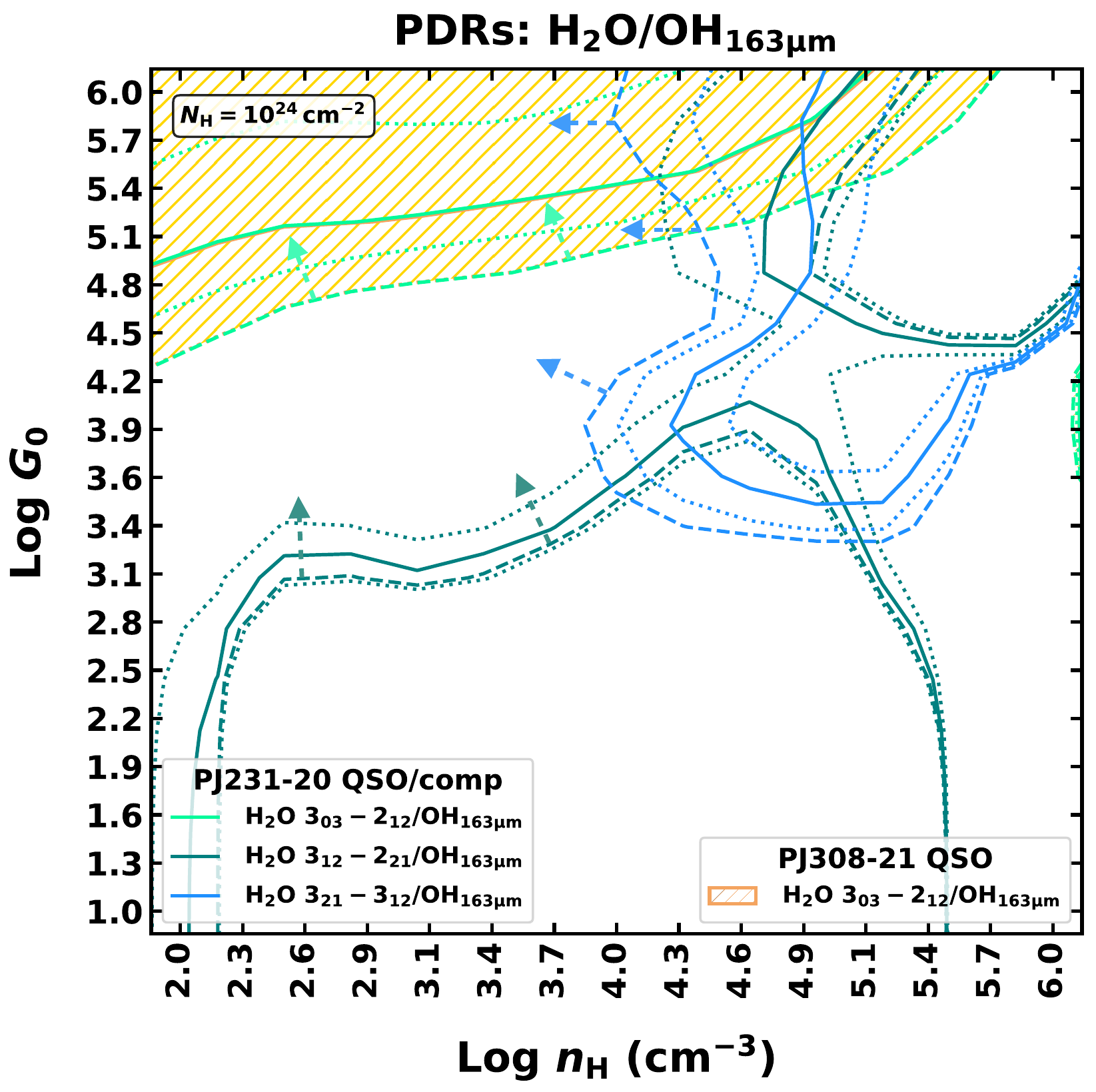}}
      	\caption{CLOUDY models of H$_2$O/OH$_{\rm 163\mu m}$ luminosity ratios. Model predictions are reported for the PJ231-20 QSO (solid and dotted lines indicate values and their uncertainties) and $3\sigma$ upper limits for its companion (dashed lines). The dashed gold area reports the constraints retrieved from PJ308-21 QSO detections. The clouds' column density of the models is $N_{\rm H}=10^{24}\,{\rm cm^{-2}}$.}
         \label{fig:h2o_oh_lines}
\end{figure}

In Fig.~\ref{fig:h2o_oh_lines} we compare the observed values of the H$_2$O/OH$_{\rm 163\mu m}$ line ratios with our CLOUDY predictions in a diagnostic plot. Here, we considered OH$_{\rm 163\mu m}$ as the sum of the luminosity of the two $\Lambda$-doubling transitions of the OH$_{\rm 163\mu m}$, including the hyper-fine structure lines. In the case of the PJ231-20 QSO, model predictions are fully consistent with what we found in Sect.~\ref{ssect:h2o_models}, pointing to a high density medium exposed to a strong FUV field. We also found similar constraints for the companion galaxy and the PJ308-21 QSO, with $G_0>10^4$ and $n_{\rm H}<1.5\times10^4-0.5\times10^6\,{\rm cm^{-3}}$, respectively. These findings, albeit for a lower column density, are also in broad agreement with the constraints obtained from [CII]$_{\rm 158\mu m}$/[CI]$_{\rm 369\mu m}$ ratios and the CO SLED modeling (see Sects.~\ref{ssect:cii_ci} and ~\ref{ssect:co_models}).

\section{Mass budgets}
\label{sect:mass_contributions}
Starting from the FSL luminosities obtained in this work (see Table~\ref{tbl:line_data}) and by adopting simplistic assumptions, we can estimate the mass contributions in our systems in the form of ionized, atomic and molecular mediums. In the 2-level approximation, assuming local thermodynamical equilibrium (LTE) and in the optically thin limit and Rayleigh-Jeans regime, the beam-averaged column density of a species $s$ in the upper level $u$ is \citep[see e.g.,][]{Goldsmith+1999}:
\eq{N_{\rm s} \approx \gamma_{\rm u}\frac{c^2}{8\pi k_{\rm B}}\nu_{\rm obs}^{-2}\,S\Delta\varv \,(1+z)^{-1},}
with
\eq{\gamma_{\rm u} = \frac{8\pi k_{\rm B} \nu_{\rm ul}^2}{hc^3A_{\rm ul}}\frac{Q(T_{\rm ex})}{g_{\rm u}}e^{E_{\rm u}/k_{\rm B} T_{\rm ex}}.}
Here, $S\Delta\varv$ is the observed line velocity-integrated flux, $A_{\rm ul}$ is the Einstein coefficient, $Q(T_{\rm ex})$ is the partition function of the species $s$ (depending on the excitation temperature $T_{\rm ex}$), and $g_{\rm u}$ and $E_{\rm u}$ are the statistical weight and the energy of the upper level, respectively. The total number of particles of species $s$ is obtained by integrating over the size of the emitting region. Then, by using the definition of line luminosity \citep{Solomon+1992}, the total mass of the species $s$ can be expressed as
\eq{M_{\rm s} = C m_{\rm s} \gamma_{\rm u} L'_{\rm line}, \label{eq:mass_species}}
where $C=9.52\times10^{41}$ is the conversion factor between ${\rm km\,pc^2}$ and ${\rm cm^{3}}$, and $m_s$ is the mass of the species $s$.

Using Eq.~\ref{eq:mass_species} we derive the CI mass following \citet{Weiss+2003} via
\eq{M_{\rm CI}[M_{\astrosun}] = 4.566\times10^{-4}Q(T_{\rm ex})\frac{1}{5}e^{62.5/T_{\rm ex}}L'_{\rm [CI]_{\rm 369\mu m}},}
where $Q(T_{\rm ex}) = 1+3e^{-T_1/T_{\rm ex}}+5e^{-T_2/T_{\rm ex}}$ is the CI partition function, and $T_1 = 23.6\,{\rm K}$, $T_2 = 62.5\,{\rm K}$ are the energies (per unit of $k_{\rm B}$) above the ground state of the two [CI] (${^3P_1}\!\!\to\!\!{^3P_0}$) and (${^3P_2}\!\!\to\!\!{^3P_1}$) transitions respectively. Assuming an excitation temperature of $T_{\rm ex}=30\,{\rm K}$ \citep[see e.g.][]{Walter+2011} we find for the PJ231-20 QSO a neutral carbon mass of $M_{\rm CI} = (0.7\pm0.3)\times10^7\,M_{\astrosun}$, and carbon mass limit of $M_{\rm CI} < 0.4-0.5\times10^7\,M_{\astrosun}$ for its companion and the sources in the PJ308-21 system. We note that if we assume a higher excitation temperature $T_{\rm ex}=50\,{\rm K}$, the derived $M_{\rm CI}$ would be $\sim 40\%$ lower. Using these mass estimates, we can put constraints on the molecular gas mass ($M_{\rm H_2}$) using the atomic carbon-to-molecular hydrogen abundance ratio derived by \citet{Walter+2011} for a sample of $z=2-3$ FIR-bright sources (SMGs and quasar host galaxies), $X[{\rm CI}]/X[{\rm H_2}] = M_{\rm CI} / (6M_{\rm H_2}) = (8.4\pm3.5)\times 10^{-5}$. Applying this abundance we obtain a molecular gas mass of $M_{\rm H_2} = 1.3^{+1.1}_{-0.6}\times10^{10}\,M_{\astrosun} $ for PJ231-20 QSO and $M_{\rm H_2} < 1.8\times10^{10}\,M_{\astrosun}$ for the companion galaxy and the other sources in the PJ308-21 system.

Following \citet{Ferkinhoff+2011}, in the high-temperature limit, we can also derive the minimum mass of ionized hydrogen (H$^+$) assuming that all nitrogen in HII regions is singly ionized as
\eq{M_{\rm min}({\rm H^+}) = \frac{L_{{\rm [NII]}{\rm 205\mu m}} m_{\rm H}}{(g_1/Q(T_{\rm ex}))A_{10}h\nu_{10}\chi({\rm N^+})}\approx 2.27\frac{L_{{\rm [NII]}{\rm 205\mu m}}}{L_{\astrosun}}M_{\astrosun},}
where $A_{10}$ is the Einstein coefficient for spontaneous emission of the [NII]$_{\rm 205\mu m}$ (${^3P_1}\!\!\to\!\!{^3P_0}$) transition ($2.08\times10^{-6}\,{\rm s^{-1}}$), $g_1=3$ is the statistical weight of the $J=1$ level, $\nu_{10} = 1461.1318\,{\rm GHz}$ the rest-frame frequency of the transition, $m_{\rm H}$ is the mass of the hydrogen atom, and $\chi({\rm N^+})$ is the ${\rm N^+/H^+}$ abundance ratio. For the minimum mass case, in our working assumption $\chi({\rm N^+})=\chi({\rm N})$. Therefore, assuming solar metallicity, the 'HII-region' gas-phase nitrogen abundance is $\chi({\rm N}) = 9.3\times10^{-5}$ \citep{Savage+1996}. This gives $M_{\rm min}({\rm H^+})= (4.1\pm1.6)\times10^8\,M_{\astrosun}$, and $(1.8\pm1.2)\times10^8\,M_{\astrosun}$ for PJ231-20 QSO host and companion galaxy, respectively. 

Starting from the dust mass budgets derived in Sect.~\ref{ssect:dust}, we can also obtain independent measurements of the gas masses adopting the local gas-to-dust ratio of $\delta_{\rm gdr}\sim70-100$ \citep[see e.g.][]{Draine+2007,Sandstrom+2013, Genzel+2015}. Indeed, similar gas-to-dust ratio of $\sim70$ have also been found in high-$z$ starburst galaxies \citep[e.g.][]{Riechers+2013, Wang+2016}. Adopting the local value we obtain (atomic and molecular) gas masses of $M_{\rm gas} = (3.6 - 5.1)\times 10^{10}\,M_{\astrosun}$ and $(1.6 - 3.4)\times 10^{10}\,M_{\astrosun}$ for the PJ231-20 QSO host and companion galaxy respectively, and $M_{\rm gas} = (3.8 - 5.4)\times 10^{9}\,M_{\astrosun}$ and $(0.7 - 2.0)\times 10^{9}\,M_{\astrosun}$ for the PJ308-20 QSO host and companion galaxy, respectively. If we further assume that $\sim 75\%$ of the dust-derived gas mass is in molecular form \citep[e.g.][]{Riechers+2013, Wang+2016}, these values are consistent with those derived above from $M_{\rm CI}$ applying the CI-to-H$_2$ abundance ratio.

Finally, from the gas mass values derived above, we obtain a lower limit on the gas ionization percentage of $M_{\rm min}(\rm{H^+})/M_{\rm H_2} > 1-3\%$. For comparison, \citet{Ferkinhoff+2011} found values $<1\%$ for nearby galaxies (in the sample of \citealt{Brauher+2008}), while high ionization percentages suggest high SFR density ($>10^2-10^3\,M_{\astrosun}\,{\rm\,yr^{-1}\,kpc^{-2}}$, see \citealt{Ferkinhoff+2011}). 

\section{Summary and conclusions}
\label{sect:conclusions}
In this work, we presented ALMA observations of two quasar host--companion galaxy pairs PJ231-20 and PJ308-21 at $z>6$, captured when the Universe is $<900\,{\rm Myr}$ old. The companion galaxies were serendipitously discovered with previous ALMA observations in the field of known quasars, but they show no evidence of AGN activity. This multi-line study aimed to investigate the effect of the star formation and AGN activity on the ISM in massive galaxies at cosmic dawn.

The covered lines in the ALMA observations include tracers of the atomic ionized/neutral and molecular medium, such as [NII]$_{\rm 205\mu m}$, [CI]$_{\rm 369\mu m}$, CO(7--6, 10--9, 15--14, 16--15), H$_2$O $3_{12}-2_{21}$, $3_{21}-3_{12}$, $3_{03}-2_{12}$, and the OH$_{\rm 163\mu m}$ doublet. However, for PJ308-21 system only half of the program was executed. This work was complemented with already collected ALMA [CII]$_{\rm 158 \mu m}$ observations \citep{Decarli+2018}, that were analyzed in a consistent way. In order to interpret our results and put quantitative constraints on the physical properties of the ISM in galaxies, we ran a set of PDR/XDR models by employing the CLOUDY radiative transfer code, with varying volume hydrogen density, radiation field strength and total hydrogen column density. The main results obtained in this work are summarized as follows:
\begin{itemize}
\item We modeled the FIR dust continuum on all the four targeted galaxies. We modeled this emission as a modified black body and inferred dust masses, spectral indexes, total infrared luminosities and SFRs (see Sect.~\ref{ssect:dust} and Table~\ref{tbl:dust_data}).
\item From the analysis of the [CII]$_{\rm 158\mu m}$ and [NII]$_{\rm 205\mu m}$ lines, we found that $\apprge 80\%$ of [CII]$_{\rm 158\mu m}$ emission in both the quasar and companion galaxies of PJ231-20 system arises from neutral gas in the photo-dissociation regions rather from the fully (diffuse) ionized medium. Furthermore, the observed line deficits are comparable with local LIRGs.
\item The [CI]$_{\rm 369\mu m}$/TIR luminosity ratios in the targeted systems are similar to those measured in $z\sim 2-4$ SMGs and quasars, revealing high star-formation efficiency in our galaxies.
\item CLOUDY models suggest that [CII]$_{\rm 158\mu m}$ and  [CI]$_{\rm 369\mu m}$ emission predominantly arise in PDRs illuminated by the FUV radiation field. In particular, constraints obtained from the [CII]$_{\rm 158\mu m}$/TIR and  [CI]$_{\rm 369\mu m}$/TIR luminosity ratios for PJ231-20 QSO, point to a high FUV field intensity of $G_0\sim 10^5$ ($\sim 1.6\times 10^2\,{\rm erg\,s^{-1}\,cm^{-2}}$) and density of $n_{\rm H}>10^3$. We found similar constraints for the other sources.
\item From the atomic fine-structure lines we retrieved constraints on atomic carbon, ionized hydrogen and molecular gas masses. The latter are consistent with the dust-derived gas masses assuming the local gas-to-dust ratio and that $\sim75\%$ of the gas mass is in molecular form. For both quasar and companion in PJ231-20 system, we also inferred gas ionization percentages of $>1-3\%$, greater than the typical values observed in nearby galaxies, suggesting high SFR density $> 10^{2}-10^{3}\,M_{\astrosun}{\rm \,yr^{-1}\,kpc^{-2}}$.
\item The multiple mid-/high-$J$ CO line detections, allowed us to study the CO SLEDs (see Sect~\ref{ssect:co_sleds}). Despite large uncertainties, the PJ231-20 QSO SLED resembles those of local AGNs. Similar result is obtained for the companion galaxy. On the other hand, by assuming a low dust temperature $35\,{\rm K}$, the CO SLED of the companion appears similar to the average SLED of local starburst galaxies. However, the mere detections of high-$J$ ($J_{\rm up} = 16, 15$) CO lines in the PJ231-20 and PJ308-21 quasars suggest the presence of additional heating source than PDRs. Indeed, CO SLED of the PJ231-20 QSO is well reproduced by a CLOUDY PDR+XDR composite model. The result indicates that PDRs dominate the low-/mid-$J$ CO line luminosities and account for $\sim90\%$ of the CO mass, while XDRs, that possibly represent the main drivers of high-$J$ CO emission lines, contribute to $\sim10\%$ of the mass. The latter likely resides in the central regions of the PJ231-20 QSO host galaxy, exposed to strong X-ray radiation due to the BH accretion. The best-fit PDR component indicates $n_{\rm H}^{\rm PDR}\sim4\times10^{2}\,{\rm cm^{-3}}$ and $G_0\sim1\times10^5$, while the XDR component is determined by $n_{\rm H}^{\rm XDR}\sim5\times10^5\,{\rm cm^{-3}}$ and $F_{\rm X}\sim 10\, {\rm erg\,s^{-1}\,cm^{-2}}$. However, due to the low number of data points, these results are affected by degeneracies. 
\item We detected three H$_2$O lines ($3_{12}-2_{21}$, $3_{21}-3_{12}$, $3_{03}-2_{12}$) in the PJ231-20 QSO, and H$_2$O  $3_{03}-2_{12}$ in the PJ308-21 QSO. The PJ231-20 QSO, lies on the $L_{\rm H_2O}\propto L_{\rm TIR}^{1.1-1.2}$ trend found for low-$z$ sources, suggesting that the H$_2$O $3_{21}-3_{12}$  line is predominantly excited by pumping of IR photons of the dust UV-reprocessed radiation field. The $L_{\rm H_2O}/L_{\rm TIR}$ ratios in the PJ231-20 and PJ308-21 systems are of order of $\sim10^{-5}$, within the range observed in starburst galaxies at low $z$. The observed H$_2$O SLED in PJ231-20 QSO indicates that collisional excitation of the upper energy levels of the observed water vapor transitions, is likely minor. Furthermore, the SLED is similar to that of other high-$z$ sources. CLOUDY best-fit models are obtained using a single PDR component with $n_{\rm H}\sim 0.8\times10^5\,{\rm cm^{-3}}$, $G_0\sim 5\times10^5$ and high hydrogen column density of $N_{\rm H}=10^{24}\,{\rm cm^{-2}}$, indicating that H$_2$O emission arises deeply in the molecular clouds with high density, exposed to a strong radiation field.
\item The H$_2$O/OH$_{\rm 163\mu m}$ ratios in the PJ231-20 QSO predicted by CLOUDY, are fully consistent with the results obtained from the H$_2$O SLED modeling, indicating that OH$_{\rm 163\mu m}$ doublet emission likely arises from similar regions traced by water vapor. We found similar constraints, but with large uncertainties, for the PJ231-20 companion galaxy and the PJ308-21 QSO, indicating that the ISM is exposed to strong FUV radiation field with $G_0>10^4$. These results are also in broad agreement with the analysis of CO and atomic fine-structure lines.
\end{itemize}

In conclusion, in both the PJ231-20 and PJ308-21 quasars the AGN activity is mainly revealed through high-$J$ CO line emission, requiring an additional source of heating than PDRs, possibly provided by a strong X-ray radiation field, while the CO SLEDs in conjunction with the non-detections of high-$J$ CO lines in companions, at least for PJ231-20, suggest a less excited medium, possibly as a consequence of the lack of a strong AGN as suggested by previous optical/UV and X-rays observations. Then, the analysis of the atomic fine-structure and molecular lines, in conjunction with the analysis of H$_2$O and OH$_{\rm 163\mu m}$ transitions, points to a strong radiation field in the quasar host and companion galaxies associated with massive episodes of star formation, as reflected by the enormous amount of dust ($M_{\rm dust}\approx0.5\times10^8-0.5\times10^9\,M_{\astrosun}$) heated at relatively high temperature and huge gas reservoir ($M_{\rm gas}\apprge 10^{10}\,M_{\astrosun}$) and high star-formation efficiencies. However, although the presented analysis of the ISM properties highlight differences in the quasars their companion galaxies, the current data do not allow us to definitively rule out the presence of an AGN in the companions, especially for the PJ308-21 companion galaxy for which there is little information due to the lack of line detections.

This work demonstrates that a large variety of investigations can be carried out on the ISM of quasar hosts and SMGs
at such high redshift by targeting emission lines arising from different gas phases and environments. Future deeper follow-up observations have to be conducted, in order to get higher S/N and sample additional FIR emission lines to provide a better understanding of the ISM in these systems. Also, more comparisons between AGN and non-AGN environments have to be done by observing more object in this rich set of lines in the same redshift, thus providing us with a better statistics and in-depth exploration of the interplay between AGN and star formation in extreme conditions at early epochs.

\begin{acknowledgements}
We thanks the anonymous referee for the useful comments that significantly improve the paper. This paper makes use of the following ALMA data: ADS/JAO.ALMA\#2015.1.01115.S, ADS/JAO.ALMA\#2017.1.00139.S, ADS/JAO.ALMA\#2019.1.00147.S. ALMA is a partnership of ESO (representing its member states), NSF (USA) and NINS (Japan), together with NRC (Canada), MOST and ASIAA (Taiwan), and KASI (Republic of Korea), in cooperation with the Republic of Chile. The Joint ALMA Observatory is operated by ESO, AUI/NRAO and NAOJ. MN and BV acknowledge support from the ERC Advanced Grant 740246 (Cosmic Gas). XF and JY acknowledge the supports from the US NSF grant AST 15-15115 and AST 19-08284. DR acknowledges support from the National Science Foundation under grant numbers AST-1614213 and AST-1910107. DR also acknowledges support from the Alexander von Humboldt Foundation through a Humboldt Research Fellowship for Experienced Researchers. RW acknowledges supports from the National Science Foundation of China (11721303, 11991052) and the National Key R\&D Program of China (2016YFA0400703). This research made use of Astropy\footnote{http://www.astropy.org}, a community-developed core Python package for Astronomy \citep{AstropyI, AstropyII}, and Matplotlib \citep{Matplotlib}.
\end{acknowledgements}

\bibliographystyle{aa.bst}
\bibliography{MyBib}

\begin{thebibliography}{268}
\expandafter\ifx\csname natexlab\endcsname\relax\def\natexlab#1{#1}\fi

\bibitem[{{Abel} {et~al.}(2009){Abel}, {Dudley}, {Fischer}, {Satyapal}, \& {van
  Hoof}}]{Abel+2009}
{Abel}, N.~P., {Dudley}, C., {Fischer}, J., {Satyapal}, S., \& {van Hoof},
  P.~A.~M. 2009, \apj, 701, 1147

\bibitem[{{Abel} {et~al.}(2008){Abel}, {van Hoof}, {Shaw}, {Ferland }, \&
  {Elwert}}]{Abel+2008}
{Abel}, N.~P., {van Hoof}, P.~A.~M., {Shaw}, G., {Ferland }, G.~J., \&
  {Elwert}, T. 2008, \apj, 686, 1125

\bibitem[{{Ag{\'u}ndez} {et~al.}(2010){Ag{\'u}ndez}, {Goicoechea},
  {Cernicharo}, {Faure}, \& {Roueff}}]{Agundez+2010}
{Ag{\'u}ndez}, M., {Goicoechea}, J.~R., {Cernicharo}, J., {Faure}, A., \&
  {Roueff}, E. 2010, \apj, 713, 662

\bibitem[{{Alaghband-Zadeh} {et~al.}(2013){Alaghband-Zadeh}, {Chapman},
  {Swinbank}, {Smail}, {Danielson}, {Decarli}, {Ivison}, {Meijerink}, {Weiss},
  \& {van der Werf}}]{Alaghband-Zadeh+2013}
{Alaghband-Zadeh}, S., {Chapman}, S.~C., {Swinbank}, A.~M., {et~al.} 2013,
  \mnras, 435, 1493

\bibitem[{{Angulo} {et~al.}(2012){Angulo}, {Springel}, {White}, {Jenkins},
  {Baugh}, \& {Frenk}}]{Angulo+2012}
{Angulo}, R.~E., {Springel}, V., {White}, S.~D.~M., {et~al.} 2012, \mnras, 426,
  2046

\bibitem[{{Astropy Collaboration} {et~al.}(2018){Astropy Collaboration},
  {Price-Whelan}, {Sip{\H{o}}cz}, {G{\"u}nther}, {Lim}, {Crawford}, {Conseil},
  {Shupe}, {Craig}, {Dencheva}, {Ginsburg}, {Vand erPlas}, {Bradley},
  {P{\'e}rez-Su{\'a}rez}, {de Val-Borro}, {Aldcroft}, {Cruz}, {Robitaille},
  {Tollerud}, {Ardelean}, {Babej}, {Bach}, {Bachetti}, {Bakanov}, {Bamford},
  {Barentsen}, {Barmby}, {Baumbach}, {Berry}, {Biscani}, {Boquien}, {Bostroem},
  {Bouma}, {Brammer}, {Bray}, {Breytenbach}, {Buddelmeijer}, {Burke},
  {Calderone}, {Cano Rodr{\'\i}guez}, {Cara}, {Cardoso}, {Cheedella}, {Copin},
  {Corrales}, {Crichton}, {D'Avella}, {Deil}, {Depagne}, {Dietrich}, {Donath},
  {Droettboom}, {Earl}, {Erben}, {Fabbro}, {Ferreira}, {Finethy}, {Fox},
  {Garrison}, {Gibbons}, {Goldstein}, {Gommers}, {Greco}, {Greenfield},
  {Groener}, {Grollier}, {Hagen}, {Hirst}, {Homeier}, {Horton}, {Hosseinzadeh},
  {Hu}, {Hunkeler}, {Ivezi{\'c}}, {Jain}, {Jenness}, {Kanarek}, {Kendrew},
  {Kern}, {Kerzendorf}, {Khvalko}, {King}, {Kirkby}, {Kulkarni}, {Kumar},
  {Lee}, {Lenz}, {Littlefair}, {Ma}, {Macleod}, {Mastropietro}, {McCully},
  {Montagnac}, {Morris}, {Mueller}, {Mumford}, {Muna}, {Murphy}, {Nelson},
  {Nguyen}, {Ninan}, {N{\"o}the}, {Ogaz}, {Oh}, {Parejko}, {Parley}, {Pascual},
  {Patil}, {Patil}, {Plunkett}, {Prochaska}, {Rastogi}, {Reddy Janga},
  {Sabater}, {Sakurikar}, {Seifert}, {Sherbert}, {Sherwood-Taylor}, {Shih},
  {Sick}, {Silbiger}, {Singanamalla}, {Singer}, {Sladen}, {Sooley},
  {Sornarajah}, {Streicher}, {Teuben}, {Thomas}, {Tremblay}, {Turner},
  {Terr{\'o}n}, {van Kerkwijk}, {de la Vega}, {Watkins}, {Weaver}, {Whitmore},
  {Woillez}, {Zabalza}, \& {Astropy Contributors}}]{AstropyII}
{Astropy Collaboration}, {Price-Whelan}, A.~M., {Sip{\H{o}}cz}, B.~M., {et~al.}
  2018, \aj, 156, 123

\bibitem[{{Astropy Collaboration} {et~al.}(2013){Astropy Collaboration},
  {Robitaille}, {Tollerud}, {Greenfield}, {Droettboom}, {Bray}, {Aldcroft},
  {Davis}, {Ginsburg}, {Price-Whelan}, {Kerzendorf}, {Conley}, {Crighton},
  {Barbary}, {Muna}, {Ferguson}, {Grollier}, {Parikh}, {Nair}, {Unther},
  {Deil}, {Woillez}, {Conseil}, {Kramer}, {Turner}, {Singer}, {Fox}, {Weaver},
  {Zabalza}, {Edwards}, {Azalee Bostroem}, {Burke}, {Casey}, {Crawford},
  {Dencheva}, {Ely}, {Jenness}, {Labrie}, {Lim}, {Pierfederici}, {Pontzen},
  {Ptak}, {Refsdal}, {Servillat}, \& {Streicher}}]{AstropyI}
{Astropy Collaboration}, {Robitaille}, T.~P., {Tollerud}, E.~J., {et~al.} 2013,
  \aap, 558, A33

\bibitem[{{Ba{\~n}ados} {et~al.}(2015){Ba{\~n}ados}, {Decarli}, {Walter},
  {Venemans}, {Farina}, \& {Fan}}]{Banados+2015}
{Ba{\~n}ados}, E., {Decarli}, R., {Walter}, F., {et~al.} 2015, \apj, 805, L8

\bibitem[{{Ba{\~n}ados} {et~al.}(2016){Ba{\~n}ados}, {Venemans}, {Decarli},
  {Farina}, {Mazzucchelli}, {Walter}, {Fan}, {Stern}, {Schlafly}, {Chambers},
  {Rix}, {Jiang}, {McGreer}, {Simcoe}, {Wang}, {Yang}, {Morganson}, {De Rosa},
  {Greiner}, {Balokovi{\'c}}, {Burgett}, {Cooper}, {Draper}, {Flewelling},
  {Hodapp}, {Jun}, {Kaiser}, {Kudritzki}, {Magnier}, {Metcalfe}, {Miller},
  {Schindler}, {Tonry}, {Wainscoat}, {Waters}, \& {Yang}}]{Banados+2016}
{Ba{\~n}ados}, E., {Venemans}, B.~P., {Decarli}, R., {et~al.} 2016, \apjs, 227,
  11

\bibitem[{{Ba{\~n}ados} {et~al.}(2018){Ba{\~n}ados}, {Venemans},
  {Mazzucchelli}, {Farina}, {Walter}, {Wang}, {Decarli}, {Stern}, {Fan},
  {Davies}, {Hennawi}, {Simcoe}, {Turner}, {Rix}, {Yang}, {Kelson}, {Rudie}, \&
  {Winters}}]{Banados+2018}
{Ba{\~n}ados}, E., {Venemans}, B.~P., {Mazzucchelli}, C., {et~al.} 2018, \nat,
  553, 473

\bibitem[{{Balmaverde} {et~al.}(2017){Balmaverde}, {Gilli}, {Mignoli},
  {Bolzonella}, {Brusa}, {Cappelluti}, {Comastri}, {Sani}, {Vanzella},
  {Vignali}, {Vito}, \& {Zamorani}}]{Balmaverde+2017}
{Balmaverde}, B., {Gilli}, R., {Mignoli}, M., {et~al.} 2017, \aap, 606, A23

\bibitem[{{Becker} {et~al.}(2015){Becker}, {Bolton}, \& {Lidz}}]{Becker+2015}
{Becker}, G.~D., {Bolton}, J.~S., \& {Lidz}, A. 2015, \pasa, 32, e045

\bibitem[{{Beelen} {et~al.}(2006){Beelen}, {Cox}, {Benford}, {Dowell},
  {Kov{\'a}cs}, {Bertoldi}, {Omont}, \& {Carilli}}]{Beelen+2006}
{Beelen}, A., {Cox}, P., {Benford}, D.~J., {et~al.} 2006, \apj, 642, 694

\bibitem[{{Begelman} {et~al.}(2006){Begelman}, {Volonteri}, \&
  {Rees}}]{Begelman+2006}
{Begelman}, M.~C., {Volonteri}, M., \& {Rees}, M.~J. 2006, \mnras, 370, 289

\bibitem[{{Bergin} {et~al.}(2003){Bergin}, {Kaufman}, {Melnick}, {Snell}, \&
  {Howe}}]{Bergin+2003}
{Bergin}, E.~A., {Kaufman}, M.~J., {Melnick}, G.~J., {Snell}, R.~L., \& {Howe},
  J.~E. 2003, \apj, 582, 830

\bibitem[{{Bertoldi} {et~al.}(2003{\natexlab{a}}){Bertoldi}, {Carilli}, {Cox},
  {Fan}, {Strauss}, {Beelen}, {Omont}, \& {Zylka}}]{Bertoldi+2003b}
{Bertoldi}, F., {Carilli}, C.~L., {Cox}, P., {et~al.} 2003{\natexlab{a}}, \aap,
  406, L55

\bibitem[{{Bertoldi} {et~al.}(2003{\natexlab{b}}){Bertoldi}, {Cox}, {Neri},
  {Carilli}, {Walter}, {Omont}, {Beelen}, {Henkel}, {Fan}, {Strauss}, \&
  {Menten}}]{Bertoldi+2003a}
{Bertoldi}, F., {Cox}, P., {Neri}, R., {et~al.} 2003{\natexlab{b}}, \aap, 409,
  L47

\bibitem[{{Birkin} {et~al.}(2020){Birkin}, {Weiss}, {Wardlow}, {Smail},
  {Swinbank}, {Dudzevi{\v{c}}i{\={u}}t{\.{e}}}, {An}, {Ao}, {Chapman}, {Chen},
  {da Cunha}, {Dannerbauer}, {Gullberg}, {Hodge}, {Ikarashi}, {Ivison},
  {Matsuda}, {Stach}, {Walter}, {Wang}, \& {van der Werf}}]{Birkin+2020}
{Birkin}, J.~E., {Weiss}, A., {Wardlow}, J.~L., {et~al.} 2020, arXiv e-prints,
  arXiv:2009.03341

\bibitem[{{Bisbas} {et~al.}(2012){Bisbas}, {Bell}, {Viti}, {Yates}, \&
  {Barlow}}]{Bisbas+2012}
{Bisbas}, T.~G., {Bell}, T.~A., {Viti}, S., {Yates}, J., \& {Barlow}, M.~J.
  2012, \mnras, 427, 2100

\bibitem[{{Blain} {et~al.}(2003){Blain}, {Barnard}, \& {Chapman}}]{Blain+2003}
{Blain}, A.~W., {Barnard}, V.~E., \& {Chapman}, S.~C. 2003, \mnras, 338, 733

\bibitem[{{Bolatto} {et~al.}(2013){Bolatto}, {Wolfire}, \&
  {Leroy}}]{Bolatto+2013}
{Bolatto}, A.~D., {Wolfire}, M., \& {Leroy}, A.~K. 2013, \araa, 51, 207

\bibitem[{{Bonoli} {et~al.}(2009){Bonoli}, {Marulli}, {Springel}, {White},
  {Branchini}, \& {Moscardini}}]{Bonoli+2009}
{Bonoli}, S., {Marulli}, F., {Springel}, V., {et~al.} 2009, \mnras, 396, 423

\bibitem[{{Bonoli} {et~al.}(2014){Bonoli}, {Mayer}, \&
  {Callegari}}]{Bonoli+2014}
{Bonoli}, S., {Mayer}, L., \& {Callegari}, S. 2014, \mnras, 437, 1576

\bibitem[{{Boogaard} {et~al.}(2020){Boogaard}, {van der Werf}, {Wei{\ss}},
  {Popping}, {Decarli}, {Walter}, {Aravena}, {Bouwens}, {Riechers},
  {Gonz{\'a}lez-L{\'o}pez}, {Carilli}, {Kaasinen}, {Daddi}, {Cox},
  {D{\'\i}az-Santos}, {Inami}, {Cortes}, \& {Wagg}}]{Boogaard+2020}
{Boogaard}, L.~A., {van der Werf}, P., {Wei{\ss}}, A., {et~al.} 2020, arXiv
  e-prints, arXiv:2009.04348

\bibitem[{{Bothwell} {et~al.}(2017){Bothwell}, {Aguirre}, {Aravena},
  {Bethermin}, {Bisbas}, {Chapman}, {De Breuck}, {Gonzalez}, {Greve},
  {Hezaveh}, {Ma}, {Malkan}, {Marrone}, {Murphy}, {Spilker}, {Strandet},
  {Vieira}, \& {Wei{\ss}}}]{Bothwell+2017}
{Bothwell}, M.~S., {Aguirre}, J.~E., {Aravena}, M., {et~al.} 2017, \mnras, 466,
  2825

\bibitem[{{Bothwell} {et~al.}(2013){Bothwell}, {Aguirre}, {Chapman}, {Marrone},
  {Vieira}, {Ashby}, {Aravena}, {Benson}, {Bock}, {Bradford}, {Brodwin},
  {Carlstrom}, {Crawford}, {de Breuck}, {Downes}, {Fassnacht}, {Gonzalez},
  {Greve}, {Gullberg}, {Hezaveh}, {Holder}, {Holzapfel}, {Ibar}, {Ivison},
  {Kamenetzky}, {Keisler}, {Lupu}, {Ma}, {Malkan}, {McIntyre}, {Murphy},
  {Nguyen}, {Reichardt}, {Rosenman}, {Spilker}, {Stalder}, {Stark}, {Strand
  et}, {Vernet}, {Wei{\ss}}, \& {Welikala}}]{Bothwell+2013}
{Bothwell}, M.~S., {Aguirre}, J.~E., {Chapman}, S.~C., {et~al.} 2013, \apj,
  779, 67

\bibitem[{{Bothwell} {et~al.}(2010){Bothwell}, {Chapman}, {Tacconi}, {Smail},
  {Ivison}, {Casey}, {Bertoldi}, {Beswick}, {Biggs}, {Blain}, {Cox}, {Genzel},
  {Greve}, {Kennicutt}, {Muxlow}, {Neri}, \& {Omont}}]{Bothwell+2010}
{Bothwell}, M.~S., {Chapman}, S.~C., {Tacconi}, L., {et~al.} 2010, \mnras, 405,
  219

\bibitem[{{Bradford} {et~al.}(2011){Bradford}, {Bolatto}, {Maloney}, {Aguirre},
  {Bock}, {Glenn}, {Kamenetzky}, {Lupu}, {Matsuhara}, {Murphy}, {Naylor},
  {Nguyen}, {Scott}, \& {Zmuidzinas}}]{Bradford+2011}
{Bradford}, C.~M., {Bolatto}, A.~D., {Maloney}, P.~R., {et~al.} 2011, \apjl,
  741, L37

\bibitem[{{Bradford} {et~al.}(1999){Bradford}, {Stacey}, {Fischer}, {Smith},
  {Cohen}, {Greenhouse}, {Lord}, {Lutz}, {Maiolino}, {Malkan}, \&
  {Rieu}}]{Bradford+1999}
{Bradford}, C.~M., {Stacey}, G.~J., {Fischer}, J., {et~al.} 1999, in ESA
  Special Publication, Vol. 427, The Universe as Seen by ISO, ed. P.~{Cox} \&
  M.~{Kessler}, 861

\bibitem[{{Brauher} {et~al.}(2008){Brauher}, {Dale}, \& {Helou}}]{Brauher+2008}
{Brauher}, J.~R., {Dale}, D.~A., \& {Helou}, G. 2008, \apjs, 178, 280

\bibitem[{{Ca{\~n}ameras} {et~al.}(2018){Ca{\~n}ameras}, {Yang}, {Nesvadba},
  {Beelen}, {Kneissl}, {Koenig}, {Le Floc'h}, {Limousin}, {Malhotra}, {Omont},
  \& {Scott}}]{Canameras+2018}
{Ca{\~n}ameras}, R., {Yang}, C., {Nesvadba}, N.~P.~H., {et~al.} 2018, \aap,
  620, A61

\bibitem[{{Carilli} \& {Walter}(2013)}]{CarilliWalter2013}
{Carilli}, C.~L. \& {Walter}, F. 2013, Annual Review of Astronomy and
  Astrophysics, 51, 105

\bibitem[{{Carniani} {et~al.}(2019){Carniani}, {Gallerani}, {Vallini},
  {Pallottini}, {Tazzari}, {Ferrara}, {Maiolino}, {Cicone}, {Feruglio}, {Neri},
  {D'Odorico}, {Wang}, \& {Li}}]{Carniani+2019}
{Carniani}, S., {Gallerani}, S., {Vallini}, L., {et~al.} 2019, \mnras, 489,
  3939

\bibitem[{{Carniani} {et~al.}(2017){Carniani}, {Marconi}, {Maiolino},
  {Feruglio}, {Brusa}, {Cresci}, {Cano-D{\'\i}az}, {Cicone}, {Balmaverde},
  {Fiore}, {Ferrara}, {Gallerani}, {La Franca}, {Mainieri}, {Mannucci},
  {Netzer}, {Piconcelli}, {Sani}, {Schneider}, {Shemmer}, \&
  {Testi}}]{Carniani+2017}
{Carniani}, S., {Marconi}, A., {Maiolino}, R., {et~al.} 2017, \aap, 605, A105

\bibitem[{{Carral} {et~al.}(1994){Carral}, {Hollenbach}, {Lord}, {Colgan},
  {Haas}, {Rubin}, \& {Erickson}}]{Carral+1994}
{Carral}, P., {Hollenbach}, D.~J., {Lord}, S.~D., {et~al.} 1994, \apj, 423, 223

\bibitem[{{Casey}(2012)}]{Casey+2012}
{Casey}, C.~M. 2012, \mnras, 425, 3094

\bibitem[{{Cernicharo} {et~al.}(2006){Cernicharo}, {Goicoechea}, {Pardo}, \&
  {Asensio-Ramos}}]{Cernicharo+2006}
{Cernicharo}, J., {Goicoechea}, J.~R., {Pardo}, J.~R., \& {Asensio-Ramos}, A.
  2006, \apj, 642, 940

\bibitem[{{Chapman} {et~al.}(2005){Chapman}, {Blain}, {Smail}, \&
  {Ivison}}]{Chapman+2005}
{Chapman}, S.~C., {Blain}, A.~W., {Smail}, I., \& {Ivison}, R.~J. 2005, \apj,
  622, 772

\bibitem[{{Charmandaris} {et~al.}(2004){Charmandaris}, {Uchida}, {Weedman},
  {Herter}, {Houck}, {Teplitz}, {Armus}, {Brandl}, {Higdon}, {Soifer},
  {Appleton}, {van Cleve}, \& {Higdon}}]{Charmandaris+2004}
{Charmandaris}, V., {Uchida}, K.~I., {Weedman}, D., {et~al.} 2004, \apjs, 154,
  142

\bibitem[{{Colbert} {et~al.}(1999){Colbert}, {Malkan}, {Clegg}, {Cox},
  {Fischer}, {Lord}, {Luhman}, {Satyapal}, {Smith}, {Spinoglio}, {Stacey}, \&
  {Unger}}]{Colbert+1999}
{Colbert}, J.~W., {Malkan}, M.~A., {Clegg}, P.~E., {et~al.} 1999, \apj, 511,
  721

\bibitem[{{Combes} {et~al.}(2012){Combes}, {Rex}, {Rawle}, {Egami}, {Boone},
  {Smail}, {Richard}, {Ivison}, {Gurwell}, {Casey}, {Omont}, {Berciano Alba},
  {Dessauges-Zavadsky}, {Edge}, {Fazio}, {Kneib}, {Okabe}, {Pell{\'o}},
  {P{\'e}rez-Gonz{\'a}lez}, {Schaerer}, {Smith}, {Swinbank}, \& {van der
  Werf}}]{Combes+2012}
{Combes}, F., {Rex}, M., {Rawle}, T.~D., {et~al.} 2012, \aap, 538, L4

\bibitem[{{Combes} \& {Wiklind}(1997)}]{Combes+1997}
{Combes}, F. \& {Wiklind}, T. 1997, \apjl, 486, L79

\bibitem[{{Conley} {et~al.}(2011){Conley}, {Cooray}, {Vieira}, {Gonz{\'a}lez
  Solares}, {Kim}, {Aguirre}, {Amblard}, {Auld}, {Baker}, {Beelen}, {Blain},
  {Blundell}, {Bock}, {Bradford}, {Bridge}, {Brisbin}, {Burgarella},
  {Carpenter}, {Chanial}, {Chapin}, {Christopher}, {Clements}, {Cox},
  {Djorgovski}, {Dowell}, {Eales}, {Earle}, {Ellsworth-Bowers}, {Farrah},
  {Franceschini}, {Frayer}, {Fu}, {Gavazzi}, {Glenn}, {Griffin}, {Gurwell},
  {Halpern}, {Ibar}, {Ivison}, {Jarvis}, {Kamenetzky}, {Krips}, {Levenson},
  {Lupu}, {Mahabal}, {Maloney}, {Maraston}, {Marchetti}, {Marsden},
  {Matsuhara}, {Mortier}, {Murphy}, {Naylor}, {Neri}, {Nguyen}, {Oliver},
  {Omont}, {Page}, {Papageorgiou}, {Pearson}, {P{\'e}rez-Fournon}, {Pohlen},
  {Rangwala}, {Rawlings}, {Raymond}, {Riechers}, {Rodighiero}, {Roseboom},
  {Rowan-Robinson}, {Schulz}, {Scott}, {Scott}, {Serra}, {Seymour}, {Shupe},
  {Smith}, {Symeonidis}, {Tugwell}, {Vaccari}, {Valiante}, {Valtchanov},
  {Verma}, {Viero}, {Vigroux}, {Wang}, {Wiebe}, {Wright}, {Xu}, {Zeimann},
  {Zemcov}, \& {Zmuidzinas}}]{Conley+2011}
{Conley}, A., {Cooray}, A., {Vieira}, J.~D., {et~al.} 2011, \apjl, 732, L35

\bibitem[{{Connor} {et~al.}(2020){Connor}, {Ba{\~n}ados}, {Mazzucchelli},
  {Stern}, {Decarli}, {Fan}, {Farina}, {Lusso}, {Neeleman}, \&
  {Walter}}]{Connor+2020}
{Connor}, T., {Ba{\~n}ados}, E., {Mazzucchelli}, C., {et~al.} 2020, arXiv
  e-prints, arXiv:2007.14571

\bibitem[{{Connor} {et~al.}(2019){Connor}, {Ba{\~n}ados}, {Stern}, {Decarli},
  {Schindler}, {Fan}, {Farina}, {Mazzucchelli}, {Mulchaey}, \&
  {Walter}}]{Connor+2019}
{Connor}, T., {Ba{\~n}ados}, E., {Stern}, D., {et~al.} 2019, \apj, 887, 171

\bibitem[{{Costa} {et~al.}(2014){Costa}, {Sijacki}, {Trenti}, \&
  {Haehnelt}}]{Costa+2014}
{Costa}, T., {Sijacki}, D., {Trenti}, M., \& {Haehnelt}, M.~G. 2014, \mnras,
  439, 2146

\bibitem[{{Cruddace} {et~al.}(1974){Cruddace}, {Paresce}, {Bowyer}, \&
  {Lampton}}]{Cruddace+1974}
{Cruddace}, R., {Paresce}, F., {Bowyer}, S., \& {Lampton}, M. 1974, \apj, 187,
  497

\bibitem[{{da Cunha} {et~al.}(2013){da Cunha}, {Groves}, {Walter}, {Decarli},
  {Weiss}, {Bertoldi}, {Carilli}, {Daddi}, {Elbaz}, {Ivison}, {Maiolino},
  {Riechers}, {Rix}, {Sargent}, \& {Smail}}]{daCunha+2013}
{da Cunha}, E., {Groves}, B., {Walter}, F., {et~al.} 2013, \apj, 766, 13

\bibitem[{{Daddi} {et~al.}(2010){Daddi}, {Bournaud}, {Walter}, {Dannerbauer},
  {Carilli}, {Dickinson}, {Elbaz}, {Morrison}, {Riechers}, {Onodera}, {Salmi},
  {Krips}, \& {Stern}}]{Daddi+2010}
{Daddi}, E., {Bournaud}, F., {Walter}, F., {et~al.} 2010, \apj, 713, 686

\bibitem[{{Daniel} {et~al.}(2012){Daniel}, {Goicoechea}, {Cernicharo},
  {Dubernet}, \& {Faure}}]{Daniel+2012}
{Daniel}, F., {Goicoechea}, J.~R., {Cernicharo}, J., {Dubernet}, M.~L., \&
  {Faure}, A. 2012, \aap, 547, A81

\bibitem[{{De Rosa} {et~al.}(2011){De Rosa}, {Decarli}, {Walter}, {Fan},
  {Jiang}, {Kurk}, {Pasquali}, \& {Rix}}]{DeRosa+2011}
{De Rosa}, G., {Decarli}, R., {Walter}, F., {et~al.} 2011, \apj, 739, 56

\bibitem[{{Decarli} {et~al.}(2019{\natexlab{a}}){Decarli}, {Dotti},
  {Ba{\~n}ados}, {Farina}, {Walter}, {Carilli}, {Fan}, {Mazzucchelli},
  {Neeleman}, {Novak}, {Riechers}, {Strauss}, {Venemans}, {Yang}, \&
  {Wang}}]{Decarli+2019}
{Decarli}, R., {Dotti}, M., {Ba{\~n}ados}, E., {et~al.} 2019{\natexlab{a}},
  \apj, 880, 157

\bibitem[{{Decarli} {et~al.}(2019{\natexlab{b}}){Decarli}, {Mignoli}, {Gilli},
  {Balmaverde}, {Brusa}, {Cappelluti}, {Comastri}, {Nanni}, {Peca},
  {Pensabene}, {Vanzella}, \& {Vignali}}]{Decarli+2019b}
{Decarli}, R., {Mignoli}, M., {Gilli}, R., {et~al.} 2019{\natexlab{b}}, \aap,
  631, L10

\bibitem[{{Decarli} {et~al.}(2014){Decarli}, {Walter}, {Carilli}, {Bertoldi},
  {Cox}, {Ferkinhoff}, {Groves}, {Maiolino}, {Neri}, {Riechers}, \&
  {Weiss}}]{Decarli+2014}
{Decarli}, R., {Walter}, F., {Carilli}, C., {et~al.} 2014, \apjl, 782, L17

\bibitem[{{Decarli} {et~al.}(2012){Decarli}, {Walter}, {Neri}, {Bertoldi},
  {Carilli}, {Cox}, {Kneib}, {Lestrade}, {Maiolino}, {Omont}, {Richard},
  {Riechers}, {Thanjavur}, \& {Weiss}}]{Decarli+2012}
{Decarli}, R., {Walter}, F., {Neri}, R., {et~al.} 2012, \apj, 752, 2

\bibitem[{{Decarli} {et~al.}(2017){Decarli}, {Walter}, {Venemans},
  {Ba{\~n}ados}, {Bertoldi}, {Carilli}, {Fan}, {Farina}, {Mazzucchelli},
  {Riechers}, {Rix}, {Strauss}, {Wang}, \& {Yang}}]{Decarli+2017}
{Decarli}, R., {Walter}, F., {Venemans}, B.~P., {et~al.} 2017, \nat, 545, 457

\bibitem[{{Decarli} {et~al.}(2018){Decarli}, {Walter}, {Venemans},
  {Ba{\~n}ados}, {Bertoldi}, {Carilli}, {Fan}, {Farina}, {Mazzucchelli},
  {Riechers}, {Rix}, {Strauss}, {Wang}, \& {Yang}}]{Decarli+2018}
{Decarli}, R., {Walter}, F., {Venemans}, B.~P., {et~al.} 2018, \apj, 854, 97

\bibitem[{{Destombes} {et~al.}(1977){Destombes}, {Marliere}, {Baudry}, \&
  {Brillet}}]{Destombes+1977}
{Destombes}, J.~L., {Marliere}, C., {Baudry}, A., \& {Brillet}, J. 1977, \aap,
  61, 769

\bibitem[{{Dewangan} {et~al.}(1987){Dewangan}, {Flower}, \&
  {Alexander}}]{Dewangan+1987}
{Dewangan}, D.~P., {Flower}, D.~R., \& {Alexander}, M.~H. 1987, \mnras, 226,
  505

\bibitem[{{D{\'\i}az-Santos} {et~al.}(2017){D{\'\i}az-Santos}, {Armus},
  {Charmandaris}, {Lu}, {Stierwalt}, {Stacey}, {Malhotra}, {van der Werf},
  {Howell}, {Privon}, {Mazzarella}, {Goldsmith}, {Murphy}, {Barcos-Mu{\~n}oz},
  {Linden}, {Inami}, {Larson}, {Evans}, {Appleton}, {Iwasawa}, {Lord},
  {Sanders}, \& {Surace}}]{Diaz-Santos+2017}
{D{\'\i}az-Santos}, T., {Armus}, L., {Charmandaris}, V., {et~al.} 2017, \apj,
  846, 32

\bibitem[{{D{\'\i}az-Santos} {et~al.}(2013){D{\'\i}az-Santos}, {Armus},
  {Charmandaris}, {Stierwalt}, {Murphy}, {Haan}, {Inami}, {Malhotra},
  {Meijerink}, {Stacey}, {Petric}, {Evans}, {Veilleux}, {van der Werf}, {Lord},
  {Lu}, {Howell}, {Appleton}, {Mazzarella}, {Surace}, {Xu}, {Schulz},
  {Sanders}, {Bridge}, {Chan}, {Frayer}, {Iwasawa}, {Melbourne}, \&
  {Sturm}}]{Diaz-Santos+2013}
{D{\'\i}az-Santos}, T., {Armus}, L., {Charmandaris}, V., {et~al.} 2013, \apj,
  774, 68

\bibitem[{{Downes} {et~al.}(1992){Downes}, {Radford}, {Greve}, {Thum},
  {Solomon}, \& {Wink}}]{Downes+1992}
{Downes}, D., {Radford}, J.~E., {Greve}, A., {et~al.} 1992, \apjl, 398, L25

\bibitem[{{Downes} \& {Solomon}(1998)}]{Downes+1998}
{Downes}, D. \& {Solomon}, P.~M. 1998, \apj, 507, 615

\bibitem[{{Draine} {et~al.}(2007){Draine}, {Dale}, {Bendo}, {Gordon}, {Smith},
  {Armus}, {Engelbracht}, {Helou}, {Kennicutt}, {Li}, {Roussel}, {Walter},
  {Calzetti}, {Moustakas}, {Murphy}, {Rieke}, {Bot}, {Hollenbach}, {Sheth}, \&
  {Teplitz}}]{Draine+2007}
{Draine}, B.~T., {Dale}, D.~A., {Bendo}, G., {et~al.} 2007, \apj, 663, 866

\bibitem[{{Dunne} {et~al.}(2000){Dunne}, {Eales}, {Edmunds}, {Ivison},
  {Alexander}, \& {Clements}}]{Dunne+2000}
{Dunne}, L., {Eales}, S., {Edmunds}, M., {et~al.} 2000, \mnras, 315, 115

\bibitem[{{Elbaz} {et~al.}(2011){Elbaz}, {Dickinson}, {Hwang},
  {D{\'\i}az-Santos}, {Magdis}, {Magnelli}, {Le Borgne}, {Galliano},
  {Pannella}, {Chanial}, {Armus}, {Charmandaris}, {Daddi}, {Aussel}, {Popesso},
  {Kartaltepe}, {Altieri}, {Valtchanov}, {Coia}, {Dannerbauer}, {Dasyra},
  {Leiton}, {Mazzarella}, {Alexander}, {Buat}, {Burgarella}, {Chary}, {Gilli},
  {Ivison}, {Juneau}, {Le Floc'h}, {Lutz}, {Morrison}, {Mullaney}, {Murphy},
  {Pope}, {Scott}, {Brodwin}, {Calzetti}, {Cesarsky}, {Charlot}, {Dole},
  {Eisenhardt}, {Ferguson}, {F{\"o}rster Schreiber}, {Frayer}, {Giavalisco},
  {Huynh}, {Koekemoer}, {Papovich}, {Reddy}, {Surace}, {Teplitz}, {Yun}, \&
  {Wilson}}]{Elbaz+2011}
{Elbaz}, D., {Dickinson}, M., {Hwang}, H.~S., {et~al.} 2011, \aap, 533, A119

\bibitem[{{Elitzur} \& {Asensio Ramos}(2006)}]{Elitzur+2006}
{Elitzur}, M. \& {Asensio Ramos}, A. 2006, \mnras, 365, 779

\bibitem[{{Elitzur} \& {de Jong}(1978)}]{Elitzur+1978}
{Elitzur}, M. \& {de Jong}, T. 1978, \aap, 67, 323

\bibitem[{{Engel} {et~al.}(2010){Engel}, {Tacconi}, {Davies}, {Neri}, {Smail},
  {Chapman}, {Genzel}, {Cox}, {Greve}, {Ivison}, {Blain}, {Bertoldi}, \&
  {Omont}}]{Engel+2010}
{Engel}, H., {Tacconi}, L.~J., {Davies}, R.~I., {et~al.} 2010, \apj, 724, 233

\bibitem[{{Faisst} {et~al.}(2020){Faisst}, {Fudamoto}, {Oesch}, {Scoville},
  {Riechers}, {Pavesi}, \& {Capak}}]{Faisst+2020}
{Faisst}, A.~L., {Fudamoto}, Y., {Oesch}, P.~A., {et~al.} 2020, \mnras, 498,
  4192

\bibitem[{{Fan} {et~al.}(2006{\natexlab{a}}){Fan}, {Carilli}, \&
  {Keating}}]{Fan+2006}
{Fan}, X., {Carilli}, C.~L., \& {Keating}, B. 2006{\natexlab{a}}, \araa, 44,
  415

\bibitem[{{Fan} {et~al.}(2006{\natexlab{b}}){Fan}, {Strauss}, {Becker},
  {White}, {Gunn}, {Knapp}, {Richards}, {Schneider}, {Brinkmann}, \&
  {Fukugita}}]{Fan+2006b}
{Fan}, X., {Strauss}, M.~A., {Becker}, R.~H., {et~al.} 2006{\natexlab{b}}, \aj,
  132, 117

\bibitem[{{Farina} {et~al.}(2019){Farina}, {Arrigoni-Battaia}, {Costa},
  {Walter}, {Hennawi}, {Drake}, {Decarli}, {Gutcke}, {Mazzucchelli},
  {Neeleman}, {Georgiev}, {Eilers}, {Davies}, {Ba{\~n}ados}, {Fan}, {Onoue},
  {Schindler}, {Venemans}, {Wang}, {Yang}, {Rabien}, \& {Busoni}}]{Farina+2019}
{Farina}, E.~P., {Arrigoni-Battaia}, F., {Costa}, T., {et~al.} 2019, \apj, 887,
  196

\bibitem[{{Farina} {et~al.}(2017){Farina}, {Venemans}, {Decarli}, {Hennawi},
  {Walter}, {Ba{\~n}ados}, {Mazzucchelli}, {Cantalupo}, {Arrigoni-Battaia}, \&
  {McGreer}}]{Farina+2017}
{Farina}, E.~P., {Venemans}, B.~P., {Decarli}, R., {et~al.} 2017, \apj, 848, 78

\bibitem[{{Farrah} {et~al.}(2013){Farrah}, {Lebouteiller}, {Spoon},
  {Bernard-Salas}, {Pearson}, {Rigopoulou}, {Smith}, {Gonz{\'a}lez-Alfonso},
  {Clements}, {Efstathiou}, {Cormier}, {Afonso}, {Petty}, {Harris}, {Hurley},
  {Borys}, {Verma}, {Cooray}, \& {Salvatelli}}]{Farrah+2013}
{Farrah}, D., {Lebouteiller}, V., {Spoon}, H.~W.~W., {et~al.} 2013, \apj, 776,
  38

\bibitem[{{Faure} {et~al.}(2007){Faure}, {Crimier}, {Ceccarelli}, {Valiron},
  {Wiesenfeld}, \& {Dubernet}}]{Faure+2007}
{Faure}, A., {Crimier}, N., {Ceccarelli}, C., {et~al.} 2007, \aap, 472, 1029

\bibitem[{{Faure} \& {Josselin}(2008)}]{Faure+2008}
{Faure}, A. \& {Josselin}, E. 2008, \aap, 492, 257

\bibitem[{{Ferkinhoff} {et~al.}(2011){Ferkinhoff}, {Brisbin}, {Nikola},
  {Parshley}, {Stacey}, {Phillips}, {Falgarone}, {Benford}, {Staguhn}, \&
  {Tucker}}]{Ferkinhoff+2011}
{Ferkinhoff}, C., {Brisbin}, D., {Nikola}, T., {et~al.} 2011, \apjl, 740, L29

\bibitem[{{Ferland} {et~al.}(2017){Ferland}, {Chatzikos}, {Guzm{\'a}n},
  {Lykins}, {van Hoof}, {Williams}, {Abel}, {Badnell}, {Keenan}, {Porter}, \&
  {Stancil}}]{Ferland+2017}
{Ferland}, G.~J., {Chatzikos}, M., {Guzm{\'a}n}, F., {et~al.} 2017, \rmxaa, 53,
  385

\bibitem[{{Feruglio} {et~al.}(2018){Feruglio}, {Fiore}, {Carniani}, {Maiolino},
  {D'Odorico}, {Luminari}, {Barai}, {Bischetti}, {Bongiorno}, {Cristiani},
  {Ferrara}, {Gallerani}, {Marconi}, {Pallottini}, {Piconcelli}, \&
  {Zappacosta}}]{Feruglio+2018}
{Feruglio}, C., {Fiore}, F., {Carniani}, S., {et~al.} 2018, \aap, 619, A39

\bibitem[{{Fischer} {et~al.}(1999){Fischer}, {Luhman}, {Satyapal},
  {Greenhouse}, {Stacey}, {Bradford}, {Lord}, {Brauher}, {Unger}, {Clegg},
  {Smith}, {Melnick}, {Colbert}, {Malkan}, {Spinoglio}, {Cox}, {Harvey},
  {Suter}, \& {Strelnitski}}]{Fischer+1999}
{Fischer}, J., {Luhman}, M.~L., {Satyapal}, S., {et~al.} 1999, \apss, 266, 91

\bibitem[{{Fischer} {et~al.}(2010){Fischer}, {Sturm}, {Gonz{\'a}lez-Alfonso},
  {Graci{\'a}-Carpio}, {Hailey-Dunsheath}, {Poglitsch}, {Contursi}, {Lutz},
  {Genzel}, {Sternberg}, {Verma}, \& {Tacconi}}]{Fischer+2010}
{Fischer}, J., {Sturm}, E., {Gonz{\'a}lez-Alfonso}, E., {et~al.} 2010, \aap,
  518, L41

\bibitem[{{Fixsen} {et~al.}(1999){Fixsen}, {Bennett}, \&
  {Mather}}]{Fixsen+1999}
{Fixsen}, D.~J., {Bennett}, C.~L., \& {Mather}, J.~C. 1999, \apj, 526, 207

\bibitem[{{Flower} \& {Pineau Des For{\^e}ts}(2010)}]{Flower+2010}
{Flower}, D.~R. \& {Pineau Des For{\^e}ts}, G. 2010, \mnras, 406, 1745

\bibitem[{{Foreman-Mackey} {et~al.}(2013){Foreman-Mackey}, {Hogg}, {Lang}, \&
  {Goodman}}]{Foreman+2013}
{Foreman-Mackey}, D., {Hogg}, D.~W., {Lang}, D., \& {Goodman}, J. 2013,
  Publications of the Astronomical Society of the Pacific, 125, 306

\bibitem[{{Gallerani} {et~al.}(2014){Gallerani}, {Ferrara}, {Neri}, \&
  {Maiolino}}]{Gallerani+2014}
{Gallerani}, S., {Ferrara}, A., {Neri}, R., \& {Maiolino}, R. 2014, \mnras,
  445, 2848

\bibitem[{{Genzel} {et~al.}(2015){Genzel}, {Tacconi}, {Lutz}, {Saintonge},
  {Berta}, {Magnelli}, {Combes}, {Garc{\'\i}a-Burillo}, {Neri}, {Bolatto},
  {Contini}, {Lilly}, {Boissier}, {Boone}, {Bouch{\'e}}, {Bournaud}, {Burkert},
  {Carollo}, {Colina}, {Cooper}, {Cox}, {Feruglio}, {F{\"o}rster Schreiber},
  {Freundlich}, {Gracia-Carpio}, {Juneau}, {Kovac}, {Lippa}, {Naab}, {Salome},
  {Renzini}, {Sternberg}, {Walter}, {Weiner}, {Weiss}, \&
  {Wuyts}}]{Genzel+2015}
{Genzel}, R., {Tacconi}, L.~J., {Lutz}, D., {et~al.} 2015, \apj, 800, 20

\bibitem[{{Gerin} \& {Phillips}(2000)}]{Gerin+2000}
{Gerin}, M. \& {Phillips}, T.~G. 2000, \apj, 537, 644

\bibitem[{{Glassgold} \& {Langer}(1974)}]{Glassgold+1974}
{Glassgold}, A.~E. \& {Langer}, W.~D. 1974, \apj, 193, 73

\bibitem[{{Godard} {et~al.}(2019){Godard}, {Pineau des For{\^e}ts}, {Lesaffre},
  {Lehmann}, {Gusdorf}, \& {Falgarone}}]{Godard+2019}
{Godard}, B., {Pineau des For{\^e}ts}, G., {Lesaffre}, P., {et~al.} 2019, \aap,
  622, A100

\bibitem[{{Goicoechea} \& {Cernicharo}(2002)}]{Goicoechea+2002}
{Goicoechea}, J.~R. \& {Cernicharo}, J. 2002, \apjl, 576, L77

\bibitem[{{Goicoechea} {et~al.}(2005){Goicoechea}, {Mart{\'\i}n-Pintado}, \&
  {Cernicharo}}]{Goicoechea+2005}
{Goicoechea}, J.~R., {Mart{\'\i}n-Pintado}, J., \& {Cernicharo}, J. 2005, \apj,
  619, 291

\bibitem[{{Goldsmith} \& {Langer}(1999)}]{Goldsmith+1999}
{Goldsmith}, P.~F. \& {Langer}, W.~D. 1999, \apj, 517, 209

\bibitem[{{Goldsmith} {et~al.}(2012){Goldsmith}, {Langer}, {Pineda}, \&
  {Velusamy}}]{Goldsmith+2012}
{Goldsmith}, P.~F., {Langer}, W.~D., {Pineda}, J.~L., \& {Velusamy}, T. 2012,
  \apjs, 203, 13

\bibitem[{{Gonz{\'a}lez-Alfonso} {et~al.}(2014){Gonz{\'a}lez-Alfonso},
  {Fischer}, {Aalto}, \& {Falstad}}]{Gonzalez-Alfonso+2014}
{Gonz{\'a}lez-Alfonso}, E., {Fischer}, J., {Aalto}, S., \& {Falstad}, N. 2014,
  \aap, 567, A91

\bibitem[{{Gonz{\'a}lez-Alfonso} {et~al.}(2013){Gonz{\'a}lez-Alfonso},
  {Fischer}, {Bruderer}, {M{\"u}ller}, {Graci{\'a}-Carpio}, {Sturm}, {Lutz},
  {Poglitsch}, {Feuchtgruber}, {Veilleux}, {Contursi}, {Sternberg},
  {Hailey-Dunsheath}, {Verma}, {Christopher}, {Davies}, {Genzel}, \&
  {Tacconi}}]{Gonzalez-Alfonso+2013}
{Gonz{\'a}lez-Alfonso}, E., {Fischer}, J., {Bruderer}, S., {et~al.} 2013, \aap,
  550, A25

\bibitem[{{Gonz{\'a}lez-Alfonso} {et~al.}(2012){Gonz{\'a}lez-Alfonso},
  {Fischer}, {Graci{\'a}-Carpio}, {Sturm}, {Hailey-Dunsheath}, {Lutz},
  {Poglitsch}, {Contursi}, {Feuchtgruber}, {Veilleux}, {Spoon}, {Verma},
  {Christopher}, {Davies}, {Sternberg}, {Genzel}, \&
  {Tacconi}}]{Gonzalez-Alfonso+2012}
{Gonz{\'a}lez-Alfonso}, E., {Fischer}, J., {Graci{\'a}-Carpio}, J., {et~al.}
  2012, \aap, 541, A4

\bibitem[{{Gonz{\'a}lez-Alfonso} {et~al.}(2010){Gonz{\'a}lez-Alfonso},
  {Fischer}, {Isaak}, {Rykala}, {Savini}, {Spaans}, {van der Werf},
  {Meijerink}, {Israel}, {Loenen}, {Vlahakis}, {Smith}, {Charmandaris},
  {Aalto}, {Henkel}, {Wei{\ss}}, {Walter}, {Greve}, {Mart{\'\i}n-Pintado},
  {Naylor}, {Spinoglio}, {Veilleux}, {Harris}, {Armus}, {Lord}, {Mazzarella},
  {Xilouris}, {Sand ers}, {Dasyra}, {Wiedner}, {Kramer}, {Papadopoulos},
  {Stacey}, {Evans}, \& {Gao}}]{Gonzalez-Alfonso+2010}
{Gonz{\'a}lez-Alfonso}, E., {Fischer}, J., {Isaak}, K., {et~al.} 2010, \aap,
  518, L43

\bibitem[{{Gonz{\'a}lez-Alfonso} {et~al.}(2017){Gonz{\'a}lez-Alfonso},
  {Fischer}, {Spoon}, {Stewart}, {Ashby}, {Veilleux}, {Smith}, {Sturm},
  {Farrah}, {Falstad}, {Mel{\'e}ndez}, {Graci{\'a}-Carpio}, {Janssen}, \&
  {Lebouteiller}}]{Gonzalez-Alfonso+2017}
{Gonz{\'a}lez-Alfonso}, E., {Fischer}, J., {Spoon}, H.~W.~W., {et~al.} 2017,
  \apj, 836, 11

\bibitem[{{Gonz{\'a}lez-Alfonso} {et~al.}(2008){Gonz{\'a}lez-Alfonso}, {Smith},
  {Ashby}, {Fischer}, {Spinoglio}, \& {Grundy}}]{Gonzalez-Alfonso+2008}
{Gonz{\'a}lez-Alfonso}, E., {Smith}, H.~A., {Ashby}, M. L.~N., {et~al.} 2008,
  \apj, 675, 303

\bibitem[{{Gonz{\'a}lez-Alfonso} {et~al.}(2004){Gonz{\'a}lez-Alfonso}, {Smith},
  {Fischer}, \& {Cernicharo}}]{Gonzalez-Alfonso+2004}
{Gonz{\'a}lez-Alfonso}, E., {Smith}, H.~A., {Fischer}, J., \& {Cernicharo}, J.
  2004, \apj, 613, 247

\bibitem[{{Graci{\'a}-Carpio} {et~al.}(2011){Graci{\'a}-Carpio}, {Sturm},
  {Hailey-Dunsheath}, {Fischer}, {Contursi}, {Poglitsch}, {Genzel},
  {Gonz{\'a}lez-Alfonso}, {Sternberg}, {Verma}, {Christopher}, {Davies},
  {Feuchtgruber}, {de Jong}, {Lutz}, \& {Tacconi}}]{Gracia-Carpio+2011}
{Graci{\'a}-Carpio}, J., {Sturm}, E., {Hailey-Dunsheath}, S., {et~al.} 2011,
  \apjl, 728, L7

\bibitem[{{Graff} \& {Dalgarno}(1987)}]{Graff+1987}
{Graff}, M.~M. \& {Dalgarno}, A. 1987, \apj, 317, 432

\bibitem[{{Habing}(1968)}]{Habing+1968}
{Habing}, H.~J. 1968, \bain, 19, 421

\bibitem[{{Habouzit} {et~al.}(2019){Habouzit}, {Volonteri}, {Somerville},
  {Dubois}, {Peirani}, {Pichon}, \& {Devriendt}}]{Habouzit+2019}
{Habouzit}, M., {Volonteri}, M., {Somerville}, R.~S., {et~al.} 2019, \mnras,
  489, 1206

\bibitem[{{Hartquist} \& {Williams}(1998)}]{Hartquist+1998}
{Hartquist}, T.~W. \& {Williams}, D.~A. 1998, The molecular astrophysics of
  stars and galaxies, 4

\bibitem[{{Hashimoto} {et~al.}(2019{\natexlab{a}}){Hashimoto}, {Inoue},
  {Mawatari}, {Tamura}, {Matsuo}, {Furusawa}, {Harikane}, {Shibuya}, {Knudsen},
  {Kohno}, {Ono}, {Zackrisson}, {Okamoto}, {Kashikawa}, {Oesch}, {Ouchi},
  {Ota}, {Shimizu}, {Taniguchi}, {Umehata}, \& {Watson}}]{Hashimoto+2019a}
{Hashimoto}, T., {Inoue}, A.~K., {Mawatari}, K., {et~al.} 2019{\natexlab{a}},
  \pasj, 71, 71

\bibitem[{{Hashimoto} {et~al.}(2019{\natexlab{b}}){Hashimoto}, {Inoue},
  {Tamura}, {Matsuo}, {Mawatari}, \& {Yamaguchi}}]{Hashimoto+2019b}
{Hashimoto}, T., {Inoue}, A.~K., {Tamura}, Y., {et~al.} 2019{\natexlab{b}},
  \pasj, 71, 109

\bibitem[{{Helou} {et~al.}(1988){Helou}, {Khan}, {Malek}, \&
  {Boehmer}}]{Helou+1988}
{Helou}, G., {Khan}, I.~R., {Malek}, L., \& {Boehmer}, L. 1988, \apjs, 68, 151

\bibitem[{{Herrera-Camus} {et~al.}(2016){Herrera-Camus}, {Bolatto}, {Smith},
  {Draine}, {Pellegrini}, {Wolfire}, {Croxall}, {de Looze}, {Calzetti},
  {Kennicutt}, {Crocker}, {Armus}, {van der Werf}, {Sand strom}, {Galametz},
  {Brandl}, {Groves}, {Rigopoulou}, {Walter}, {Leroy}, {Boquien}, {Tabatabaei},
  \& {Beirao}}]{Herrera-Camus+2016}
{Herrera-Camus}, R., {Bolatto}, A., {Smith}, J.~D., {et~al.} 2016, \apj, 826,
  175

\bibitem[{{Herrera-Camus} {et~al.}(2018{\natexlab{a}}){Herrera-Camus}, {Sturm},
  {Graci{\'a}-Carpio}, {Lutz}, {Contursi}, {Veilleux}, {Fischer},
  {Gonz{\'a}lez-Alfonso}, {Poglitsch}, {Tacconi}, {Genzel}, {Maiolino},
  {Sternberg}, {Davies}, \& {Verma}}]{Herrera-Camus+2018}
{Herrera-Camus}, R., {Sturm}, E., {Graci{\'a}-Carpio}, J., {et~al.}
  2018{\natexlab{a}}, \apj, 861, 94

\bibitem[{{Herrera-Camus} {et~al.}(2018{\natexlab{b}}){Herrera-Camus}, {Sturm},
  {Graci{\'a}-Carpio}, {Lutz}, {Contursi}, {Veilleux}, {Fischer},
  {Gonz{\'a}lez-Alfonso}, {Poglitsch}, {Tacconi}, {Genzel}, {Maiolino},
  {Sternberg}, {Davies}, \& {Verma}}]{Herrera-Camus+2018b}
{Herrera-Camus}, R., {Sturm}, E., {Graci{\'a}-Carpio}, J., {et~al.}
  2018{\natexlab{b}}, \apj, 861, 95

\bibitem[{{Herrera-Camus} {et~al.}(2020){Herrera-Camus}, {Sturm},
  {Graci{\'a}-Carpio}, {Veilleux}, {Shimizu}, {Lutz}, {Stone},
  {Gonz{\'a}lez-Alfonso}, {Davies}, {Fischer}, {Genzel}, {Maiolino},
  {Sternberg}, {Tacconi}, \& {Verma}}]{Herrera-Camus+2020}
{Herrera-Camus}, R., {Sturm}, E., {Graci{\'a}-Carpio}, J., {et~al.} 2020, \aap,
  633, L4

\bibitem[{{Hinshaw} {et~al.}(2013){Hinshaw}, {Larson}, {Komatsu}, {Spergel},
  {Bennett}, {Dunkley}, {Nolta}, {Halpern}, {Hill}, {Odegard}, {Page}, {Smith},
  {Weiland}, {Gold}, {Jarosik}, {Kogut}, {Limon}, {Meyer}, {Tucker}, {Wollack},
  \& {Wright}}]{Hinshaw+2013}
{Hinshaw}, G., {Larson}, D., {Komatsu}, E., {et~al.} 2013, The Astrophysical
  Journal Supplement Series, 208, 19

\bibitem[{{Hollenbach} {et~al.}(2009){Hollenbach}, {Kaufman}, {Bergin}, \&
  {Melnick}}]{Hollenbach+2009}
{Hollenbach}, D., {Kaufman}, M.~J., {Bergin}, E.~A., \& {Melnick}, G.~J. 2009,
  \apj, 690, 1497

\bibitem[{{Hollenbach} \& {McKee}(1989)}]{Hollenbach+1989}
{Hollenbach}, D. \& {McKee}, C.~F. 1989, \apj, 342, 306

\bibitem[{{Hollenbach} \& {Tielens}(1997)}]{Hollenbach+1997}
{Hollenbach}, D.~J. \& {Tielens}, A.~G.~G.~M. 1997, \araa, 35, 179

\bibitem[{{Hollenbach} \& {Tielens}(1999)}]{Hollenbach+1999}
{Hollenbach}, D.~J. \& {Tielens}, A.~G.~G.~M. 1999, Reviews of Modern Physics,
  71, 173

\bibitem[{{Hunter}(2007)}]{Matplotlib}
{Hunter}, J.~D. 2007, Computing in Science Engineering, 9, 90

\bibitem[{{Ikeda} {et~al.}(2002){Ikeda}, {Oka}, {Tatematsu}, {Sekimoto}, \&
  {Yamamoto}}]{Ikeda+2002}
{Ikeda}, M., {Oka}, T., {Tatematsu}, K., {Sekimoto}, Y., \& {Yamamoto}, S.
  2002, \apjs, 139, 467

\bibitem[{{Indriolo} {et~al.}(2007){Indriolo}, {Geballe}, {Oka}, \&
  {McCall}}]{Indriolo+2007}
{Indriolo}, N., {Geballe}, T.~R., {Oka}, T., \& {McCall}, B.~J. 2007, \apj,
  671, 1736

\bibitem[{{Israel} {et~al.}(2015){Israel}, {Rosenberg}, \& {van der
  Werf}}]{Israel+2015}
{Israel}, F.~P., {Rosenberg}, M.~J.~F., \& {van der Werf}, P. 2015, \aap, 578,
  A95

\bibitem[{{Ivison} {et~al.}(2011){Ivison}, {Papadopoulos}, {Smail}, {Greve},
  {Thomson}, {Xilouris}, \& {Chapman}}]{Ivison+2011}
{Ivison}, R.~J., {Papadopoulos}, P.~P., {Smail}, I., {et~al.} 2011, \mnras,
  412, 1913

\bibitem[{{Jiang} {et~al.}(2007){Jiang}, {Fan}, {Vestergaard}, {Kurk},
  {Walter}, {Kelly}, \& {Strauss}}]{Jiang+2007}
{Jiang}, L., {Fan}, X., {Vestergaard}, M., {et~al.} 2007, \aj, 134, 1150

\bibitem[{{Jiang} {et~al.}(2016){Jiang}, {McGreer}, {Fan}, {Strauss},
  {Ba{\~n}ados}, {Becker}, {Bian}, {Farnsworth}, {Shen}, {Wang}, {Wang},
  {Wang}, {White}, {Wu}, {Wu}, {Yang}, \& {Yang}}]{Jiang+2016}
{Jiang}, L., {McGreer}, I.~D., {Fan}, X., {et~al.} 2016, \apj, 833, 222

\bibitem[{{Jiao} {et~al.}(2017){Jiao}, {Zhao}, {Zhu}, {Lu}, {Gao}, \&
  {Zhang}}]{Jiao+2017}
{Jiao}, Q., {Zhao}, Y., {Zhu}, M., {et~al.} 2017, \apjl, 840, L18

\bibitem[{{Kaufman} {et~al.}(2006){Kaufman}, {Wolfire}, \&
  {Hollenbach}}]{Kaufman+2006}
{Kaufman}, M.~J., {Wolfire}, M.~G., \& {Hollenbach}, D.~J. 2006, \apj, 644, 283

\bibitem[{{Kaufman} {et~al.}(1999){Kaufman}, {Wolfire}, {Hollenbach}, \&
  {Luhman}}]{Kaufman+1999}
{Kaufman}, M.~J., {Wolfire}, M.~G., {Hollenbach}, D.~J., \& {Luhman}, M.~L.
  1999, \apj, 527, 795

\bibitem[{{Kegel} {et~al.}(1999){Kegel}, {Hertenstein}, \&
  {Quirrenbach}}]{Kegel+1999}
{Kegel}, W.~H., {Hertenstein}, T., \& {Quirrenbach}, A. 1999, \aap, 351, 472

\bibitem[{{Kirkpatrick} {et~al.}(2015){Kirkpatrick}, {Pope}, {Sajina},
  {Roebuck}, {Yan}, {Armus}, {D{\'\i}az-Santos}, \&
  {Stierwalt}}]{Kirkpatrick+2015}
{Kirkpatrick}, A., {Pope}, A., {Sajina}, A., {et~al.} 2015, \apj, 814, 9

\bibitem[{{Korista} {et~al.}(1997){Korista}, {Baldwin}, {Ferland}, \&
  {Verner}}]{Korista+1997}
{Korista}, K., {Baldwin}, J., {Ferland}, G., \& {Verner}, D. 1997, \apjs, 108,
  401

\bibitem[{{Kov{\'a}cs} {et~al.}(2006){Kov{\'a}cs}, {Chapman}, {Dowell},
  {Blain}, {Ivison}, {Smail}, \& {Phillips}}]{Kovacs+2006}
{Kov{\'a}cs}, A., {Chapman}, S.~C., {Dowell}, C.~D., {et~al.} 2006, \apj, 650,
  592

\bibitem[{{Krolik} \& {Begelman}(1988)}]{Krolik+1988}
{Krolik}, J.~H. \& {Begelman}, M.~C. 1988, \apj, 329, 702

\bibitem[{{Krolik} \& {Kallman}(1983)}]{Krolik+1983}
{Krolik}, J.~H. \& {Kallman}, T.~R. 1983, \apj, 267, 610

\bibitem[{{Kroupa}(2001)}]{Kroupa2001}
{Kroupa}, P. 2001, \mnras, 322, 231

\bibitem[{{Langer}(1978)}]{Langer+1978}
{Langer}, W.~D. 1978, \apj, 225, 860

\bibitem[{{Leipski} {et~al.}(2013){Leipski}, {Meisenheimer}, {Walter}, {Besel},
  {Dannerbauer}, {Fan}, {Haas}, {Klaas}, {Krause}, \& {Rix}}]{Leipski+2013}
{Leipski}, C., {Meisenheimer}, K., {Walter}, F., {et~al.} 2013, \apj, 772, 103

\bibitem[{{Leipski} {et~al.}(2014){Leipski}, {Meisenheimer}, {Walter}, {Klaas},
  {Dannerbauer}, {De Rosa}, {Fan}, {Haas}, {Krause}, \& {Rix}}]{Leipski+2014}
{Leipski}, C., {Meisenheimer}, K., {Walter}, F., {et~al.} 2014, \apj, 785, 154

\bibitem[{{Lepp} \& {Dalgarno}(1996)}]{Lepp+1996}
{Lepp}, S. \& {Dalgarno}, A. 1996, \aap, 306, L21

\bibitem[{{Lepp} \& {McCray}(1983)}]{Lepp+1983}
{Lepp}, S. \& {McCray}, R. 1983, \apj, 269, 560

\bibitem[{{Li} {et~al.}(2020{\natexlab{a}}){Li}, {Wang}, {Cox}, {Gao},
  {Walter}, {Wagg}, {Menten}, {Bertoldi}, {Shao}, {Venemans}, {Decarli},
  {Riechers}, {Neri}, {Fan}, {Omont}, \& {Narayanan}}]{Li+2020b}
{Li}, J., {Wang}, R., {Cox}, P., {et~al.} 2020{\natexlab{a}}, \apj, 900, 131

\bibitem[{{Li} {et~al.}(2020{\natexlab{b}}){Li}, {Wang}, {Riechers}, {Walter},
  {Decarli}, {Venamans}, {Neri}, {Shao}, {Fan}, {Gao}, {Carilli}, {Omont},
  {Cox}, {Menten}, {Wagg}, {Bertoldi}, \& {Narayanan}}]{Li+2020}
{Li}, J., {Wang}, R., {Riechers}, D., {et~al.} 2020{\natexlab{b}}, \apj, 889,
  162

\bibitem[{{Liang} {et~al.}(2019){Liang}, {Feldmann}, {Kere{\v{s}}}, {Scoville},
  {Hayward}, {Faucher-Gigu{\`e}re}, {Schreiber}, {Ma}, {Hopkins}, \&
  {Quataert}}]{Liang+2019}
{Liang}, L., {Feldmann}, R., {Kere{\v{s}}}, D., {et~al.} 2019, \mnras, 489,
  1397

\bibitem[{{Lis} {et~al.}(2011){Lis}, {Neufeld}, {Phillips}, {Gerin}, \&
  {Neri}}]{Lis+2011}
{Lis}, D.~C., {Neufeld}, D.~A., {Phillips}, T.~G., {Gerin}, M., \& {Neri}, R.
  2011, \apjl, 738, L6

\bibitem[{{Liu} {et~al.}(2017){Liu}, {Wei{\ss}}, {Perez-Beaupuits},
  {G{\"u}sten}, {Liu}, {Gao}, {Menten}, {van der Werf}, {Israel}, {Harris},
  {Martin-Pintado}, {Requena-Torres}, \& {Stutzki}}]{Liu+2017}
{Liu}, L., {Wei{\ss}}, A., {Perez-Beaupuits}, J.~P., {et~al.} 2017, \apj, 846,
  5

\bibitem[{{Lord} {et~al.}(1996){Lord}, {Hollenbach}, {Haas}, {Rubin}, {Colgan},
  \& {Erickson}}]{Lord+1996}
{Lord}, S.~D., {Hollenbach}, D.~J., {Haas}, M.~R., {et~al.} 1996, \apj, 465,
  703

\bibitem[{{Luhman} {et~al.}(2003){Luhman}, {Satyapal}, {Fischer}, {Wolfire},
  {Sturm}, {Dudley}, {Lutz}, \& {Genzel}}]{Luhman+2003}
{Luhman}, M.~L., {Satyapal}, S., {Fischer}, J., {et~al.} 2003, \apj, 594, 758

\bibitem[{{Lupu} {et~al.}(2012){Lupu}, {Scott}, {Aguirre}, {Aretxaga}, {Auld},
  {Barton}, {Beelen}, {Bertoldi}, {Bock}, {Bonfield}, {Bradford},
  {Buttiglione}, {Cava}, {Clements}, {Cooke}, {Cooray}, {Dannerbauer},
  {Dariush}, {De Zotti}, {Dunne}, {Dye}, {Eales}, {Frayer}, {Fritz}, {Glenn},
  {Hughes}, {Ibar}, {Ivison}, {Jarvis}, {Kamenetzky}, {Kim}, {Lagache},
  {Leeuw}, {Maddox}, {Maloney}, {Matsuhara}, {Murphy}, {Naylor}, {Negrello},
  {Nguyen}, {Omont}, {Pascale}, {Pohlen}, {Rigby}, {Rodighiero}, {Serjeant},
  {Smith}, {Temi}, {Thompson}, {Valtchanov}, {Verma}, {Vieira}, \&
  {Zmuidzinas}}]{Lupu+2012}
{Lupu}, R.~E., {Scott}, K.~S., {Aguirre}, J.~E., {et~al.} 2012, \apj, 757, 135

\bibitem[{{Ma} {et~al.}(2019){Ma}, {Hayward}, {Casey}, {Hopkins}, {Quataert},
  {Liang}, {Faucher-Gigu{\`e}re}, {Feldmann}, \& {Kere{\v{s}}}}]{Ma+2019}
{Ma}, X., {Hayward}, C.~C., {Casey}, C.~M., {et~al.} 2019, \mnras, 487, 1844

\bibitem[{{Maiolino} {et~al.}(2015){Maiolino}, {Carniani}, {Fontana},
  {Vallini}, {Pentericci}, {Ferrara}, {Vanzella}, {Grazian}, {Gallerani},
  {Castellano}, {Cristiani}, {Brammer}, {Santini}, {Wagg}, \&
  {Williams}}]{Maiolino+2015}
{Maiolino}, R., {Carniani}, S., {Fontana}, A., {et~al.} 2015, \mnras, 452, 54

\bibitem[{{Maiolino} {et~al.}(2009){Maiolino}, {Caselli}, {Nagao}, {Walmsley},
  {De Breuck}, \& {Meneghetti}}]{Maiolino+2009}
{Maiolino}, R., {Caselli}, P., {Nagao}, T., {et~al.} 2009, \aap, 500, L1

\bibitem[{{Maiolino} {et~al.}(2005){Maiolino}, {Cox}, {Caselli}, {Beelen},
  {Bertoldi}, {Carilli}, {Kaufman}, {Menten}, {Nagao}, {Omont}, {Wei{\ss}},
  {Walmsley}, \& {Walter}}]{Maiolino+2005}
{Maiolino}, R., {Cox}, P., {Caselli}, P., {et~al.} 2005, \aap, 440, L51

\bibitem[{{Malhotra} {et~al.}(1997){Malhotra}, {Helou}, {Stacey}, {Hollenbach},
  {Lord}, {Beichman}, {Dinerstein}, {Hunter}, {Lo}, {Lu}, {Rubin},
  {Silbermann}, {Thronson}, \& {Werner}}]{Malhotra+1997}
{Malhotra}, S., {Helou}, G., {Stacey}, G., {et~al.} 1997, \apjl, 491, L27

\bibitem[{{Maloney} {et~al.}(1996){Maloney}, {Hollenbach}, \&
  {Tielens}}]{Maloney+1996}
{Maloney}, P.~R., {Hollenbach}, D.~J., \& {Tielens}, A.~G.~G.~M. 1996, \apj,
  466, 561

\bibitem[{{Mashian} {et~al.}(2015){Mashian}, {Sturm}, {Sternberg}, {Janssen},
  {Hailey-Dunsheath}, {Fischer}, {Contursi}, {Gonz{\'a}lez-Alfonso},
  {Graci{\'a}-Carpio}, {Poglitsch}, {Veilleux}, {Davies}, {Genzel}, {Lutz},
  {Tacconi}, {Verma}, {Wei{\ss}}, {Polisensky}, \& {Nikola}}]{Mashian+2015}
{Mashian}, N., {Sturm}, E., {Sternberg}, A., {et~al.} 2015, \apj, 802, 81

\bibitem[{{Mathis}(1997)}]{Mathis+1997}
{Mathis}, J.~S. 1997, Astronomical Society of the Pacific Conference Series,
  Vol. 122, {Composition and Size of Interstellar Dust}, ed. Y.~J. {Pendleton},
  87

\bibitem[{{Matsuoka} {et~al.}(2018){Matsuoka}, {Strauss}, {Kashikawa}, {Onoue},
  {Iwasawa}, {Tang}, {Lee}, {Imanishi}, {Nagao}, {Akiyama}, {Asami}, {Bosch},
  {Furusawa}, {Goto}, {Gunn}, {Harikane}, {Ikeda}, {Izumi}, {Kawaguchi},
  {Kato}, {Kikuta}, {Kohno}, {Komiyama}, {Lupton}, {Minezaki}, {Miyazaki},
  {Murayama}, {Niida}, {Nishizawa}, {Noboriguchi}, {Oguri}, {Ono}, {Ouchi},
  {Price}, {Sameshima}, {Schulze}, {Shirakata}, {Silverman}, {Sugiyama},
  {Tait}, {Takada}, {Takata}, {Tanaka}, {Toba}, {Utsumi}, {Wang}, \&
  {Yamashita}}]{Matsuoka+2018}
{Matsuoka}, Y., {Strauss}, M.~A., {Kashikawa}, N., {et~al.} 2018, \apj, 869,
  150

\bibitem[{{Mazzucchelli} {et~al.}(2017{\natexlab{a}}){Mazzucchelli},
  {Ba{\~n}ados}, {Decarli}, {Farina}, {Venemans}, {Walter}, \&
  {Overzier}}]{Mazzucchelli+2017b}
{Mazzucchelli}, C., {Ba{\~n}ados}, E., {Decarli}, R., {et~al.}
  2017{\natexlab{a}}, \apj, 834, 83

\bibitem[{{Mazzucchelli} {et~al.}(2017{\natexlab{b}}){Mazzucchelli},
  {Ba{\~n}ados}, {Venemans}, {Decarli}, {Farina}, {Walter}, {Eilers}, {Rix},
  {Simcoe}, {Stern}, {Fan}, {Schlafly}, {De Rosa}, {Hennawi}, {Chambers},
  {Greiner}, {Burgett}, {Draper}, {Kaiser}, {Kudritzki}, {Magnier}, {Metcalfe},
  {Waters}, \& {Wainscoat}}]{Mazzucchelli+2017}
{Mazzucchelli}, C., {Ba{\~n}ados}, E., {Venemans}, B.~P., {et~al.}
  2017{\natexlab{b}}, \apj, 849, 91

\bibitem[{{Mazzucchelli} {et~al.}(2019){Mazzucchelli}, {Decarli}, {Farina},
  {Ba{\~n}ados}, {Venemans}, {Strauss}, {Walter}, {Neeleman}, {Bertoldi},
  {Fan}, {Riechers}, {Rix}, \& {Wang}}]{Mazzucchelli+2019}
{Mazzucchelli}, C., {Decarli}, R., {Farina}, E.~P., {et~al.} 2019, \apj, 881,
  163

\bibitem[{{McKee} \& {Ostriker}(1977)}]{McKee+1977}
{McKee}, C.~F. \& {Ostriker}, J.~P. 1977, \apj, 218, 148

\bibitem[{{McMullin} {et~al.}(2007){McMullin}, {Waters}, {Schiebel}, {Young},
  \& {Golap}}]{McMullin+2007}
{McMullin}, J.~P., {Waters}, B., {Schiebel}, D., {Young}, W., \& {Golap}, K.
  2007, in Astronomical Data Analysis Software and Systems XVI, Vol. 376, 127

\bibitem[{{Meijerink} {et~al.}(2013){Meijerink}, {Kristensen}, {Wei{\ss}}, {van
  der Werf}, {Walter}, {Spaans}, {Loenen}, {Fischer}, {Israel}, {Isaak},
  {Papadopoulos}, {Aalto}, {Armus}, {Charmand aris}, {Dasyra}, {Diaz-Santos},
  {Evans}, {Gao}, {Gonz{\'a}lez-Alfonso}, {G{\"u}sten}, {Henkel}, {Kramer},
  {Lord}, {Mart{\'\i}n-Pintado}, {Naylor}, {Sanders}, {Smith}, {Spinoglio},
  {Stacey}, {Veilleux}, \& {Wiedner}}]{Meijerink+2013}
{Meijerink}, R., {Kristensen}, L.~E., {Wei{\ss}}, A., {et~al.} 2013, \apjl,
  762, L16

\bibitem[{{Meijerink} \& {Spaans}(2005)}]{Meijerink+2005}
{Meijerink}, R. \& {Spaans}, M. 2005, \aap, 436, 397

\bibitem[{{Meijerink} {et~al.}(2007){Meijerink}, {Spaans}, \&
  {Israel}}]{Meijerink+2007}
{Meijerink}, R., {Spaans}, M., \& {Israel}, F.~P. 2007, \aap, 461, 793

\bibitem[{{Meijerink} {et~al.}(2011){Meijerink}, {Spaans}, {Loenen}, \& {van
  der Werf}}]{Meijerink+2011}
{Meijerink}, R., {Spaans}, M., {Loenen}, A.~F., \& {van der Werf}, P.~P. 2011,
  \aap, 525, A119

\bibitem[{{Meixner} \& {Tielens}(1993)}]{Meixner+1993}
{Meixner}, M. \& {Tielens}, A.~G.~G.~M. 1993, \apj, 405, 216

\bibitem[{{Menten} {et~al.}(2008){Menten}, {Lundgren}, {Belloche}, {Thorwirth},
  \& {Reid}}]{Menten+2008}
{Menten}, K.~M., {Lundgren}, A., {Belloche}, A., {Thorwirth}, S., \& {Reid},
  M.~J. 2008, \aap, 477, 185

\bibitem[{{Mingozzi} {et~al.}(2018){Mingozzi}, {Vallini}, {Pozzi}, {Vignali},
  {Mignano}, {Gruppioni}, {Talia}, {Cimatti}, {Cresci}, \&
  {Massardi}}]{Mingozzi+2018}
{Mingozzi}, M., {Vallini}, L., {Pozzi}, F., {et~al.} 2018, \mnras, 474, 3640

\bibitem[{{Morselli} {et~al.}(2014){Morselli}, {Mignoli}, {Gilli}, {Vignali},
  {Comastri}, {Sani}, {Cappelluti}, {Zamorani}, {Brusa}, {Gallozzi}, \&
  {Vanzella}}]{Morselli+2014}
{Morselli}, L., {Mignoli}, M., {Gilli}, R., {et~al.} 2014, \aap, 568, A1

\bibitem[{{Murphy} {et~al.}(2011){Murphy}, {Condon}, {Schinnerer}, {Kennicutt},
  {Calzetti}, {Armus}, {Helou}, {Turner}, {Aniano}, {Beir{\~a}o}, {Bolatto},
  {Brandl}, {Croxall}, {Dale}, {Donovan Meyer}, {Draine}, {Engelbracht},
  {Hunt}, {Hao}, {Koda}, {Roussel}, {Skibba}, \& {Smith}}]{Murphy+2011}
{Murphy}, E.~J., {Condon}, J.~J., {Schinnerer}, E., {et~al.} 2011, \apj, 737,
  67

\bibitem[{{Narayanan} \& {Krumholz}(2014)}]{Narayanan+2014}
{Narayanan}, D. \& {Krumholz}, M.~R. 2014, \mnras, 442, 1411

\bibitem[{{Neeleman} {et~al.}(2019){Neeleman}, {Ba{\~n}ados}, {Walter},
  {Decarli}, {Venemans}, {Carilli}, {Fan}, {Farina}, {Mazzucchelli}, {Novak},
  {Riechers}, {Rix}, \& {Wang}}]{Neeleman+2019}
{Neeleman}, M., {Ba{\~n}ados}, E., {Walter}, F., {et~al.} 2019, \apj, 882, 10

\bibitem[{{Neufeld} \& {Kaufman}(1993)}]{Neufeld+1993}
{Neufeld}, D.~A. \& {Kaufman}, M.~J. 1993, \apj, 418, 263

\bibitem[{{Neufeld} {et~al.}(2002){Neufeld}, {Kaufman}, {Goldsmith},
  {Hollenbach}, \& {Plume}}]{Neufeld+2002}
{Neufeld}, D.~A., {Kaufman}, M.~J., {Goldsmith}, P.~F., {Hollenbach}, D.~J., \&
  {Plume}, R. 2002, \apj, 580, 278

\bibitem[{{Neufeld} {et~al.}(1995){Neufeld}, {Lepp}, \&
  {Melnick}}]{Neufeld+1995}
{Neufeld}, D.~A., {Lepp}, S., \& {Melnick}, G.~J. 1995, \apjs, 100, 132

\bibitem[{{Noeske} {et~al.}(2007){Noeske}, {Weiner}, {Faber}, {Papovich},
  {Koo}, {Somerville}, {Bundy}, {Conselice}, {Newman}, {Schiminovich}, {Le
  Floc'h}, {Coil}, {Rieke}, {Lotz}, {Primack}, {Barmby}, {Cooper}, {Davis},
  {Ellis}, {Fazio}, {Guhathakurta}, {Huang}, {Kassin}, {Martin}, {Phillips},
  {Rich}, {Small}, {Willmer}, \& {Wilson}}]{Noeske+2007}
{Noeske}, K.~G., {Weiner}, B.~J., {Faber}, S.~M., {et~al.} 2007, \apjl, 660,
  L43

\bibitem[{{Novak} {et~al.}(2019){Novak}, {Ba{\~n}ados}, {Decarli}, {Walter},
  {Venemans}, {Neeleman}, {Farina}, {Mazzucchelli}, {Carilli}, {Fan}, {Rix}, \&
  {Wang}}]{Novak+2019}
{Novak}, M., {Ba{\~n}ados}, E., {Decarli}, R., {et~al.} 2019, \apj, 881, 63

\bibitem[{{Oberst} {et~al.}(2006){Oberst}, {Parshley}, {Stacey}, {Nikola},
  {L{\"o}hr}, {Harnett}, {Tothill}, {Lane}, {Stark}, \& {Tucker}}]{Oberst+2006}
{Oberst}, T.~E., {Parshley}, S.~C., {Stacey}, G.~J., {et~al.} 2006, \apjl, 652,
  L125

\bibitem[{{Obreschkow} {et~al.}(2009){Obreschkow}, {Heywood}, {Kl{\"o}ckner},
  \& {Rawlings}}]{Obreschkow+2009}
{Obreschkow}, D., {Heywood}, I., {Kl{\"o}ckner}, H.~R., \& {Rawlings}, S. 2009,
  \apj, 702, 1321

\bibitem[{{Omont} {et~al.}(2011){Omont}, {Neri}, {Cox}, {Lupu}, {Gu{\'e}lin},
  {van der Werf}, {Wei{\ss}}, {Ivison}, {Negrello}, {Leeuw}, {Lehnert},
  {Smail}, {Verma}, {Baker}, {Beelen}, {Aguirre}, {Baes}, {Bertoldi},
  {Clements}, {Cooray}, {Coppin}, {Dannerbauer}, {de Zotti}, {Dye}, {Fiolet},
  {Frayer}, {Gavazzi}, {Hughes}, {Jarvis}, {Krips}, {Micha{\l}owski}, {Murphy},
  {Riechers}, {Serjeant}, {Swinbank}, {Temi}, {Vaccari}, {Vieira}, {Auld},
  {Buttiglione}, {Cava}, {Dariush}, {Dunne}, {Eales}, {Fritz}, {Gomez}, {Ibar},
  {Maddox}, {Pascale}, {Pohlen}, {Rigby}, {Smith}, {Bock}, {Bradford}, {Glenn},
  {Scott}, \& {Zmuidzinas}}]{Omont+2011}
{Omont}, A., {Neri}, R., {Cox}, P., {et~al.} 2011, \aap, 530, L3

\bibitem[{{Omont} {et~al.}(2013){Omont}, {Yang}, {Cox}, {Neri}, {Beelen},
  {Bussmann}, {Gavazzi}, {van der Werf}, {Riechers}, {Downes}, {Krips}, {Dye},
  {Ivison}, {Vieira}, {Wei{\ss}}, {Aguirre}, {Baes}, {Baker}, {Bertoldi},
  {Cooray}, {Dannerbauer}, {De Zotti}, {Eales}, {Fu}, {Gao}, {Gu{\'e}lin},
  {Harris}, {Jarvis}, {Lehnert}, {Leeuw}, {Lupu}, {Menten}, {Micha{\l}owski},
  {Negrello}, {Serjeant}, {Temi}, {Auld}, {Dariush}, {Dunne}, {Fritz},
  {Hopwood}, {Hoyos}, {Ibar}, {Maddox}, {Smith}, {Valiante}, {Bock},
  {Bradford}, {Glenn}, \& {Scott}}]{Omont+2013}
{Omont}, A., {Yang}, C., {Cox}, P., {et~al.} 2013, \aap, 551, A115

\bibitem[{{Onoue} {et~al.}(2020){Onoue}, {Ba{\~n}ados}, {Mazzucchelli},
  {Venemans}, {Schindler}, {Walter}, {Hennawi}, {Andika}, {Davies}, {Decarli},
  {Farina}, {Jahnke}, {Nagao}, {Tominaga}, \& {Wang}}]{Onoue+2020}
{Onoue}, M., {Ba{\~n}ados}, E., {Mazzucchelli}, C., {et~al.} 2020, \apj, 898,
  105

\bibitem[{{Overzier} {et~al.}(2009){Overzier}, {Guo}, {Kauffmann}, {De Lucia},
  {Bouwens}, \& {Lemson}}]{Overzier+2009}
{Overzier}, R.~A., {Guo}, Q., {Kauffmann}, G., {et~al.} 2009, \mnras, 394, 577

\bibitem[{{Papadopoulos} {et~al.}(2004){Papadopoulos}, {Thi}, \&
  {Viti}}]{Papadopoulos+2004}
{Papadopoulos}, P.~P., {Thi}, W.~F., \& {Viti}, S. 2004, \mnras, 351, 147

\bibitem[{{Pavesi} {et~al.}(2016){Pavesi}, {Riechers}, {Capak}, {Carilli},
  {Sharon}, {Stacey}, {Karim}, {Scoville}, \&
  {Smol{\v{c}}i{\'c}}}]{Pavesi+2016}
{Pavesi}, R., {Riechers}, D.~A., {Capak}, P.~L., {et~al.} 2016, \apj, 832, 151

\bibitem[{{Pentericci} {et~al.}(2002){Pentericci}, {Fan}, {Rix}, {Strauss},
  {Narayanan}, {Richards}, {Schneider}, {Krolik}, {Heckman}, {Brinkmann},
  {Lamb}, \& {Szokoly}}]{Pentericci+2002}
{Pentericci}, L., {Fan}, X., {Rix}, H.-W., {et~al.} 2002, \aj, 123, 2151

\bibitem[{{Pereira-Santaella} {et~al.}(2013){Pereira-Santaella}, {Spinoglio},
  {Busquet}, {Wilson}, {Glenn}, {Isaak}, {Kamenetzky}, {Rangwala}, {Schirm},
  {Baes}, {Barlow}, {Boselli}, {Cooray}, \& {Cormier}}]{Pereira-Santaella+2013}
{Pereira-Santaella}, M., {Spinoglio}, L., {Busquet}, G., {et~al.} 2013, \apj,
  768, 55

\bibitem[{{Poelman} \& {Spaans}(2005)}]{Poelman+2005}
{Poelman}, D.~R. \& {Spaans}, M. 2005, \aap, 440, 559

\bibitem[{{Pound} \& {Wolfire}(2008)}]{Pound+2008}
{Pound}, M.~W. \& {Wolfire}, M.~G. 2008, Astronomical Society of the Pacific
  Conference Series, Vol. 394, {The Photo Dissociation Region Toolbox}, ed.
  R.~W. {Argyle}, P.~S. {Bunclark}, \& J.~R. {Lewis}, 654

\bibitem[{{Priddey} \& {McMahon}(2001)}]{Priddey+2001}
{Priddey}, R.~S. \& {McMahon}, R.~G. 2001, \mnras, 324, L17

\bibitem[{{Rangwala} {et~al.}(2011){Rangwala}, {Maloney}, {Glenn}, {Wilson},
  {Rykala}, {Isaak}, {Baes}, {Bendo}, {Boselli}, {Bradford}, {Clements},
  {Cooray}, {Fulton}, {Imhof}, {Kamenetzky}, {Madden}, {Mentuch}, {Sacchi},
  {Sauvage}, {Schirm}, {Smith}, {Spinoglio}, \& {Wolfire}}]{Rangwala+2011}
{Rangwala}, N., {Maloney}, P.~R., {Glenn}, J., {et~al.} 2011, \apj, 743, 94

\bibitem[{{Riechers} {et~al.}(2013){Riechers}, {Bradford}, {Clements},
  {Dowell}, {P{\'e}rez-Fournon}, {Ivison}, {Bridge}, {Conley}, {Fu}, {Vieira},
  {Wardlow}, {Calanog}, {Cooray}, {Hurley}, {Neri}, {Kamenetzky}, {Aguirre},
  {Altieri}, {Arumugam}, {Benford}, {B{\'e}thermin}, {Bock}, {Burgarella},
  {Cabrera-Lavers}, {Chapman}, {Cox}, {Dunlop}, {Earle}, {Farrah}, {Ferrero},
  {Franceschini}, {Gavazzi}, {Glenn}, {Solares}, {Gurwell}, {Halpern},
  {Hatziminaoglou}, {Hyde}, {Ibar}, {Kov{\'a}cs}, {Krips}, {Lupu}, {Maloney},
  {Martinez-Navajas}, {Matsuhara}, {Murphy}, {Naylor}, {Nguyen}, {Oliver},
  {Omont}, {Page}, {Petitpas}, {Rangwala}, {Roseboom}, {Scott}, {Smith},
  {Staguhn}, {Streblyanska}, {Thomson}, {Valtchanov}, {Viero}, {Wang},
  {Zemcov}, \& {Zmuidzinas}}]{Riechers+2013}
{Riechers}, D.~A., {Bradford}, C.~M., {Clements}, D.~L., {et~al.} 2013, \nat,
  496, 329

\bibitem[{{Riechers} {et~al.}(2014){Riechers}, {Carilli}, {Capak}, {Scoville},
  {Smol{\v{c}}i{\'c}}, {Schinnerer}, {Yun}, {Cox}, {Bertoldi}, {Karim}, \&
  {Yan}}]{Riechers+2014}
{Riechers}, D.~A., {Carilli}, C.~L., {Capak}, P.~L., {et~al.} 2014, \apj, 796,
  84

\bibitem[{{Riechers} {et~al.}(2011){Riechers}, {Carilli}, {Walter}, {Weiss},
  {Wagg}, {Bertoldi}, {Downes}, {Henkel}, \& {Hodge}}]{Riechers+2011}
{Riechers}, D.~A., {Carilli}, L.~C., {Walter}, F., {et~al.} 2011, \apjl, 733,
  L11

\bibitem[{{Riechers} {et~al.}(2020){Riechers}, {Hodge}, {Pavesi}, {Daddi},
  {Decarli}, {Ivison}, {Sharon}, {Smail}, {Walter}, {Aravena}, {Capak},
  {Carilli}, {Cox}, {Cunha}, {Dannerbauer}, {Dickinson}, {Neri}, \&
  {Wagg}}]{Riechers+2020}
{Riechers}, D.~A., {Hodge}, J.~A., {Pavesi}, R., {et~al.} 2020, \apj, 895, 81

\bibitem[{{Riechers} {et~al.}(2009){Riechers}, {Walter}, {Bertoldi}, {Carilli},
  {Aravena}, {Neri}, {Cox}, {Wei{\ss}}, \& {Menten}}]{Riechers+2009}
{Riechers}, D.~A., {Walter}, F., {Bertoldi}, F., {et~al.} 2009, \apj, 703, 1338

\bibitem[{{Rosenberg} {et~al.}(2015){Rosenberg}, {van der Werf}, {Aalto},
  {Armus}, {Charmandaris}, {D{\'\i}az-Santos}, {Evans}, {Fischer}, {Gao},
  {Gonz{\'a}lez-Alfonso}, {Greve}, {Harris}, {Henkel}, {Israel}, {Isaak},
  {Kramer}, {Meijerink}, {Naylor}, {Sanders}, {Smith}, {Spaans}, {Spinoglio},
  {Stacey}, {Veenendaal}, {Veilleux}, {Walter}, {Wei{\ss}}, {Wiedner}, {van der
  Wiel}, \& {Xilouris}}]{Rosenberg+2015}
{Rosenberg}, M.~J.~F., {van der Werf}, P.~P., {Aalto}, S., {et~al.} 2015, \apj,
  801, 72

\bibitem[{{Rowan-Robinson}(1995)}]{Rowan-Robinson+1995}
{Rowan-Robinson}, M. 1995, \mnras, 272, 737

\bibitem[{{Rybak} {et~al.}(2019){Rybak}, {Calistro Rivera}, {Hodge}, {Smail},
  {Walter}, {van der Werf}, {da Cunha}, {Chen}, {Dannerbauer}, {Ivison},
  {Karim}, {Simpson}, {Swinbank}, \& {Wardlow}}]{Rybak+2019}
{Rybak}, M., {Calistro Rivera}, G., {Hodge}, J.~A., {et~al.} 2019, \apj, 876,
  112

\bibitem[{{Saintonge} {et~al.}(2017){Saintonge}, {Catinella}, {Tacconi},
  {Kauffmann}, {Genzel}, {Cortese}, {Dav{\'e}}, {Fletcher},
  {Graci{\'a}-Carpio}, {Kramer}, {Heckman}, {Janowiecki}, {Lutz}, {Rosario},
  {Schiminovich}, {Schuster}, {Wang}, {Wuyts}, {Borthakur}, {Lamperti}, \&
  {Roberts-Borsani}}]{Saintonge+2017}
{Saintonge}, A., {Catinella}, B., {Tacconi}, L.~J., {et~al.} 2017, \apjs, 233,
  22

\bibitem[{{Saintonge} {et~al.}(2018){Saintonge}, {Wilson}, {Xiao}, {Lin},
  {Hwang}, {Tosaki}, {Bureau}, {Cigan}, {Clark}, {Clements}, {De Looze},
  {Dharmawardena}, {Gao}, {Gear}, {Greenslade}, {Lamperti}, {Lee}, {Li},
  {Micha{\l}owski}, {Mok}, {Pan}, {Sansom}, {Sargent}, {Smith}, {Williams},
  {Yang}, {Zhu}, {Accurso}, {Barmby}, {Brinks}, {Bourne}, {Brown}, {Chung},
  {Chung}, {Cibinel}, {Coppin}, {Davies}, {Davis}, {Eales}, {Fanciullo},
  {Fang}, {Gao}, {Glass}, {Gomez}, {Greve}, {He}, {Ho}, {Huang}, {Jeong},
  {Jiang}, {Jiao}, {Kemper}, {Kim}, {Kim}, {Kim}, {Ko}, {Kong}, {Lacaille},
  {Lacey}, {Lee}, {Lee}, {Lee}, {Masters}, {Oh}, {Papadopoulos}, {Park},
  {Park}, {Parsons}, {Rowland s}, {Scicluna}, {Scudder}, {Sethuram},
  {Serjeant}, {Shao}, {Sheen}, {Shi}, {Shim}, {Smith}, {Spekkens}, {Tsai},
  {Verma}, {Urquhart}, {Violino}, {Viti}, {Wake}, {Wang}, {Wouterloot}, {Yang},
  {Yim}, {Yuan}, \& {Zheng}}]{Saintonge+2018}
{Saintonge}, A., {Wilson}, C.~D., {Xiao}, T., {et~al.} 2018, \mnras, 481, 3497

\bibitem[{{Sanders} {et~al.}(2003){Sanders}, {Mazzarella}, {Kim}, {Surace}, \&
  {Soifer}}]{Sanders+2003}
{Sanders}, D.~B., {Mazzarella}, J.~M., {Kim}, D.~C., {Surace}, J.~A., \&
  {Soifer}, B.~T. 2003, \aj, 126, 1607

\bibitem[{{Sandstrom} {et~al.}(2013){Sandstrom}, {Leroy}, {Walter}, {Bolatto},
  {Croxall}, {Draine}, {Wilson}, {Wolfire}, {Calzetti}, {Kennicutt}, {Aniano},
  {Donovan Meyer}, {Usero}, {Bigiel}, {Brinks}, {de Blok}, {Crocker}, {Dale},
  {Engelbracht}, {Galametz}, {Groves}, {Hunt}, {Koda}, {Kreckel}, {Linz},
  {Meidt}, {Pellegrini}, {Rix}, {Roussel}, {Schinnerer}, {Schruba}, {Schuster},
  {Skibba}, {van der Laan}, {Appleton}, {Armus}, {Brandl}, {Gordon}, {Hinz},
  {Krause}, {Montiel}, {Sauvage}, {Schmiedeke}, {Smith}, \&
  {Vigroux}}]{Sandstrom+2013}
{Sandstrom}, K.~M., {Leroy}, A.~K., {Walter}, F., {et~al.} 2013, \apj, 777, 5

\bibitem[{{Savage} \& {Sembach}(1996)}]{Savage+1996}
{Savage}, B.~D. \& {Sembach}, K.~R. 1996, \araa, 34, 279

\bibitem[{{Schleicher} {et~al.}(2010){Schleicher}, {Spaans}, \&
  {Klessen}}]{Schleicher+2010}
{Schleicher}, D.~R.~G., {Spaans}, M., \& {Klessen}, R.~S. 2010, \aap, 513, A7

\bibitem[{{Schreiber} {et~al.}(2018){Schreiber}, {Elbaz}, {Pannella}, {Ciesla},
  {Wang}, \& {Franco}}]{Schreiber+2018}
{Schreiber}, C., {Elbaz}, D., {Pannella}, M., {et~al.} 2018, \aap, 609, A30

\bibitem[{{Shao} {et~al.}(2019){Shao}, {Wang}, {Carilli}, {Wagg}, {Walter},
  {Li}, {Fan}, {Jiang}, {Riechers}, {Bertoldi}, {Strauss}, {Cox}, {Omont}, \&
  {Menten}}]{Shao+2019}
{Shao}, Y., {Wang}, R., {Carilli}, C.~L., {et~al.} 2019, \apj, 876, 99

\bibitem[{{Sijacki} {et~al.}(2015){Sijacki}, {Vogelsberger}, {Genel},
  {Springel}, {Torrey}, {Snyder}, {Nelson}, \& {Hernquist}}]{Sijacki+2015}
{Sijacki}, D., {Vogelsberger}, M., {Genel}, S., {et~al.} 2015, \mnras, 452, 575

\bibitem[{{Smith} {et~al.}(2004){Smith}, {Gonzalez-Alfonso}, {Fischer},
  {Ashby}, {Dudley}, \& {Spinoglio}}]{Smith+2004}
{Smith}, H., {Gonzalez-Alfonso}, E., {Fischer}, J., {et~al.} 2004, in
  Astronomical Society of the Pacific Conference Series, Vol. 320, The Neutral
  ISM in Starburst Galaxies, ed. S.~{Aalto}, S.~{Huttemeister}, \& A.~{Pedlar},
  49

\bibitem[{{Solomon} {et~al.}(1992){Solomon}, {Radford}, \&
  {Downes}}]{Solomon+1992}
{Solomon}, P.~M., {Radford}, S.~J.~E., \& {Downes}, D. 1992, \nat, 356, 318

\bibitem[{{Spaans}(1996)}]{Spaans+1996}
{Spaans}, M. 1996, \aap, 307, 271

\bibitem[{{Spinoglio} {et~al.}(2005){Spinoglio}, {Malkan}, {Smith},
  {Gonz{\'a}lez-Alfonso}, \& {Fischer}}]{Spinoglio+2005}
{Spinoglio}, L., {Malkan}, M.~A., {Smith}, H.~A., {Gonz{\'a}lez-Alfonso}, E.,
  \& {Fischer}, J. 2005, \apj, 623, 123

\bibitem[{{Spinoglio} {et~al.}(2012){Spinoglio}, {Pereira-Santaella},
  {Busquet}, {Schirm}, {Wilson}, {Glenn}, {Kamenetzky}, {Rangwala}, {Maloney},
  {Parkin}, {Bendo}, {Madden}, {Wolfire}, {Boselli}, {Cooray}, \&
  {Page}}]{Spinoglio+2012}
{Spinoglio}, L., {Pereira-Santaella}, M., {Busquet}, G., {et~al.} 2012, \apj,
  758, 108

\bibitem[{{Spoon} {et~al.}(2013){Spoon}, {Farrah}, {Lebouteiller},
  {Gonz{\'a}lez-Alfonso}, {Bernard-Salas}, {Urrutia}, {Rigopoulou},
  {Westmoquette}, {Smith}, {Afonso}, {Pearson}, {Cormier}, {Efstathiou},
  {Borys}, {Verma}, {Etxaluze}, \& {Clements}}]{Spoon+2013}
{Spoon}, H.~W.~W., {Farrah}, D., {Lebouteiller}, V., {et~al.} 2013, \apj, 775,
  127

\bibitem[{{Stefan} {et~al.}(2015){Stefan}, {Carilli}, {Wagg}, {Walter},
  {Riechers}, {Bertoldi}, {Green}, {Fan}, {Menten}, \& {Wang}}]{Stefan+2015}
{Stefan}, I.~I., {Carilli}, C.~L., {Wagg}, J., {et~al.} 2015, \mnras, 451, 1713

\bibitem[{{Strandet} {et~al.}(2017){Strandet}, {Weiss}, {De Breuck}, {Marrone},
  {Vieira}, {Aravena}, {Ashby}, {B{\'e}thermin}, {Bothwell}, {Bradford},
  {Carlstrom}, {Chapman}, {Cunningham}, {Chen}, {Fassnacht}, {Gonzalez},
  {Greve}, {Gullberg}, {Hayward}, {Hezaveh}, {Litke}, {Ma}, {Malkan}, {Menten},
  {Miller}, {Murphy}, {Narayanan}, {Phadke}, {Rotermund}, {Spilker}, \&
  {Sreevani}}]{Strandet+2017}
{Strandet}, M.~L., {Weiss}, A., {De Breuck}, C., {et~al.} 2017, \apjl, 842, L15

\bibitem[{{Sturm} {et~al.}(2011){Sturm}, {Gonz{\'a}lez-Alfonso}, {Veilleux},
  {Fischer}, {Graci{\'a}-Carpio}, {Hailey-Dunsheath}, {Contursi}, {Poglitsch},
  {Sternberg}, {Davies}, {Genzel}, {Lutz}, {Tacconi}, {Verma}, {Maiolino}, \&
  {de Jong}}]{Sturm+2011}
{Sturm}, E., {Gonz{\'a}lez-Alfonso}, E., {Veilleux}, S., {et~al.} 2011, \apjl,
  733, L16

\bibitem[{{Tacconi} {et~al.}(2018){Tacconi}, {Genzel}, {Saintonge}, {Combes},
  {Garc{\'\i}a-Burillo}, {Neri}, {Bolatto}, {Contini}, {F{\"o}rster Schreiber},
  {Lilly}, {Lutz}, {Wuyts}, {Accurso}, {Boissier}, {Boone}, {Bouch{\'e}},
  {Bournaud}, {Burkert}, {Carollo}, {Cooper}, {Cox}, {Feruglio}, {Freundlich},
  {Herrera-Camus}, {Juneau}, {Lippa}, {Naab}, {Renzini}, {Salome}, {Sternberg},
  {Tadaki}, {{\"U}bler}, {Walter}, {Weiner}, \& {Weiss}}]{Tacconi+2018}
{Tacconi}, L.~J., {Genzel}, R., {Saintonge}, A., {et~al.} 2018, \apj, 853, 179

\bibitem[{{Tacconi} {et~al.}(2008){Tacconi}, {Genzel}, {Smail}, {Neri},
  {Chapman}, {Ivison}, {Blain}, {Cox}, {Omont}, {Bertoldi}, {Greve},
  {F{\"o}rster Schreiber}, {Genel}, {Lutz}, {Swinbank}, {Shapley}, {Erb},
  {Cimatti}, {Daddi}, \& {Baker}}]{Tacconi+2008}
{Tacconi}, L.~J., {Genzel}, R., {Smail}, I., {et~al.} 2008, \apj, 680, 246

\bibitem[{{Tacconi} {et~al.}(2020){Tacconi}, {Genzel}, \&
  {Sternberg}}]{Tacconi+2020}
{Tacconi}, L.~J., {Genzel}, R., \& {Sternberg}, A. 2020, arXiv e-prints,
  arXiv:2003.06245

\bibitem[{{Tielens} \& {Hollenbach}(1985{\natexlab{a}})}]{Tielens+1985a}
{Tielens}, A.~G.~G.~M. \& {Hollenbach}, D. 1985{\natexlab{a}}, \apj, 291, 722

\bibitem[{{Tielens} \& {Hollenbach}(1985{\natexlab{b}})}]{Tielens+1985b}
{Tielens}, A.~G.~G.~M. \& {Hollenbach}, D. 1985{\natexlab{b}}, \apj, 291, 747

\bibitem[{{Tomassetti} {et~al.}(2014){Tomassetti}, {Porciani}, {Romano-Diaz},
  {Ludlow}, \& {Papadopoulos}}]{Tomassetti+2014}
{Tomassetti}, M., {Porciani}, C., {Romano-Diaz}, E., {Ludlow}, A.~D., \&
  {Papadopoulos}, P.~P. 2014, \mnras, 445, L124

\bibitem[{{Trakhtenbrot} {et~al.}(2017){Trakhtenbrot}, {Lira}, {Netzer},
  {Cicone}, {Maiolino}, \& {Shemmer}}]{Trakhtenbrot+2017}
{Trakhtenbrot}, B., {Lira}, P., {Netzer}, H., {et~al.} 2017, \apj, 836, 8

\bibitem[{{Valentino} {et~al.}(2018){Valentino}, {Magdis}, {Daddi}, {Liu},
  {Aravena}, {Bournaud}, {Cibinel}, {Cormier}, {Dickinson}, {Gao}, {Jin},
  {Juneau}, {Kartaltepe}, {Lee}, {Madden}, {Puglisi}, {Sanders}, \&
  {Silverman}}]{Valentino+2018}
{Valentino}, F., {Magdis}, G.~E., {Daddi}, E., {et~al.} 2018, \apj, 869, 27

\bibitem[{{Valentino} {et~al.}(2020){Valentino}, {Magdis}, {Daddi}, {Liu},
  {Aravena}, {Bournaud}, {Cortzen}, {Gao}, {Jin}, {Juneau}, {Kartaltepe},
  {Kokorev}, {Lee}, {Madden}, {Narayanan}, {Popping}, \&
  {Puglisi}}]{Valentino+2020}
{Valentino}, F., {Magdis}, G.~E., {Daddi}, E., {et~al.} 2020, \apj, 890, 24

\bibitem[{{Vallini} {et~al.}(2018){Vallini}, {Pallottini}, {Ferrara},
  {Gallerani}, {Sobacchi}, \& {Behrens}}]{Vallini+2018}
{Vallini}, L., {Pallottini}, A., {Ferrara}, A., {et~al.} 2018, \mnras, 473, 271

\bibitem[{{Vallini} {et~al.}(2019){Vallini}, {Tielens}, {Pallottini},
  {Gallerani}, {Gruppioni}, {Carniani}, {Pozzi}, \& {Talia}}]{Vallini+2019}
{Vallini}, L., {Tielens}, A.~G.~G.~M., {Pallottini}, A., {et~al.} 2019, \mnras,
  490, 4502

\bibitem[{{van der Tak} {et~al.}(2007){van der Tak}, {Black}, {Sch{\"o}ier},
  {Jansen}, \& {van Dishoeck}}]{vanderTak+2007}
{van der Tak}, F.~F.~S., {Black}, J.~H., {Sch{\"o}ier}, F.~L., {Jansen}, D.~J.,
  \& {van Dishoeck}, E.~F. 2007, \aap, 468, 627

\bibitem[{{van der Vlugt} \& {Costa}(2019)}]{vanderVlugt+2019}
{van der Vlugt}, D. \& {Costa}, T. 2019, \mnras, 490, 4918

\bibitem[{{van der Werf} {et~al.}(2011){van der Werf}, {Berciano Alba},
  {Spaans}, {Loenen}, {Meijerink}, {Riechers}, {Cox}, {Wei{\ss}}, \&
  {Walter}}]{vanderWerf+2011}
{van der Werf}, P.~P., {Berciano Alba}, A., {Spaans}, M., {et~al.} 2011, \apjl,
  741, L38

\bibitem[{{van der Werf} {et~al.}(2010){van der Werf}, {Isaak}, {Meijerink},
  {Spaans}, {Rykala}, {Fulton}, {Loenen}, {Walter}, {Wei{\ss}}, {Armus},
  {Fischer}, {Israel}, {Harris}, {Veilleux}, {Henkel}, {Savini}, {Lord},
  {Smith}, {Gonz{\'a}lez-Alfonso}, {Naylor}, {Aalto}, {Charmand aris},
  {Dasyra}, {Evans}, {Gao}, {Greve}, {G{\"u}sten}, {Kramer},
  {Mart{\'\i}n-Pintado}, {Mazzarella}, {Papadopoulos}, {Sanders}, {Spinoglio},
  {Stacey}, {Vlahakis}, {Wiedner}, \& {Xilouris}}]{vanderWerf+2010}
{van der Werf}, P.~P., {Isaak}, K.~G., {Meijerink}, R., {et~al.} 2010, \aap,
  518, L42

\bibitem[{{Veilleux} {et~al.}(2013){Veilleux}, {Mel{\'e}ndez}, {Sturm},
  {Gracia-Carpio}, {Fischer}, {Gonz{\'a}lez-Alfonso}, {Contursi}, {Lutz},
  {Poglitsch}, {Davies}, {Genzel}, {Tacconi}, {de Jong}, {Sternberg}, {Netzer},
  {Hailey-Dunsheath}, {Verma}, {Rupke}, {Maiolino}, {Teng}, \&
  {Polisensky}}]{Veilleux+2013}
{Veilleux}, S., {Mel{\'e}ndez}, M., {Sturm}, E., {et~al.} 2013, \apj, 776, 27

\bibitem[{{Venemans} {et~al.}(2018){Venemans}, {Decarli}, {Walter},
  {Ba{\~n}ados}, {Bertoldi}, {Fan}, {Farina}, {Mazzucchelli}, {Riechers},
  {Rix}, {Wang}, \& {Yang}}]{Venemans+2018}
{Venemans}, B.~P., {Decarli}, R., {Walter}, F., {et~al.} 2018, \apj, 866, 159

\bibitem[{{Venemans} {et~al.}(2019){Venemans}, {Neeleman}, {Walter}, {Novak},
  {Decarli}, {Hennawi}, \& {Rix}}]{Venemans+2019}
{Venemans}, B.~P., {Neeleman}, M., {Walter}, F., {et~al.} 2019, \apj, 874, L30

\bibitem[{{Venemans} {et~al.}(2017{\natexlab{a}}){Venemans}, {Walter},
  {Decarli}, {Ba{\~n}ados}, {Carilli}, {Winters}, {Schuster}, {da Cunha},
  {Fan}, \& {Farina}}]{Venemans+2017b}
{Venemans}, B.~P., {Walter}, F., {Decarli}, R., {et~al.} 2017{\natexlab{a}},
  \apjl, 851, L8

\bibitem[{{Venemans} {et~al.}(2017{\natexlab{b}}){Venemans}, {Walter},
  {Decarli}, {Ba{\~n}ados}, {Hodge}, {Hewett}, {McMahon}, {Mortlock}, \&
  {Simpson}}]{Venemans+2017c}
{Venemans}, B.~P., {Walter}, F., {Decarli}, R., {et~al.} 2017{\natexlab{b}},
  \apj, 837, 146

\bibitem[{{Venemans} {et~al.}(2017{\natexlab{c}}){Venemans}, {Walter},
  {Decarli}, {Ferkinhoff}, {Wei{\ss}}, {Findlay}, {McMahon}, {Sutherland}, \&
  {Meijerink}}]{Venemans+2017a}
{Venemans}, B.~P., {Walter}, F., {Decarli}, R., {et~al.} 2017{\natexlab{c}},
  \apj, 845, 154

\bibitem[{{Vito} {et~al.}(2019){Vito}, {Brandt}, {Bauer}, {Gilli}, {Luo},
  {Zamorani}, {Calura}, {Comastri}, {Mazzucchelli}, {Mignoli}, {Nanni},
  {Shemmer}, {Vignali}, {Brusa}, {Cappelluti}, {Civano}, \&
  {Volonteri}}]{Vito+2019}
{Vito}, F., {Brandt}, W.~N., {Bauer}, F.~E., {et~al.} 2019, \aap, 628, L6

\bibitem[{{Volonteri}(2012)}]{Volonteri2012}
{Volonteri}, M. 2012, Science, 337, 544

\bibitem[{{Wagner} \& {Graff}(1987)}]{Wagner+1987}
{Wagner}, A.~F. \& {Graff}, M.~M. 1987, \apj, 317, 423

\bibitem[{{Walter} {et~al.}(2003){Walter}, {Bertoldi}, {Carilli}, {Cox}, {Lo},
  {Neri}, {Fan}, {Omont}, {Strauss}, \& {Menten}}]{Walter+2003}
{Walter}, F., {Bertoldi}, F., {Carilli}, C., {et~al.} 2003, \nat, 424, 406

\bibitem[{{Walter} {et~al.}(2004){Walter}, {Carilli}, {Bertoldi}, {Menten},
  {Cox}, {Lo}, {Fan}, \& {Strauss}}]{Walter+2004}
{Walter}, F., {Carilli}, C., {Bertoldi}, F., {et~al.} 2004, \apj, 615, L17

\bibitem[{{Walter} {et~al.}(2009){Walter}, {Riechers}, {Cox}, {Neri},
  {Carilli}, {Bertoldi}, {Weiss}, \& {Maiolino}}]{Walter+2009nat}
{Walter}, F., {Riechers}, D., {Cox}, P., {et~al.} 2009, \nat, 457, 699

\bibitem[{{Walter} {et~al.}(2018){Walter}, {Riechers}, {Novak}, {Decarli},
  {Ferkinhoff}, {Venemans}, {Ba{\~n}ados}, {Bertoldi}, {Carilli}, {Fan},
  {Farina}, {Mazzucchelli}, {Neeleman}, {Rix}, {Strauss}, {Uzgil}, \&
  {Wang}}]{Walter+2018}
{Walter}, F., {Riechers}, D., {Novak}, M., {et~al.} 2018, \apjl, 869, L22

\bibitem[{{Walter} {et~al.}(2011){Walter}, {Wei{\ss}}, {Downes}, {Decarli}, \&
  {Henkel}}]{Walter+2011}
{Walter}, F., {Wei{\ss}}, A., {Downes}, D., {Decarli}, R., \& {Henkel}, C.
  2011, \apj, 730, 18

\bibitem[{{Wang} {et~al.}(2019){Wang}, {Wang}, {Fan}, {Wu}, {Yang}, {Neri}, \&
  {Yue}}]{WangF+2019}
{Wang}, F., {Wang}, R., {Fan}, X., {et~al.} 2019, \apj, 880, 2

\bibitem[{{Wang} {et~al.}(2021){Wang}, {Yang}, {Fan}, {Hennawi}, {Barth},
  {Banados}, {Bian}, {Boutsia}, {Connor}, {Davies}, {Decarli}, {Eilers},
  {Farina}, {Green}, {Jiang}, {Li}, {Mazzucchelli}, {Nanni}, {Schindler},
  {Venemans}, {Walter}, {Wu}, \& {Yue}}]{Wang+2021}
{Wang}, F., {Yang}, J., {Fan}, X., {et~al.} 2021, arXiv e-prints,
  arXiv:2101.03179

\bibitem[{{Wang} {et~al.}(2010){Wang}, {Carilli}, {Neri}, {Riechers}, {Wagg},
  {Walter}, {Bertoldi}, {Menten}, {Omont}, {Cox}, \& {Fan}}]{Wang+2010}
{Wang}, R., {Carilli}, C.~L., {Neri}, R., {et~al.} 2010, \apj, 714, 699

\bibitem[{{Wang} {et~al.}(2008){Wang}, {Wagg}, {Carilli}, {Benford}, {Dowell},
  {Bertoldi}, {Walter}, {Menten}, {Omont}, {Cox}, {Strauss}, {Fan}, \&
  {Jiang}}]{Wang+2008}
{Wang}, R., {Wagg}, J., {Carilli}, C.~L., {et~al.} 2008, \aj, 135, 1201

\bibitem[{{Wang} {et~al.}(2013){Wang}, {Wagg}, {Carilli}, {Walter}, {Lentati},
  {Fan}, {Riechers}, {Bertoldi}, {Narayanan}, {Strauss}, {Cox}, {Omont},
  {Menten}, {Knudsen}, {Neri}, \& {Jiang}}]{Wang+2013}
{Wang}, R., {Wagg}, J., {Carilli}, C.~L., {et~al.} 2013, \apj, 773, 44

\bibitem[{{Wang} {et~al.}(2016){Wang}, {Wu}, {Neri}, {Fan}, {Walter},
  {Carilli}, {Momjian}, {Bertoldi}, {Strauss}, {Li}, {Wang}, {Riechers},
  {Jiang}, {Omont}, {Wagg}, \& {Cox}}]{Wang+2016}
{Wang}, R., {Wu}, X.-B., {Neri}, R., {et~al.} 2016, \apj, 830, 53

\bibitem[{{Wei{\ss}} {et~al.}(2005){Wei{\ss}}, {Downes}, {Henkel}, \&
  {Walter}}]{Weiss+2005}
{Wei{\ss}}, A., {Downes}, D., {Henkel}, C., \& {Walter}, F. 2005, \aap, 429,
  L25

\bibitem[{{Wei{\ss}} {et~al.}(2003){Wei{\ss}}, {Henkel}, {Downes}, \&
  {Walter}}]{Weiss+2003}
{Wei{\ss}}, A., {Henkel}, C., {Downes}, D., \& {Walter}, F. 2003, \aap, 409,
  L41

\bibitem[{{Wei{\ss}} {et~al.}(2010){Wei{\ss}}, {Requena-Torres}, {G{\"u}sten},
  {Garc{\'\i}a-Burillo}, {Harris}, {Israel}, {Klein}, {Kramer}, {Lord},
  {Martin-Pintado}, {R{\"o}llig}, {Stutzki}, {Szczerba}, {van der Werf},
  {Philipp-May}, {Yorke}, {Akyilmaz}, {Gal}, {Higgins}, {Marston}, {Roberts},
  {Schl{\"o}der}, {Schultz}, {Teyssier}, {Whyborn}, \& {Wunsch}}]{Weiss+2010}
{Wei{\ss}}, A., {Requena-Torres}, M.~A., {G{\"u}sten}, R., {et~al.} 2010, \aap,
  521, L1

\bibitem[{{Willott} {et~al.}(2015{\natexlab{a}}){Willott}, {Bergeron}, \&
  {Omont}}]{Willott+2015mbh}
{Willott}, C.~J., {Bergeron}, J., \& {Omont}, A. 2015{\natexlab{a}}, \apj, 801,
  123

\bibitem[{{Willott} {et~al.}(2017){Willott}, {Bergeron}, \&
  {Omont}}]{Willott+2017}
{Willott}, C.~J., {Bergeron}, J., \& {Omont}, A. 2017, \apj, 850, 108

\bibitem[{{Willott} {et~al.}(2015{\natexlab{b}}){Willott}, {Carilli}, {Wagg},
  \& {Wang}}]{Willott+2015}
{Willott}, C.~J., {Carilli}, C.~L., {Wagg}, J., \& {Wang}, R.
  2015{\natexlab{b}}, \apj, 807, 180

\bibitem[{{Yang} {et~al.}(2013){Yang}, {Gao}, {Omont}, {Liu}, {Isaak},
  {Downes}, {van der Werf}, \& {Lu}}]{Yang+2013}
{Yang}, C., {Gao}, Y., {Omont}, A., {et~al.} 2013, \apjl, 771, L24

\bibitem[{{Yang} {et~al.}(2019{\natexlab{a}}){Yang}, {Gavazzi}, {Beelen},
  {Cox}, {Omont}, {Lehnert}, {Gao}, {Ivison}, {Swinbank}, {Barcos-Mu{\~n}oz},
  {Neri}, {Cooray}, {Dye}, {Eales}, {Fu}, {Gonz{\'a}lez-Alfonso}, {Ibar},
  {Micha{\l}owski}, {Nayyeri}, {Negrello}, {Nightingale}, {P{\'e}rez-Fournon},
  {Riechers}, {Smail}, \& {van der Werf}}]{Yang+2019}
{Yang}, C., {Gavazzi}, R., {Beelen}, A., {et~al.} 2019{\natexlab{a}}, \aap,
  624, A138

\bibitem[{{Yang} {et~al.}(2020{\natexlab{a}}){Yang}, {Gonz{\'a}lez-Alfonso},
  {Omont}, {Pereira-Santaella}, {Fischer}, {Beelen}, \& {Gavazzi}}]{Yang+2020}
{Yang}, C., {Gonz{\'a}lez-Alfonso}, E., {Omont}, A., {et~al.}
  2020{\natexlab{a}}, \aap, 634, L3

\bibitem[{{Yang} {et~al.}(2017){Yang}, {Omont}, {Beelen}, {Gao}, {van der
  Werf}, {Gavazzi}, {Zhang}, {Ivison}, {Lehnert}, {Liu}, {Oteo},
  {Gonz{\'a}lez-Alfonso}, {Dannerbauer}, {Cox}, {Krips}, {Neri}, {Riechers},
  {Baker}, {Micha{\l}owski}, {Cooray}, \& {Smail}}]{Yang+2017}
{Yang}, C., {Omont}, A., {Beelen}, A., {et~al.} 2017, \aap, 608, A144

\bibitem[{{Yang} {et~al.}(2016){Yang}, {Omont}, {Beelen},
  {Gonz{\'a}lez-Alfonso}, {Neri}, {Gao}, {van der Werf}, {Wei{\ss}}, {Gavazzi},
  {Falstad}, {Baker}, {Bussmann}, {Cooray}, {Cox}, {Dannerbauer}, {Dye},
  {Gu{\'e}lin}, {Ivison}, {Krips}, {Lehnert}, {Micha{\l}owski}, {Riechers},
  {Spaans}, \& {Valiante}}]{Yang+2016}
{Yang}, C., {Omont}, A., {Beelen}, A., {et~al.} 2016, \aap, 595, A80

\bibitem[{{Yang} {et~al.}(2019{\natexlab{b}}){Yang}, {Venemans}, {Wang}, {Fan},
  {Novak}, {Decarli}, {Walter}, {Yue}, {Momjian}, {Keeton}, {Wang},
  {Zabludoff}, {Wu}, \& {Bian}}]{YangJ+2019}
{Yang}, J., {Venemans}, B., {Wang}, F., {et~al.} 2019{\natexlab{b}}, \apj, 880,
  153

\bibitem[{{Yang} {et~al.}(2020{\natexlab{b}}){Yang}, {Wang}, {Fan}, {Hennawi},
  {Davies}, {Yue}, {Banados}, {Wu}, {Venemans}, {Barth}, {Bian}, {Boutsia},
  {Decarli}, {Farina}, {Green}, {Jiang}, {Li}, {Mazzucchelli}, \&
  {Walter}}]{YangJ+2020}
{Yang}, J., {Wang}, F., {Fan}, X., {et~al.} 2020{\natexlab{b}}, \apjl, 897, L14

\bibitem[{{Zhao} {et~al.}(2013){Zhao}, {Lu}, {Xu}, {Gao}, {Lord}, {Howell},
  {Isaak}, {Charmandaris}, {Diaz-Santos}, {Appleton}, {Evans}, {Iwasawa},
  {Leech}, {Mazzarella}, {Petric}, {Sand ers}, {Schulz}, {Surace}, \& {van der
  Werf}}]{Zhao+2013}
{Zhao}, Y., {Lu}, N., {Xu}, C.~K., {et~al.} 2013, \apjl, 765, L13

\bibitem[{{Zhao} {et~al.}(2016){Zhao}, {Lu}, {Xu}, {Gao}, {Lord},
  {Charmandaris}, {Diaz-Santos}, {Evans}, {Howell}, {Petric}, {van der Werf},
  \& {Sand ers}}]{Zhao+2016}
{Zhao}, Y., {Lu}, N., {Xu}, C.~K., {et~al.} 2016, \apj, 819, 69

\end{thebibliography}

\end{document}